\documentstyle[11pt]{report}

\renewcommand{\theequation}{\thesection.\arabic{equation}}

\newlength{\extraspace}
\newlength{\extraspaces}
\newcounter{dummy}

\newcommand{\baa}{\begin{eqnarray}
\addtocounter{equation}{1}
\setcounter{dummy}{\value{equation}}
\setcounter{equation}{0}
\renewcommand{\theequation}{\thesection.\arabic{dummy}\alph{equation}}
\addtolength{\abovedisplayskip}{\extraspaces}
\addtolength{\belowdisplayskip}{\extraspaces}
\addtolength{\abovedisplayshortskip}{\extraspace}
\addtolength{\belowdisplayshortskip}{\extraspace}}

\newcommand{\eaa}{\end{eqnarray}
\setcounter{equation}{\value{dummy}}
\renewcommand{\theequation}{\thesection.\arabic{equation}}}

\newcommand{\be}{\begin{equation}
\addtolength{\abovedisplayskip}{\extraspaces}
\addtolength{\belowdisplayskip}{\extraspaces}
\addtolength{\abovedisplayshortskip}{\extraspace}
\addtolength{\belowdisplayshortskip}{\extraspace}}
\newcommand{\ee}{\end{equation}}

\newcommand{\ba}{\begin{eqnarray}
\addtolength{\abovedisplayskip}{\extraspaces}
\addtolength{\belowdisplayskip}{\extraspaces}
\addtolength{\abovedisplayshortskip}{\extraspace}
\addtolength{\belowdisplayshortskip}{\extraspace}}
\newcommand{\ea}{\end{eqnarray}}

\newcommand{\bd}{\begin{displaymath}
\addtolength{\abovedisplayskip}{\extraspaces}
\addtolength{\belowdisplayskip}{\extraspaces}
\addtolength{\abovedisplayshortskip}{\extraspace}
\addtolength{\belowdisplayshortskip}{\extraspace}}
\newcommand{\ed}{\end{displaymath}}

\newcommand{\ban}{\begin{eqnarray*}
\addtolength{\abovedisplayskip}{\extraspaces}
\addtolength{\belowdisplayskip}{\extraspaces}
\addtolength{\abovedisplayshortskip}{\extraspace}
\addtolength{\belowdisplayshortskip}{\extraspace}}

\newcommand{\nonu}{\nonumber \\[.5mm]}

\newcommand{\deel}[2]{{\textstyle{#1 \over #2}}}
\newcommand{\hf}{{\textstyle{1\over 2}}}
\newcommand{\hv}{{\textstyle{1\over 4}}}

\newcommand{\re}{\mbox{I}\!\mbox{R}}

\newtheorem{thm}{Theorem}
\newtheorem{lmm}{Lemma}
\newtheorem{exam}{Example}
\newcommand{\bth}{\begin{thm}}
\newcommand{\eth}{\end{thm}}
\newcommand{\bl}{\begin{lmm}}
\newcommand{\el}{\end{lmm}}
\newcommand{\bex}{\begin{exam}}
\newcommand{\eex}{\end{exam}}
\newcommand{\hj}{\hat{J}}
\newcommand{\Whj}[1]{W(\hat{J}^{#1})}
\newcommand{\www}[4]{\deel{#1}{#2}\hj^{#3}\hj^{#4}}

\def\inbar{\,\vrule height1.5ex width.4pt depth0pt}
\font\rms=cmr12 at 12pt
\def\ce{\relax\ifmmode\mathchoice
{\hbox{$\inbar\kern-.3em{\rm C}$}}
{\hbox{$\inbar\kern-.3em{\rm C}$}}
{\lower.9pt\hbox{\rms $\inbar\kern-.3em{\rm C}$}}
{\lower1.2pt\hbox{\rms $\inbar\kern-.3em{\rm C}$}}
\else{$\inbar\kern-.3em{\rm C}$}\fi}
\font\cmss=cmss12 \font\cmsss=cmss12 at 12pt
\def\ze{\relax\ifmmode\mathchoice
{\hbox{\cmss Z\kern-.4em Z}}{\hbox{\cmss Z\kern-.4em Z}}
{\lower.9pt\hbox{\cmsss Z\kern-.4em Z}}
{\lower1.2pt\hbox{\cmsss Z\kern-.4em Z}}\else{\cmss Z\kern-.4em Z}\fi}

\newcommand{\tr}{\mbox{Tr}}

\newcommand{\actie}[1]{\deel{1}{2\pi}\int d^2z \, }

\newcommand{\mats}[9]{\left( \begin{array}{ccc}
				#1 & #2 & #3 \\
				#4 & #5 & #6 \\
				#7 & #8 & #9
                             \end{array} \right) }

\newcommand{\mat}[4]{\left( \begin{array}{cc}
				#1 & #2 \\
				#3 & #4
                             \end{array} \right) }

\newcommand{\tripel}{{({\cal G},{\cal L},\chi)}}

\newcommand{\np}[1]{Nucl. Phys. {\bf B#1}}
\newcommand{\cmp}[1]{Comm. Math. Phys. {\bf #1}}

\newcommand{\plb}[1]{Phys. Lett. {\bf B#1}}

\newcommand{\ad}[1]{\mbox{\rm ad}_{#1}}

\renewcommand{\theequation}{\thesection.\arabic{equation}}
\newcommand{\sectiona}[1]{\setcounter{equation}{0}\section{#1}}

\newcommand{\al}{\alpha}
\newcommand{\bt}{\beta}

\newcommand{\dl}{\delta}
\newcommand{\ep}{\epsilon}

\newcommand{\lm}{\lambda}

\newcommand{\sg}{\sigma}

\newcommand{\ph}{\phi}

\newcommand{\om}{\omega}
\newcommand{\Om}{\Omega}

\newcommand{\nn}{\nonumber}

\newcommand{\bea}{\begin{eqnarray}}
\newcommand{\eea}{\end{eqnarray}}
\newcommand{\p}{\partial}

\newcommand{\W}{\cal W}
\newcommand{\G}{\cal G}
\begin{document}

\thispagestyle{empty}
\begin{flushright}
{\sc ITP-SB}-95-07\\
ITFA-03-2/95
\end{flushright}
\vspace{1cm}
\setcounter{footnote}{3}
\begin{center}
{\LARGE\sc{Non-linear Finite $W$-Symmetries and Applications in
    Elementary Systems}}\\[1cm]

\sc{Jan de Boer\footnote{e-mail: deboer@max.physics.sunysb.edu} }
\\
{\it Institute for Theoretical Physics\\
State University of New York at Stony Brook\\
Stony Brook, NY 11794-3840, USA}\\[4mm]
\sc{Frederique Harmsze\footnote{e-mail: harmsze@phys.uva.nl}}\\
{\it Institute for Theoretical Physics\\
Valckenierstraat 65\\
1018 XE Amsterdam, the Netherlands}\\[4mm]
and\\[4mm]
\sc{Tjark Tjin\footnote{e-mail: tjin1@ksla.nl} }\\
{\it Koninklijke/Shell-Laboratorium, Amsterdam \\
P.O.~Box 38000\\
1030 BN Amsterdam, The Netherlands}\\[10mm]

{\sc Abstract}\\[2mm]
\end{center}

\noindent
In this paper it is stressed that there is no {\em physical} reason for
symmetries to be linear and that Lie group theory is therefore too
restrictive. We illustrate this with some simple examples. Then we give a
readable review on the theory finite $W$-algebras, which is an important
class of non-linear symmetries. In particular, we discuss both the
classical and quantum theory and elaborate on several aspects of their
representation theory. Some new results are presented. These include
finite $W$ coadjoint orbits, real forms and unitary representation of
finite $W$-algebras and Poincare-Birkhoff-Witt theorems for
finite $W$-algebras. Also we present some new finite $W$-algebras
that are not related to $sl(2)$ embeddings. At the end of the paper
we investigate how one could construct physical theories, for example
gauge field theories, that are based on non-linear algebras.

\vfill

\newpage

\tableofcontents

\vspace{12mm}

\chapter{Introduction}
The notion of symmetry is one of the most fundamental concepts in physics.
Relativity theory for example is based on symmetry, namely Lorentz
invariance. This symmetry is the mathematical expression of the postulate
that physical laws are the same for all inertial observers. In particle
physics the principles of symmetry provide a powerful overall framework.
This allows to successfully classify and interpret an overwhelming amount
of experimental data concerning the spectrum of elementary particles.
But moreover, the interactions between elementary particles are
completely determined through the
principle of {\em local gauge invariance}, a central paradigm of
modern particle theory. The highly successful standard model,
unifying the electro-weak and strong interactions between
elementary particles, is based on the local gauge group
$U(1)\times SU(2) \times SU(3)$.

The mathematical theory concerning symmetry transformations in a
physical system is group theory. It has been developed
independently in different fields of mathematics. In the context of
algebraic equations the idea of groups was used already by Lagrange in 1771,
though the name `group' was only introduced in 1830 by Galois.
The second
area in which it appeared was number theory, with
Euler (1761) and Gauss (1801) as the most important contributors.
The concept found its place in geometry in the middle of the 19th
century, when Klein proposed it as a tool to classify certain new
geometrical structures that had been discovered at the time.
At the end of the 19th century it was realized, that these three group
concepts were the same and this insight led to the formation of
modern abstract group theory by Lie (1870). The name of Lie has been
associated to continuous linear groups, now called Lie groups.
It was Cartan who subsequently almost fully developed this subject,
though he was rather
isolated for a period of about thirty years.

The simplest symmetries are those in which the physical system
is symmetric under a finite number of transformations or when it
is invariant under displacement in a finite number of `directions'.
The natural mathematical structures describing such symmetries are
discrete groups and finite-dimensional Lie groups respectively.

Group theory has been introduced in physics in the 1920s
mostly through the work of Weyl and Wigner. Apart from the
obvious description of symmetries in crystals, they realized that
group theory is of the utmost importance in quantum physics.
Weyl writes in \cite{Weyl} on group theory in
quantum physics: "It reveals essential features which are not
contingent on a special form of the dynamics laws nor on special
assumptions concerning the forces involved. We may well
expect that it is just this part of quantum physics which is
most certain of a lasting place."

Lie groups and their Lie algebras have a wide range of
applications in physics. In fact most symmetry considerations of
physical systems have up to recently been based on the application
of the theory of Lie groups and algebras. This is true for the
Lorentz group in the theory of special relativity, as well as for the
groups underlying gauge theories of the fundamental interactions
between the basic constituents of matter. It is worth noting that
these theories, based on linear symmetries, in fact have highly
non-linear dynamics.

In mathematics the development of the theory of Lie groups was
continued with the work of Chevalley (1950), Serre (1966) and Dynkin.
In 1967 Kac \cite{Kac} and Moody independently generalized the
theory of finite-dimensional Lie algebras to the infinite-dimensional
case. These so-called affine Lie algebras have found remarkable
applications in two dimensional field theory and
string theory \cite{Goddard&Olive}.

For a long time all efforts to develop the theory of symmetry in
physics were restricted to the {\em linear} case, i.e. Lie groups and
Lie algebras. However, it was realized recently that the `Lie algebra'
might be too narrow a concept from the physical point of view.
Unfortunately the extensive machinery developed for the analysis of
linear symmetries largely breaks down in the non-linear domain.
Nevertheless, in the eighties, led by developments in string
theory and solvable models, non-linear symmetries gained
importance in physics. In fact this may be seen as part of the
boom in the field of non-linear science in mathematics and physics which
started roughly in the middle of the nineteenth century.
Non-linear symmetries seem a logical next step in this field
which has led to the discovery of theories concerning for
example solitary waves (solitons), chaos and turbulence,
but also non-linear gauge theories.
\vspace{3mm}

One type of non-linear algebra that has received much attention in
recent years are the so called quantum groups. They are obtained from
ordinary Lie algebras by deforming (quantizing) the co-Poisson
structure present on any Lie algebra, or, equivalently, by
deforming the space of functions on the group manifold.
These algebras have some very interesting applications in
conformal field theory but it is not this type that we will be
concerned with in this paper.

Surprisingly enough, when in 1985 Zamolodchikov \cite{Zamol}
took up the subject of non-linear algebras, he
considered {\em infinite-dimensional} W-algebras.
In his work on conformal field theory, he generalized the well-known
Virasoro algebra, which is the infinite-dimensional Lie algebra
associated with the conformal symmetries in two-dimensional space-time.
What he found were {\em non-linear} infinite-dimensional algebras,
that were called W-algebras. Subsequently W-algebras were studied
in the context of string theory, the theory of integrable systems
and the theory of two-dimensional critical phenomena.
For more details we refer the reader to the
reviews \cite{Bouwknegt,thesis,mythesis} on W-algebras in
conformal field theory.

One important question was and still is the classification of W-algebras,
in other words, to make a complete list of all W-algebras.
The most profitable approach to date is to apply the Drinfeld-Sokolov
construction \cite{Drinfeld&Sokolov}
to derive $W$-algebras starting from affine Lie  algebras.
It has been shown in \cite{BaisTjinvanDriel} that this
construction gives rise to a large class of W-algebras.
This relation between W-algebras and affine Lie algebras
in principle enables one construct $W$-algebra theory
from the theory of affine Lie algebras.

At this point the question arose whether finite-dimensional
analogues of the infinite-dimensional W-algebras exist as well.
The relevance of this question is apparent if one considers
that the theory of the infinite-dimensional so called-loop groups
and algebras can be derived from the finite-dimensional Lie groups and
algebras underlying them. Loop groups are special cases of
infinite-dimensional groups of smooth maps from some
spacetime manifold $X$ to a finite-dimensional Lie group $G$,
namely for $X=S^1$. The study infinite-dimensional groups of
 smooth maps is a natural consequence of the combination of
symmetry principles with locality or causality.

The group multiplication in such a group is just pointwise
multiplication, i.e. if $f,g\in Map(X,G)$ and $a\in G$
then $(f.g)(a)=f(a)g(a)$.
In quantum field theory groups of the
form $Map(X,G)$ and their Lie algebras $Map(X,g)$ (where ${\cal G}$
is the Lie algebra of $G$) arise essentially in two different ways:
through the principle of local gauge invariance,
which is at the heart of modern high energy physics, and
through the theory of current groups and algebras.

Unfortunately for generic manifolds $X$ surprisingly little is
known about the group $Map(X,G)$. Especially the representation
theory of these groups is still almost unexplored. The
exception to this is the case $X=S^1$ we mentioned above,
where $Map(S^1,G)$ and $Map(S^1,g)$ are called `loop groups'
and `loop algebras'. Loop groups and algebras arise
in simplified models of quantum field theory in which space
is taken to be 1-dimensional and therefore also in string models
of elementary particles. The study of loop groups and algebras is
much simpler than when $X$ is some more complicated manifold.
This is caused by the fact that they behave much like the
ordinary finite-dimensional Lie groups and algebras that underlie them.
This remarkable fact makes knowledge of the finite-dimensional theory
essential for the study of the infinite-dimensional theory.

For the infinite-dimensional W-algebras it was not entirely
clear what the finite algebras underlying $W$-algebras were and
whether there was a finite version of $W$-theory at all.
Considering the way in which $W$-algebras were first introduced
into physics they really don't seem to have any relation to
the theory of loop groups and algebras. Nevertheless the
theory of W-algebras does have a finite counterpart as has been
shown in \cite{Finite,deBoerTjin}.In fact the finite theory is
remarkably rich and, as with loop algebras, contains already
several of the essential features of infinite-dimensional $W$-algebra theory.

The main objective of this paper is twofold. On the one hand we wish to
convey to the reader our view that non-linear symmetries are not only
interesting in relation to theories that have a mathematical
sophistication comparable to that of string theory, but that they play an
important role throughout physics. In fact, finitely generated
non-linear symmetries already show up in very elementary and
famous physical systems.

Our second objective is to give a readable account of the classical
and quantum theory of finite $W$-algebras and to present some new results.

The outline of the paper is as follows. After some general remarks
 on dynamical systems and symmetries, we illustrate by means of
known examples, that {\em many basic physical systems have non-linear
symmetry algebras}. In particular, the symmetry algebras of the
two-dimensional anisotropic harmonic oscillator with frequency
ratio $2:1$ and the Kepler problem are shown to be finite W-algebras.

Finite W-algebras can be constructed from finite Lie algebras by a
procedure resembling the constraint formalism. Precisely how this
works will be the subject of chapter 3. In this chapter we construct
and develop the classical theory of finite $W$-algebras. Also we
show that many finite $W$-algebras contain known Lie algebras as subalgebras.
The question therefore naturally arizes whether there exist
non-linear extensions of $su(3)\times su(2) \times u(1)$.
The answer to this question turns out to be affirmative.

Having developed the classical theory we turn to the quantum case
and show how to quantize finite $W$-algebras using the BRST formalism.
In some specific cases we explicitly construct the  BRST cohomologies.
Next we consider the representation theory of quantum finite $W$-algebras.
In order to define unitary highest weight representations
for finite $W$-algebras it is necessary to consider real forms and
Poincare-Birkhoff-Witt theorems for these algebras. Having developed
this part of the theory we describe some results on Kac-determinants
and character formulas that were recently conjectured.
Finally, we present some ideas about possible constructions of
theories with finite $W$-symmetries. In particular, we show how
finite $W$-algebras are realized in one-dimensional generalized
Toda theories. We conlude with yet another simple quantum
mechanical example, which has a finite $W$-algebra as a spectrum
generating algebra.

\chapter{Symmetries in Simple Physical Systems}
In this chapter we first briefly review some basic facts
concerning algebras of
conserved quantities in elementary mechanical systems. It is explicitly
stressed
that there is no reason, neither physical nor mathematical,
why these algebras have
to be linear. In fact  they are non-linear in general. We then
illustrate this
explicitly in some examples, namely the harmonic oscillator
and the Kepler problem.
\sectiona{Symmetry algebras}
During the motion of a mechanical system, the generalized coordinates $q^i$
and the velocities $\dot{q}^i$ vary in time. Nevertheless, there exist
functions of these quantities whose values do not change but depend only on
the initial conditions. Such functions are called `integrals of motion'.
Integrals of motion which do not depend explicitly on time are called
conserved
quantities. From now on we will restrict ourselves to conserved quantities.

Noether's theorem states that conserved quantities are related to symmetries.
Symmetry transformations leave the action invariant, i.e. $\dl S = 0$.
Therefore the variation of the Lagrangian can only be equal to a total
derivative $\dl L = \frac{d}{dt}  \Lambda(q,t)$.
Let us consider the symmetry transformation $q' = q + \dl q$. Then we find that
\be
\dl L = \frac{\p L}{\p q} \dl q + \frac{\p L}{\p \dot{q}} \dl \dot{q} =
\frac{d}{dt}  \Lambda(q,t) ,
\ee
if the Lagrangian only depends implicitly on time.
Partial integration yields
\be
 \frac{d}{dt} \left( \Lambda(q,t)-\frac{\p L}{\p \dot{q}} \dl q \right)
\dl L = \left( \frac{\p L}{\p q} - \frac{d}{dt} \frac{\p L}{\p \dot{q}} \right)
\dl q .
\ee
The right hand side is equal to zero because of the Euler-Lagrange equations.
Therefore if $q$ satisfies the equations of motion, then
\be
\frac{d}{dt} \left( \Lambda(q,t)-\frac{\p L}{\p \dot{q}} \dl q \right) = 0.
\ee
The expression between the brackets is a conserved quantity.
This is Noether's theorem.

In the above we considered a general transformation $q'(q)$. More particularly,
this
transformation can consist of a set of independent transformations in
different `directions', labeled by a parameter $\ep_a$:
\be
\dl q^i = \ep_a \dl_a q^i
\ee
The conserved quantities $Q_a$, also called `Noether charges', that can be
associated to all these symmetry transformations are defined as
\be
\label{Q_a}
Q_a = \Lambda_a (q,t)-\frac{\p L}{\p \dot{q}} \dl_a q,
\ee
such that
\be
\frac{d Q_a}{dt} = 0.
\ee

Now, classically, the time derivative of a quantity
$Q$ is given by
\be
\label{dt}
\frac{d Q}{dt} = \{H,Q\},
\ee
where the Poisson bracket of the functions $f$ and $g$ is defined by
\be {}
\{f,g\} =
 \frac{\p f}{\p p_i}
\frac{\p g}{\p q^i} -
\frac{\p f}{\p q^i} \frac{\p g}{\p p_i}.
\ee
Conserved quantities are therefore characterized by the fact that they
Poisson-commute with the Hamiltonian:
\be
\{Q,H\} = 0.
\ee
Let $\{Q_a\}$ be a set of independent
\footnote{Two conserved quantities $Q_1(\vec{q},\vec{p})$ and
$Q_2(\vec{q},\vec{p})$ are called  independent if the vectors $\left( \frac{\p
Q_1}{\p \vec{q}}, \frac{\p Q_1}{\p \vec{p}} \right)$ and $ \left( \frac{\p
Q_2}{\p \vec{q}}, \frac{\p Q_2}{\p \vec{p}} \right)$ are
linearly independent.}
conserved quantities, i.e. $\{Q_a,H\} = 0$. It is clear that any polynomial
$P(\{Q_a\})$ in the conserved quantities is conserved as well. Also it follows
from the Jacobi identity
\be
\{f,\{g,h\}\} + \{g, \{h,f\}\} + \{h,\{f,g\}\} = 0,
\ee
that
\be
\{H, \{Q_a,Q_b\}\} = 0.
\ee
This means that the Poisson bracket of two conserved quantities is
conserved. We call the set $\{Q_a\}$ of conserved quantities {\em closed} if
\be
\label{Poissonalgebra}
\{Q_a,Q_b\} = P_{ab},
\ee
where $P_{ab}$ is some function of the quantities $\{Q_a\}$. If the Poisson
bracket contains some conserved quantities which are {\em not} present in the
set $\{Q_a\}$, i.e. if this set is not closed, we can add these quantities
to the set and thus {\em make} it closed.

The Poisson algebra (\ref{Poissonalgebra}) of conserved quantities will from
now on be called the {\em classical symmetry algebra}. The quantities
$\{Q_a\}$ are called the {\em generators} of the algebra. If an algebra has a
finite number of generators, it is said to be `finitely generated'.
The transformations associated with the Noether charge $Q_a$ of the coordinates
$q^i$ and momenta $p_i$
are of course given by
\bea
\dl_a q^i &=& \{q^i,Q_a\} \nn \\
\dl_a p_i &=& \{p_i,Q_a\}.
\eea

In quantum mechanics, the equivalent of equation (\ref{dt}) is
\be
\frac{d Q}{d t} = \frac{i}{\hbar} [H,Q],
\ee
where $Q$ and $H$ are now operators. As in the classical case, we
define the symmetry algebra as the set of independent operators $\{Q_a\}$
which commute with the Hamiltonian and have the property that the commutation
relations of $Q_a$ and $Q_b$ can again be expressed in terms of the operators
$\{Q_a\}$.

As $[Q_a,H] = 0$ for a conserved quantity, the operator $Q_a$ will
transform eigenstates of $H$ into (possibly different) eigenstates with equal
energy. The Hilbert space of the system therefore decomposes into a
direct sum of irreducible representations of the symmetry algebra
generated by $\{Q_a\}$.

In the next section we shall introduce {\em linear} and {\em non-linear}
symmetries using
some very simple and well-known physical systems.
\sectiona{Harmonic oscillators.}
Harmonic oscillators constitute a category of very basic systems in physics.
They show up in virtually all problems with a finite or infinite number of
degrees of freedom. The reason is that, upon linearizing a generic dynamical
problem
harmonic oscillators approximate any arbitrary potential in the neighbourhood
of a stable
equilibrium position, describing for example small vibrations of an atom in a
crystalline lattice or a nucleon in a nucleus. On the other hand the
behavior of most continuous physical systems, such as the vibrations of
an elastic medium or the electromagnetic field in a cavity, can be
described as a superposition of an infinite number of harmonic oscillators.
In this section the idea of both linear and non-linear symmetries will be
illustrated using these elementary systems.

\subsection{The isotropic case}
Let us consider a particle of mass $M$ moving in a quadratic potential
in the $(x_1,x_2)$-plane.
The Hamiltonian of this system is given by
\be
H = \frac{1}{2M} ( p_1^2 + p_2^2 ) + \frac{1}{2}M \om^2 (x_1^2 + x_2^2),
\ee
where $\om$ is the angular frequency of the oscillator.
In the usual quantum description of this system one introduces so called
`raising and `lowering'
operators
\bea
a^{\dagger} &=& \sqrt{\frac{M \om}{2 \hbar} } x_1 -
i \sqrt{ \frac{1}{2M\om \hbar} } p_1 \hspace{0.5cm}
a =  \sqrt{ \frac{M \om}{2 \hbar} } x_1 +
i \sqrt{ \frac{1}{2M\om \hbar} } p_1 \nn \\
b^{\dagger} &=& \sqrt{ \frac{M \om}{2 \hbar} } x_2 -
i \sqrt{ \frac{1}{2M\om \hbar} } p_2 \hspace{0.5cm}
b = \sqrt{ \frac{M \om}{2 \hbar} } x_2 +
i \sqrt{ \frac{1}{2M\om \hbar} } p_2,
\eea
where $\hbar$ is Planck's constant.
 From the canonical commutation relations between coordinates and momenta
$[x_i,p_j] = i \hbar \dl_{ij}$, one can easily derive
\be \label{commrel}
[a,a^{\dagger}] = [b,b^{\dagger}] = 1,
\ee
and all other commutators are zero.

In terms of raising and lowering operators the Hamiltonian
reads $H = \left( a^{\dagger} a + b^{\dagger} b + 1 \right) \hbar \om$.
The Hilbert space of the system is spanned by the states
\be
|p,q> = \left( a^{\dagger} \right)^p \left( b^{\dagger} \right)^q |\Om>,
\ee
where $p,q$ are nonnegative integers and $|\Om>$ is the `ground state'
(which has the property that $a|\Om> = b|\Om> = 0)$. The energy of the
eigenstate $|p,q>$ is $E_{p,q} = (p + q + 1) \hbar \om$.  From this we see
that the states $|p - r,r>$, where $r = 0,1,...,p$ all have the same
energy, i.e. the energy eigenvalue $E_{p,q}$ has a $(p + q + 1)$-fold
degeneracy. This leads one to conjecture the existence of a symmetry
supplying extra quantum numbers and transforming eigenstates with the
same energy into each other. We shall now discuss this symmetry. For
notational convenience we take from now on $M = \om = \hbar = 1$.

Consider the operators
\be
\label{S}
S_+ = a b^{\dagger}; \hspace{1cm}
S_- = a^{\dagger} b; \hspace{1cm}
S_0 = b^{\dagger}b - a^{\dagger}a.
\ee
These quantities are conserved as is expressed by
the equation
$i \p_t S_i = [S_i,H] = 0$
for $i = \pm, 0$.
The commutation relations between the operators $S_i$ themselves can be
easily calculated using (\ref{commrel}) and read
\be \label{su(2)}
[S_0,S_{\pm}] = \pm 2 S_{\pm}; \hspace{1cm}
[S_+,S_-] = S_0.
\ee
Notice that the commutation relations of the operators $S_i$ are again
expressions in terms of $S_i$, so the algebra is closed. Furthermore these
expressions are {\em linear}, such that the algebra is linear. In fact, the
commutator algebra (\ref{su(2)}) is nothing but $su(2)$, the simplest example
of a non-abelian simple Lie algebra.

The theory of Lie algebras is well known and extensively described in the
literature.
Lie algebras can be related to symmetry groups, called {\em Lie groups}, by an
exponential map. All elements $g$ in the component connected to the unit
element of a Lie group can be written as $g = exp(\al^a t_a)$,
where $t_a$ are the generators of the Lie algebras,
$\al^a$ are numbers and summation over the index $a$ is understood.

The action of the generators $S_+$ and $S_-$ of the $su(2)$ symmetry can
be interpreted as follows. The state of the (quantum mechanical)
particle in the plane is composed of oscillations in two
directions. The operator $S_+$ decreases the oscillation in the $x_2$-direction
and increases oscillation in the other one. It can continue this action, until
the state of the particle consists only of oscillation in the
$x_1$-direction.
The operator $S_-$ in its turn, squeezes the orbit of the particle towards
oscillation in the $x_2$-direction.

What we can conclude from this section, is that the isotropic
oscillator in two dimensions has $su(2)$ symmetry, which is larger
than the obvious $so(2)$ rotation symmetry in the $xy$ plane.
Similarly it can be shown that the symmetry of an n-dimensional
isotropic harmonic oscillator is the Lie algebra $su(n)$ (or
$u(n)$, if one also views the Hamiltonian itself as part of the
symmetry algebra).

\subsection{The anisotropic case}
\label{aniso}
We will now consider a slightly more complicated case: the
two-dimensional {\em anisotropic}
harmonic oscillator.
The Hamiltonian of the anisotropic oscillator is given by
\be
H = \frac{1}{2M} ( p_1^2 + p_2^2) + \frac{1}{2}M \om_1^2 x_1^2+
\frac{1}{2}M \om_2^2 x_2^2.
\ee
Again we will take $M = \hbar = 1$.
In order to have degeneration in the energy levels of $E$, we take
$\om_1 = \frac{1}{m}$ and $\om_2 = \frac{1}{n}$, where $m$ and $n$ are
positive integers.
In terms of the raising and lowering operators, which satisfy the commutation
relations (\ref{commrel}),
the Hamiltonian reads
\be
H =  \frac{1}{m} \left( a^{\dagger}a + \frac{1}{2} \right) +
\frac{1}{n} \left(b^{\dagger}b + \frac{1}{2} \right).
\ee
If we consider the analogues of the operators (\ref{S}) in this system, we
find that
the generators $S_{\pm}$ no longer commute with the Hamiltonian $H$.
In other words, $S_{\pm}$ are not conserved. Therefore the algebra $su(2)$
is {\em not} a symmetry algebra of the anisotropic oscillator. \\
\vspace{3mm}

In order to arrive at the true symmetry algebra of the anisotropic oscillator,
we
consider, like \cite{Hill} and \cite{Grieken}, the following operators
that
{\em do} commute with $H$:
\bea
\label{jtilde}
\tilde{j}_+ &=& a^m (b^{\dagger})^n; \hspace{1cm}
\tilde{j}_- = (a^{\dagger})^m b^n \\ \nn
\tilde{j}_0 &=& \frac{1}{n} \left( b^{\dagger}b + \frac{1}{2} \right) -
\frac{1}{m} \left ( a^{\dagger}a + \frac{1}{2} \right).
\eea
These are the generators of the symmetry algebra of the anisotropic
quantum
harmonic oscillator with frequencies $m$ and $n$ positive integers, as was
shown in \cite{Grieken}.

For simplicity let us consider the case that the frequencies have the fixed
value $m = 2$ and $n = 1$.
Calculation of the commutation relations of the generators (\ref{jtilde})
for these values of $m$ and $n$ produces:
\bea
[\tilde{j}_i,H] &=& 0 \hspace{1cm} {}
[\tilde{j}_0, \tilde{j}_{\pm}] = {\pm} 2 \tilde{j}_{\pm} \nn \\ {}
[\tilde{j}_+,\tilde{j}_-] &=& -3 \tilde{j}_0^2 + 2H\tilde{j}_0 + H^2 -
\frac{3}{4},
\eea
with $i = 0, \pm$. After an invertible basis transformation, given by
\be
j_+ = \frac{1}{\sqrt{3}} \tilde{j}_+;
\hspace{1cm}
j_- = \frac{-1}{\sqrt{3}} \tilde{j}_-;  \hspace{1cm}
j_0 = \tilde{j}_0 - \frac{1}{3} H,
\ee
we obtain
\be
\label{W_3^(2)}
[H,j_i] = 0 \hspace{1cm}
[j_0,j_{\pm}] = \pm 2 j_{\pm} \hspace{1cm}
[j_+,j_-] = j_0^2 + C
\ee
where $C = \frac{1}{4} - \frac{4}{9}H^2$.
This commutator algebra is known in the literature as $W_3^{(2)}$
\cite{Finite}.
We conclude that the symmetry algebra of a two dimensional
anisotropic harmonic
oscillator with a frequency ratio $m:n = 2:1$ is the non-linear finite
W-algebra $W_3^{(2)}$. We will return to this algebra
later in the paper. More generally, if we take $m$ and $n$ arbitrary positive
integers we find that the symmetry algebra is non-linear whenever $m
\neq n$ \cite{Grieken}.

$W_3^{(2)}$ is an example of a {\em non-linear algebra}. The commutators can
not be written as linear combinations of the generators, but as linear
combinations of products of generators. As it has a finite set of
generators, it is finitely-generated.

In contrast to Lie algebras, one cannot associate a finite group to
a non-linear algebra. The only group $G$ which can be obtained with an
exponential map from a non-linear algebra is generated not only by the
generators of this algebra, but also by their polynomials and therefore
is infinitely generated. Only then is the requirement met,
that $\forall g_i \in G g_1g_2 = g_3$, because of the
Campbell-Baker-Hausdorff formula for the multiplication of
exponential maps. While the elements of an algebra give rise to
infinitesimal transformations, group elements correspond to finite
symmetry transformations.
Consequently we cannot easily
tell in the non-linear case how the transformation works
globally, though we know the infinitesimal transformations on the local level.
 From the latter one can in principle derive a differential equation
whose solution for finite time corresponds to transformations of the
`finite $W$-group', but these differential equations are typically
non-linear and very hard to solve explicitly.

Though the problem we considered is an elementary and a linear one, meaning
that the equations of motion are linear, its symmetry algebra turns
out to be non-linear. Surprisingly enough the study of these non-linear
symmetry algebras has only been taken up recently. Most attention has
focused on non-linear extensions of $su(2)$, see for example
\cite{Ro,Smith,moresu2-1,moresu2-2,moresu2-3,moresu2-4,moresu2-5,moresu2-6,moresu2-7}.

\sectiona{Coulomb potential}
A large class of well-known dynamical systems have a Coulomb
potential. Planetary motion and the motion of a charged particle in
in a Coulomb field are of this type. In this section we will consider the
symmetry algebra of this class of systems.

The Hamiltonian is given by
\be
H = \frac{\vec{p}^2}{2m} - \frac{\mu}{r},
\ee
where $\mu$ is a fixed constant.
Due to the spherical symmetry of the potential, this system is invariant under
rotations in three dimensions. This invariance leads to the conservation of
angular momentum $\vec{L}$.
We start on the classical level by considering the Poisson brackets
of the dynamical variables, which are functions on the phase space with
coordinates $(\vec{r},\vec{p})$.
The algebra is given by $\label{so(3)} \{L_i,L_j\} = \epsilon_{ijk}L_k$
and is called $so(3)$ ( or $su(2)$). It is a simple Lie algebra.

However, the angular momentum $\vec{L}$ is not the
only conserved quantity in this system. Let us consider the vector $\vec{R}$
defined
by
\be
\vec{R}= \vec{L} \times \vec{p} + \mu \frac{\vec{r}}{r},
\ee
with $\mu$ again the fixed constant proportional to the central potential.
Straightforward calculation shows that this, so called `Runge-Lenz vector,'
also
(Poisson) commutes with the
Hamiltonian: $\{L_i,H\} = \{R_i,H\} = 0$. This additional
symmetry differs from the $so(3)$
symmetry, which is a geometric symmetry, i.e. which can be expressed as
mappings of configuration space alone. The symmetry transformations
associated to the
Runge-Lenz vector do not transform the co-ordinates and momenta
separately.
They act on the entire phase space. The term `dynamical symmetries' is
sometimes used for this type of symmetries.

In the example of planetary motion, the Runge-Lenz
vector points
along the major axis of the
ellipsoid orbit of the planet and its magnitude is proportional to the
eccentricity of the orbit.

In an analogous quantum case, i.e. the hydrogen atom, which is
discussed in detail in \cite{Itzykson}, the fact that the
angular momentum is not the only conserved quantity is reflected by the
degeneracy in the spectrum. While the states of the Hamiltonian depend
on
three quantum numbers, $n$, $l$ and $m$, the
energy depends only on $n$: $E_n \sim 1/n^2$.
The energy is independent of $m$ because of the $so(3)$ symmetry,
which corresponds to the
conservation
of angular momentum. The degeneracy with
respect to this magnetic quantum number is present for any central
potential. The absence of dependence on $l$ suggests that there is
another conserved
quantity, which turns out to be the Runge-Lenz vector. This degeneracy
only
occurs if the potential is of the form $\frac{1}{r}$ and is thereby
particular to the Coulomb potential.

Let us return to the classical symmetry algebra of $L_i$
and $R_i$. In the first place the algebra is only closed if
we include $H$ in the set of generators. Secondly, we see that it is
actually non-linear:
\bea
\label{RungeEuclid}
\{L_i,L_j\} &=& \epsilon_{ijk}L_k; \hspace{1cm} \{R_i,R_j\} =
-2 \epsilon_{ijk} HL_k \nn \\
\{R_i,L_j\} &=& \epsilon_{ijk} R_k; \hspace{1cm} \{L_i,H\} = \{R_i,H\} = 0.
\eea
Usually one linearizes (by a non-linear basis transformation) this
algebra giving rise to the well-known hidden $so(4)$ symmetry in the
hydrogen atom \cite{Itzykson}.
The algebra above describes the Kepler orbit in a three dimensional
Euclidean space. Now consider the same Kepler problem on a
three-sphere $S^3$. In \cite{Higgs,Higgs2}
 it is shown that the Poisson brackets
between the components of the $R$ vector become
\be
\label{RungeS3}
\{R_i,R_j\} = \epsilon_{ijk} \left( -2H + \lambda \vec{L}^2 \right)
L_k
\ee
where $\lambda$ is the curvature of the sphere, which is equal to the inverse
radius $R$ of the sphere $\lambda = 1/R$. We will call this algebra the
the {\em Runge-Lenz algebra}. Later it will be shown that it is a finite
$W$-algebra (which was first remarked in \cite{bowc}).

In the previous sections we have shown that even simple and
well-known physical
systems may have non-linear W-symmetries. In section~\ref{sect62}
we will see yet another example, namely Toda systems. It
seems therefore to be justified to embark on a more systematic study
of these algebras and their representation theory. This will lead us
to the theory of finite W-algebras.

\chapter{Classical Finite $W$-Algebras}
As we have seen in the first chapter, algebras of conserved quantities
are always closed, but not necessarily linear. This is not merely an
abstract mathematical possibility, we have seen that some of the
simplest and most fundamental systems in physics exhibit non-linear
symmetries. When trying to analyze these symmetry algebras however, or
perhaps to construct their irreducible representations, one inevitably
runs into trouble due to the fact that they are non-linear.
New methods are therefore needed.

In this chapter it will be shown that many non-linear algebras,
including the ones we encountered in the previous chapter, can in fact be
seen as `reductions' of Lie algebras. This result clearly opens up a
new possibility of analyzing them since the theory of Lie algebras is
well developed. In the present chapter we restrict ourselves to the
classical case and leave quantization to the next chapter.

The method we use in this chapter is to construct non-linear Poisson
algebras from a canonical linear Poisson algebra associated to any Lie
algebra, the Kirillov Poisson algebra. First we give a description in
simple terms of the Kirillov Poisson algebra. Then we use a procedure
very similar to the  constraint formalism called `Poisson reduction',
to construct linear and non-linear algebras from this Poisson algebra.
After that we discuss in detail a very interesting class of algebras
which are derived using $sl(2)$ embeddings. These algebras are in
general non-linear but they may contain linear subalgebras which one
can predict rather easily. Using this we discuss how to obtain
non-linear extensions of $SU(3)\times SU(2) \times U(1)$.

\sectiona{Kirillov Poisson structures}
Consider a system with $SO(3)$ rotation invariance. The conserved
quantities associated to this symmetry are the three components of the
angular momentum $\vec{L}$. If one calculates the Poisson brackets
between two components of $\vec{L}$, one finds that they satisfy
\be
\label{Poissonso(3)}
\{L_i,L_j\} = \ep_{ijk} L_k.
\ee
Now, let $\{I_a\}_{a = 1}^3$ be the generators of $SO(3)$, i.e. $I_a$
is the infinitesimal generator of rotations around the $x^a$-axis.
The commutator between $I_a$ and $I_b$ is given by
\be
\label{commso(3)}
[I_a,I_b] = \ep_{abc}I_c.
\ee
Note that even though (\ref{Poissonso(3)}) is a Poisson relation
and (\ref{commso(3)}) is a commutator algebra, the {\em structure
constants} in (\ref{Poissonso(3)}) and (\ref{commso(3)}) are the same.
That is, the components of the angular momentum together generate a
Poisson algebra which is isomorphic to the Lie algebra of $SO(3)$.

This example illustrates a general principle: if $G$ is a
symmetry group of some physical system, then the Poisson brackets
between the (Noether)
conserved quantities $\{J_a\}$ have
the same structure constants as the
commutator brackets between the generators
$\{t_a\}$ of $G$, i.e. if $[t_a,t_b] = f_{ab}^c t_c$, then
\be
\label{KirillovPoisson}
\{ J_a,J_b \} = f_{ab}^c J_c,
\ee
which is called the Kirillov Poisson algebra.
The generators $\{t_a\}$ span the so-called `Lie algebra' ${\cal G}$
of $G$. We conclude that {\em any} system with $G$ symmetry contains the
Kirillov Poisson structure as a subalgebra.

For later use we will now give a somewhat more formal definition of the
Kirillov Poisson structure associated to ${\cal G}$. The reader
may wish to skip this part at first reading.

Let ${\cal G}$ be a Lie algebra, ${\cal G}^*$ its
dual and $C^{\infty}({\cal G}^*)$ the set of
smooth functions on ${\cal G}^*$. The Kirillov
Poisson bracket between $F,G \in C^{\infty}({\cal G}^*)$ is
defined for all $\xi \in {\cal G}^*$ by
\begin{equation}
\{F,G\}(\xi )=\langle \xi,[\mbox{grad}_
{\xi}F,\mbox{grad}_{\xi}G] \rangle  \label{kir}
\end{equation}
where $\langle.,. \rangle$ denotes the usual contraction
between ${\cal G}^*$ and ${\cal G}$ and $grad_{\xi}F$ is uniquely defined by
\begin{equation} \label{df}
\frac{d}{d\epsilon}F(\xi + \epsilon \xi') |_{\epsilon =0}=
\langle \xi',\mbox{grad}_{\xi}F\rangle ,
\end{equation}
for all $\xi' \in {\cal G}^*$.
Note that $grad_{\xi}F$ is therefore an element of ${\cal G}$, which
means that $[\mbox{grad}_{\xi}F,\mbox{grad}_{\xi}G]$ is well defined.

We can recover formula (\ref{Poissonso(3)}) as follows. Let $\{t_a\}$
be the basis of ${\cal G}$ and $J_a$ the element
of $C^{\infty}({\cal G}^*)$ given by
$J_a(\xi) \equiv \langle \xi,t_a \rangle$,
then these functions satisfy (\ref{Poissonso(3)}).

\sectiona{Poisson reduction of the Kirillov Poisson structure}
Finite W-algebras are constructed by applying a procedure very similar to the
Dirac
constraint formalism, called  `Poisson reduction',
to the Kirillov Poisson structure. Essentially what one does is
impose a set of first class constraints on the system. As usual these
first class constraints will generate gauge invariances. One therefore
looks for gauge invariant quantities. In general the set of gauge
invariant quantities will be generated by a certain finite subset.
These are the generators of the finite $W$-algebra.
Calculating the Poisson brackets between these generators we
find that the algebra of gauge invariant quantities is in general
non-linear, i.e. the Poisson brackets close on polynomials of the
generators, not on linear combinations. This is then the
finite $W$-algebra. Let us now come to a more
precise definition of finite $W$-algebras.

Let again ${\cal G}$ be some Lie algebra, ${\cal L} \subset {\cal G}$
some subalgebra and $\chi: {\cal L} \rightarrow {\bf C}$ a one-dimensional
representation\footnote{One can in principle also consider higher dimensional
representations of ${\cal L}$ in some auxiliary algebra, see
section~\ref{sect434}. Here we will for simplicity restrict ourselves to
the one-dimensional case.}
 of ${\cal L}$. Let $\{t_a\}$ and $\{t_{\al}\}$ be
bases of ${\cal G}$ and ${\cal L}$ respectively,
such that $\{t_{\al}\} \subset \{t_a\}$, and
let $J_a \in C^{\infty}({\cal G}^*)$ be defined
by $J_a(\xi) = \langle \xi , t_a \rangle, \forall \xi \in {\cal G}^*$.
Again let $K({\G})=(C^{\infty}({\cal G}^*),\{.,.\})$ denote the
Kirillov Poisson algebra associated to ${\cal G}$.

The first step is to constrain the functions $J_{\al}$, corresponding
to the subalgebra ${\cal L}$, to constant values:
\be
\label{constraints}
\ph_{\al} \equiv J_{\al} - \chi(t_{\al}) = 0.
\ee
Denote the hyper-surface in ${\cal G}^*$ determined by $\phi_{\al} = 0$
for all $\al$ by $C$. The set of functions on $C$ is equal to the
set of functions on ${\cal G}^*$ up to functions which are zero on $C$.
Any function that is zero on all of $C$ has the form $f^\al \phi_\al$.
Let us denote the set of all these `zero functions' by I:
\be
\label{idealconstraints}
I = \{ f^{\al} \ph_{\al} | f^{\al} \in C^{\infty}({\cal G}) \}.
\ee
Since all elements of $I$ are zero on all of $C$, there is no way
somebody living on $C$ can distinguish between the functions $g$
and $g+f$ if $f$ is an element of $I$. Mathematically this is expressed
by the equality
\be
C^{\infty}(C) = C^{\infty}({\cal G})/I.
\ee
This formula means that we are to identify all functions
in $C^{\infty}({\cal G}^*)$ that differ by an element of $I$.

It is easy to show that the constraints (\ref{constraints}) are
all `first class'. That means that $\{ \ph_{\al},\ph_{\bt} \} \in I$,
for all $\al, \bt$. For this remember that $\chi$ is a one dimensional
representation of ${\cal L}$, which means
that $\chi([t_{\al},t_{\bt}]) = \chi(t_{\alpha}) \chi(t_{\bt})
- \chi(t_{\bt}) \chi(t_{\al}) = 0$. On the other
hand $\chi([t_{\al},t_{\bt}]) = f_{\al \bt}^{\gamma} \chi(t_{\gamma})$.
 From this follows that $\{ \ph_{\al}, \ph_{\bt} \} =
f_{\al \bt}^{\gamma} \ph_{\gamma}$.

Obviously $I$ is an ideal with respect to the (abelian)
multiplication map in $K({\G})$, since $hf^{\alpha}\phi_{\alpha}=
\tilde{f}_{\alpha}\phi_{\alpha} \in I$ for all $h \in C^{\infty}({\G})$,
where $\tilde{f}_{\alpha}=hf^{\alpha}$.

$I$ is a also Poisson subalgebra of $K({\G})$. In order to
see this let $f=f^{\alpha}\phi_{\alpha}$ and $h=h_{\alpha} \phi_{\alpha}$
be elements of $I$. Then
\ba
\{ f,g \} & = &
\{ f^{\alpha},h^{\beta} \} \phi_{\alpha} \phi_{\beta} +
\{ f^{\alpha},\phi_{\beta} \} h^{\beta} \phi_{\alpha} +
\{ \phi_{\alpha}, h^{\beta}\} f^{\alpha} \phi_{\beta}  \nonumber \\
&   & + f^{\alpha} h^{\beta} \{ \phi_{\alpha},\phi_{\beta} \}.
\ea
Obviously the first three terms are again elements of $I$. That the
last term is an element of $I$ follows from the fact that the
constraints $\{\phi_{\alpha}\}$ are all first class.

Nevertheless $I$ is not an ideal with respect to the Poisson bracket.
Consequently, the Poisson bracket is {\em not} preserved if we
divide out $I$. That is, the Poisson bracket on ${\cal G}$ does {\em not}
induce one on $C$. Physically this is equivalent to the statement
that first class constraints induce non-physical gauge invariances,
that have to be eliminated from the theory. Mathematically
one proceeds as follows. Define the maps
\be
X_{\alpha}:C^{\infty}({\G}^*) \rightarrow C^{\infty}({\G}^*)
\ee
by
\be
X_{\alpha}(f)=\{\phi_{\alpha},f\}
\ee
In geometric terms the $X_{\alpha}$ are the 'Hamiltonian
vector fields' associated to the constraints $\{\phi_{\alpha}\}$, which
can be interpreted as the derivative of
some function $f \in C^{\infty}({\cal G}^*)$ along
the direction of the gauge invariance.

Now let $[f]$ denote the equivalence
class $f+I$, i.e. $[f] \in C^{\infty}(C)$.  Then
\be
X_{\alpha}(f+I) = \{\phi_{\alpha},f\}+\{\phi_{\alpha},I\}
\subset  X_{\alpha}(f)+I,
\ee
where we used the fact that $I$ is a Poisson subalgebra of $K({\G})$.
Put differently we therefore have
\be
X_{\alpha}[f]=[X_{\alpha}(f)],
\ee
which means that the maps $X_{\alpha}$ descend to well defined maps
\be
X_{\alpha}:C^{\infty}(C) \rightarrow C^{\infty}(C).
\ee
In geometric terms this is nothing but the statement that the Hamiltonian
vector fields of the constraints are tangent to $C$.

Now define the set of the functions which are constant under the
flow of $X_{\al}$, i.e. gauge invariant, as
\be
{\W} =\{ { [f] \in C^{\infty}(C) \mid X_{\alpha}[f]=0, \;\;
}\mbox{for all }\alpha \}.
\ee

The point is now that the Kirillov Poisson structure on $C^{\infty}({\G}^*)$
induces naturally a Poisson structure $\{.,.\}^*$ on $\W$.
Let $[f],[h] \in \W$, then this Poisson structure is simply given
by
\be
\{[f],[h]\}^*=[\{f,h\}] \label{alt}.
\ee
Of course one has to show that this Poisson structure is well defined.
In order to do so we have to check two things: firstly we have to check whether
$[\{f,h\}] \in \W$ whenever $[f],[h] \in \W$, and secondly we
have to show that the definition does not depend on the choice of the
representatives $f$  and $h$. Both these checks are easily carried
out and one finds that indeed the Poisson bracket $\{.,.\}^*$
turns $\W$ into a Poisson algebra. The Poisson algebra found by
reducing a Kirillov Poison algebra is called a {\em finite W-algebra}
\be
W({\G},{\cal L},\chi )\equiv (\W,\{.,.\}^*).
\ee
In more physical terms, $\cal W$ is the set of gauge invariant
quantities and the Poisson bracket between them is obtained by
first calculating the Poisson bracket and then putting the constraints to zero.

It is clear that the number of finite $W$-algebras is huge.
However, it is by no means clear that all non-linear algebras are of
this type. In fact this is almost certainly not the case. However,
as we shall see, several interesting non-linear symmetry algebras
encountered in physics, including the ones described in the previous
chapter, {\em are} finite $W$-algebras. Anyway, in this paper we
restrict our attention to finite $W$-algebras.

\subsection{Examples}
Let us now consider some examples in order to clarify the construction.
First take ${\cal G} = sl(2)$, the set of traceless 2x2 matrices.
This Lie algebra can be described as follows. It is the (complex or real)
span of three generators $t_+$, $t_-$ and $t_0$ with
commutation relations $[t_0,t_{\pm}] = \pm 2 t_{\pm}$ and $[t_+,t_-] = t_0$.
The Kirillov Poisson structure therefore reads
\be
\{ J_0,J_{\pm} \} = \pm 2 J_{\pm}; \hspace{2cm} \{ J_+,J_- \} = J_0.
\ee
The Lie algebra $sl(2)$ has several subalgebras. The most obvious one
is the so called Cartan subalgebra spanned by $t_0$. If we take ${\cal L}$
to be the Cartan subalgebra and $\chi = 0$, then
from equation (\ref{constraints}) we find $\ph = J_0$.
The ideal $I$ (see (\ref{idealconstraints}))
consists of elements of the form $f(J_0,J_+,J_-) \ph$, where $f$ is
an arbitrary smooth function in three variables.
Since dividing out $I$ corresponds to putting
the constraints to zero, we find that $C^{\infty}(C)$
is isomorphic to the set of arbitrary smooth functions in $J_+$ and $J_-$
\be
C^{\infty}(C) = \{ f(J_+,J_-) \mid \mbox{f is smooth} \}.
\ee
The next step in the construction is to find the set ${\cal W}$ of
all elements in $C^{\infty}(C)$ that Poisson commute with $\ph$
(after imposing the constraint). As $\{ \ph,J_+J_- \} =
J_+ \{J_0,J_- \} + J_- \{ J_0,J_+ \} = - J_-J_+ + J_-J_+ = 0$,
we find that ${\cal W}$ consists of functions that depend only on
the combination $J_+J_-$.
However, as ${\cal W}$ has only one generator, $J_+J_-$, it
turns out to be an abelian algebra.

In fact this is what generically happens when we choose ${\cal G} = sl(2)$.
If ${\cal L}$ is one of the so called `Borel subalgebras' $b_{\pm}$
spanned by $t_0$ and $t_{\pm}$, then there are effectively no
degrees of freedom left, that is ${\cal W} = 0$. If one chooses ${\cal L}$
to be the span of $t_+$ or $t_-$ ( and $\chi = 0$), then ${\cal W}$ will
again be an abelian algebra with one generator, $J_0$.
The situation doesn't change for $\chi \neq 0$,
because the numbers of generators stay the same.
We conclude that there are no interesting finite W-algebras
that can be derived from $sl(2)$.

Let us therefore turn to ${\cal G} = sl(3)$. This Lie algebra is
spanned by eight elements
$\{ t_1,t_2,t_{{\al}_1},t_{{\al}_2},t_{{\al}_3},t_{-{\al}_1},
t_{-{\al}_2},t_{-{\al}_3} \}$, where ${\al}_1,{\al}_2$ and ${\al}_3$
denote the three positive root vectors of $sl(3)$.
The Kirillov Poisson algebra reads
\bea {}
\{ J_1,J_{\pm {\al}_1} \} &=& \pm 2 J_{\pm {\al}_1}, \hspace{0.8cm} {}
\{ J_2,J_{\pm \al_1} \} = \mp J_{\pm \al_1} \nn \\ {}
\{ J_1,J_{\pm \al_2} \} &=& \mp J_{\pm \al_2}, \hspace{1cm} {}
\{ J_2,J_{\pm \al_2} \} = \pm 2 J_{\pm \al_2} \nn \\ {}
\{ J_1,J_{\pm \al_3} \} &=& \pm J_{\pm \al_3}, \hspace{1cm} {}
\{ J_2,J_{\pm \al_3} \} = \pm J_{\pm \al_3} \nn \\ {}
\{ J_{\pm \al_1},J_{\pm \al_2} \} &=& \pm J_{\pm \al_3}, \hspace{1cm} {}
\{ J_{\pm \al_1},J_{\mp \al_3} \} = \mp  J_{\mp \al_2 } \nn \\ {}
\{ J_{\pm \al_2},J_{\mp \al_3} \} &=& \pm J_{\mp \al_1}, \hspace{1cm} {}
\{ J_{+\al_3}, J_{-\al_3} \} = J_1 + J_2 \nn \\ {}
\{ J_{+\al_1}, J_{-\al_1} \} & = &  J_1, \hspace{1.7cm} {}
\{ J_{+\al_2}, J_{-\al_2} \} = J_2.
\eea
The most obvious choice for the subalgebra ${\cal L}$ is the Cartan
subalgebra spanned by $t_1$ and $t_2$. Again take $\chi = 0$.
According to (\ref{constraints}) the constraints are then $\ph_1 = J_1$
and $\ph_2 = J_2$. Reasoning as before $C^{\infty}(C)$ is shown to
consist of smooth functions in the variables
$J_{\al_1},J_{- \al_1},J_{\al_2},J_{- \al_2},J_{\al_3}$ and $J_{- \al_3}$.
In order to construct ${\cal W}$, we need to find functions of
these variables that Poisson commute (after imposing the constraints)
with $\ph_1$ and $\ph_2$. In principle $sl_3$ has dimension eight.
We have imposed two first class constraints, which brings the
dimension of $C$ down to six. As $\ph_1$ and $\ph_2$ are first class,
they generate gauge invariances, or in other words, two of the
six dimensions will correspond to gauge degrees of freedom.
On eliminating these, which is essentially what constructing ${\cal W}$
amounts to, we are left with four dimensions. We now look for four
independent elements of $C^{\infty}(C)$
that commute with $\ph_1$ and $\ph_2$. It is easy to construct them. They read
\bea
\label{reducsl3}
A_1&=& J_{\al_1}J_{-\al_1} \\ \nn
A_2&=& J_{\al_2}J_{-\al_2} \\ \nn
A_1&=& J_{\al_3}J_{-\al_3} \\ \nn
B &=& \frac{1}{2} ( J_{\al_3}J_{- \al_1}J_{- \al_2}
- J_{- \al_3}J_{\al_1} J_{\al_2} ) \\ \nn
C &=& \frac{1}{2} ( J_{\al_3}J_{- \al_1}J_{- \al_2}
+ J_{- \al_3}J_{\al_1} J_{\al_2} ).
\eea
Note that there are five invariant quantities. However, the relation
\be
C^2 = A_1A_2A_3 + B^2
\ee
between the generators in (\ref{reducsl3}) brings the number of
independent dimensions back to four. In fact this relation
defines the 4 dimensional surface
in 5 dimensional Euclidean space (with coordinates $C,B,A_1,A_2,A_3$)
on which the finite $W$-algebra lives.

The non-zero Poisson brackets between the generators (\ref{reducsl3}) read
\be
\{ A_i,A_{i + 1} \}^* = 2B \hspace{2cm} \{ A_i,B \}^*
= A_i (A_{i + 1} - A_{i - 1}),
\ee
where $i$ is a cyclic index, i.e. $i \in \{1,2,3\}$ and $i + 3 = i$.

Another reduction of $sl(3)$, and one that leads to a linear
finite W-algebra, is the following. Take ${\cal L}$ to be the
span of $t_{\al_2}$ and $t_{\al_3}$ and $\chi = 0$. The constraints
read $J_{\al_2} = J_{\al_3} = 0$ and $C^{\infty}(C)$ consists of
functions of $J_{\al_1},J_{- \al_1},J_1,J_2,J_{-\al_2}$ and $J_{- \al_3}$.
Again a simple counting argument similar to the one given  above leads
us to look for four independent generators of ${\cal W}$. It is easy to
check that $J_{\al_1},J_{- \al_1},J_1$ and $J_2$ satisfy the requirement
that they commute with the constraints. Now $J_{\al_1},J_{- \al_1}$
and $J_1$ form and $sl_2$ algebra
\be
\{ J_1,J_{\pm \al_1} \}^* = \pm 2 J_{\pm \al_1};
\hspace{1cm} \{ J_{\al_1},J_{- \al_1} \}^* = J_1,
\ee
while $s = J_1 + 2 J_2$ commutes with all the other generators and
thus forms a $u(1)$. Thus we have found that {\em $sl(2) \oplus u(1)$
is a reduction of $sl(3)$}. As we shall see later when we discuss
real forms of finite $W$-algebras this also means
that $su(2)\oplus u(1)$ is a reduction of $su(3)$:
\be
su(3) \longrightarrow su(2) \oplus u(1)
\ee

This example is actually a special case of a more general class
of reductions.
Let $\dl$ be an arbitrary element of the Cartan subalgebra of a (semi)
simple Lie algebra ${\cal G}$. We can use $\dl$ to define a so
called `grading' on ${\cal G}$. Define ${\cal G}_n = \{ x \in {\cal G}
\mid [\dl,x] = nx \}$. It then follows that
\be
\label{grading1}
{\cal G} = \oplus_n {\cal G}_n; \hspace{2cm} [{\cal G}_n,{\cal G}_m]
\subset {\cal G}_{n + m}.
\ee
If we now take ${\cal L} = \oplus_{n>0} {\cal G}_n$ and $\chi = 0$,
then all elements $J_a$ such that $t_a \in {\cal G}_0$ will
commute (after imposing the constraints) with all $\ph_{\al}
\equiv J_{\al}$. This can be seen as follows. Let $t_a \in {\cal G}_0$,
then $[t_{\al},t_a] \in {\cal L}$, as follows from
(\ref{grading1}). As all $J_{\al}$ are constrained to zero,
we find that $\{ \ph_{\al},J_a \} = \{ J_{\al},J_a \}$ is a
linear combination of $J_{\bt}$ which after imposing the constraints
become zero. We conclude therefore that this type of reduction always
leads to a finite W-algebra which is isomorphic to ${\cal G}_0$.
The ones that are most closely related to the infinite-dimensional
$W$-algebras of conformal field theory.

The finite $W$-algebras considered in this section are associated
to $sl(2)$ embeddings into $\cal G$. A given $sl(2)$ embedding fixes
both $\cal L$ and $\chi$ which means that there is one finite $W$-algebra
for every $sl(2)$ embedding. For $sl(n)$ the number of
inequivalent $sl(2)$ embeddings is equal to the number of partitions
of the number $n$ and this is therefore the number of finite $W$-algebras
of this type that one is able to extract from $sl(n)$.

\subsection{$sl_2$-embeddings}
In this section we always take ${\cal G}$ to be a simple Lie algebra.
This implies that the inner product $(x,y) =
Tr({\rm ad}(x){\rm ad}(y))$ ($x,y \in {\cal G}$),
the so called Cartan Killing form, is non-degenerate, i.e. there does
not exist an element $y \in {\cal G}$ such that $(x,y) = 0$
for all $x \in {\cal G}$. Therefore any element $f$ of ${\cal G}^*$
can be written as $f(.) = (x,.)$ for some $x \in {\cal G}$, or in
other words we can identify ${\cal G}$ and ${\cal G}^*$. Due to
this fact we can define the Kirillov Poisson structure on ${\cal G}$
instead of on ${\cal G}^*$ which makes life easier in some respects.
Consider for this $C^{\infty}({\cal G})$ instead
of $C^{\infty}({\cal G}^*)$ and define the functions $J^a$ on ${\cal G}$
by $J^a(t_b) = {\dl}^a_b$. The Poisson algebra satisfied by
these quantities is
\be
\{ J^a,J^b \} = f^{ab}_c J^c,
\ee
where we have raised and lowered indices with the metric $g_{ab} = Tr(t_at_b)$.

Now let there be given an $sl(2)$ subalgebra $\{ t_0,t_+,t_- \}$ of ${\cal G}$,
\be
[t_0,t_{\pm}] = 2 t_{\pm}, \hspace{2cm}  [t_+,t_-] = t_0.
\ee
If we define the spaces
\be
{\cal G}^{(n)} = \{ x \in {\cal G} \mid [t_0,x] = nx \},
\ee
then
\be
\label{grading2}
{\cal G} = \oplus_n {\cal G}^{(n)}.
\ee
The
decomposition (\ref{grading2}) is again a `grading' of ${\cal G}$, because
\be
[ {\cal G}^{(n)},{\cal G}^{(m)} ] \subset {\cal G}^{(n + m)}.
\ee
In general the numbers $n$ can be integers or half integers,
that is $n \in \frac{1}{2} {\bf Z}$. As in the infinite-dimensional case,
we would like to take
\be
{\cal L} = {\cal G}^{(+)} \equiv \oplus _{n>0} {\cal G}^{(n)}
\ee
together with $\chi(t_+) = 1$ and all others zero. However there is a
slight problem with this choice, because the commutator of two elements
of ${\cal G}^{(\frac{1}{2})}$ may contain $t_+$, which is an element
of ${\cal G}^{(1)}$. This means that $\chi$ would no longer be a
one-dimensional representation of ${\cal G}^{(+)}$ and not all
constraints would be first class. Classically this need not be a
problem, because one can eliminate second class constraints with the
so called Dirac bracket \cite{BT}. However,
quantization becomes much more involved if not all
constraints are first class. Fortunately it is possible for many
Lie algebras, among which all $sl(n)$, to replace ${\cal G}^{(+)}$
and $\chi$ by a new algebra ${\cal G}_+$ and a new one-dimensional
representation $\chi : {\cal G}_+ \rightarrow {\bf C}$ that lead
to the same finite W-algebra without having to resort to
Dirac brackets. What one does is replace the `grading element'
$t_0$ by a new one, $\dl$, which is also an element
of the Cartan subalgebra \cite{FORTW,BT}.
The element $\dl$ defines a grading
of ${\cal G}$ which is different from the one in (\ref{grading2})
\be
\label{grading}
{\cal G} = \oplus_n {\cal G}_n,
\ee
where ${\cal G}_n = \{ x \in {\cal G} \mid [\dl,x] = nx \}$.
Essentially $\dl$ has the property that it splits ${\cal G}^{(\frac{1}{2})}$
and ${\cal G}^{(- \frac{1}{2})}$ into two pieces:
\be
{\cal G}^{(\frac{1}{2})} = {\cal G}^{(\frac{1}{2})}_L
\oplus {\cal G}^{(\frac{1}{2})}_R; \hspace{1cm}
{\cal G}^{(- \frac{1}{2})} = {\cal G}^{(- \frac{1}{2})}_L
\oplus {\cal G}^{(- \frac{1}{2})}_R
\ee
and that
\bea
{\cal G}_0 &=& {\cal G}^{(- \frac{1}{2})}_L \oplus {\cal G}^{(0)}
\oplus {\cal G}^{(+ \frac{1}{2})}_L \\ \nn
{\cal G}_{\pm 1} &=& {\cal G}^{(\pm 1)} \oplus {\cal G}^{(\pm \frac{1}{2})}_R,
\eea
i.e. `half' of ${\cal}^{(\pm \frac{1}{2})}$ gets
added to ${\cal G}^{(0)}$, which becomes ${\cal G}_0$ and the
other half gets added to ${\cal G}^{(\pm 1)}$, which becomes ${\cal G}{\pm 1}$.
Obviously
\be
{\cal G}_+ \oplus {\cal G}^{(\frac{1}{2})}_L = {\cal G}^{(+)}.
\ee
Furthermore $\delta$ is chosen such that the grading
of $\cal G$ is integral, i.e.
all $n$ in (\ref{grading}) are integers.
The one-dimensional representation $\chi: {\cal G}_+ \rightarrow {\bf C}$
is taken to be the same as before, i.e. $\chi(t_{\pm}) = 1$ and
zero everywhere else. What has been done here, effectively, is that
the set of second class constraints (corresponding
to ${\cal G}^{(\frac{1}{2})}$) has been split into two halves.
The constraints in one half we still put to zero. One can choose $\delta$
such that they will be first class.
The constraints in the other half
are kept free. However, the degrees of freedom they represent can be
gauged away by the gauge invariance generated by the first class
constraints in the first half.  It is not possible to find an
element $\dl$ satisfying the requirements for all
Lie algebras \cite{FORTW}. For $sl(n)$ however, there is no problem.

We now proceed as before. Let again $\{ t_a \} = \{ t_{\al} \}
\cup \{ t_{\bar{\al}} \}$ be a basis of ${\cal G}$, where $\{t_a\}$ is a
basis of ${\cal L} = {\cal G}_+$ and $\{t_{\bar{\al}}\}$ is a
basis of ${\cal G}_0 \oplus {\cal G}_-$. The constraints are
given as usual by $\ph^{\al} = J^{\al} - \chi(t_{\al})$.

It is clear that elements of $C^{\infty}(C)$ are smooth functions in
the variables $\{ J^{\bar{\al}} \}$, as all the $J^{\al}$ have been
constrained to constants. The next step is to
find ${\cal W}$. For this we need to look for
elements $W$ of $C^{\infty}(C)$ such that
\be
\label{W}
\{ \ph^{\al},W \} = 0,
\ee
after imposing the constraints. This will then be the W-algebra associated
to the $sl(2)$ embedding. In general it is no easy task to find the
complete set of elements $W$ such that (\ref{W}) holds. However, in the
case of finite  W-algebras derived from $sl(2)$ embeddings there turns
out to be an algorithmic procedure for doing so. For this we have to
consider the gauge transformations generated by the first class
constraints $\ph^{\al}$. Let $x = x^at_a$ be an arbitrary element
of ${\cal G}$. The gauge transformations generated by $\ph^{\al}$
then act as follows on $x$:
\begin{eqnarray}
\delta_{\alpha}x & \equiv & ({\dl}_{\al} J^a)(x)t_a \equiv
\epsilon \{\phi^{\alpha},J^{a}\}(x)t_a \nonumber \\
& = & \epsilon \{J^{\alpha},J^a\}(x)t_a \nonumber \\
& = & \epsilon f^{\al a}_b J^b(x) t_a \nn \\
& = & \epsilon f^{\al a}_b x^c \dl^b_c t_a = \epsilon f^{\al a}_b x^bt_a \nn \\
& = & \epsilon g^{\alpha c}f^a_{cb}x^bt_a = [
\epsilon g^{\alpha c}t_c,x^bt_b ] \nonumber \\
& = & [\epsilon t^{\alpha},x].  \label{ga}
\end{eqnarray}
Now, if $t_{\al} \in {\cal G}_+$, then $t^{\al} \in {\cal G}_-$,
because ${\cal G}_+$ and ${\cal G}_-$ are non-degenerately paired
by the Cartan Killing form \cite{BT}. It therefore follows from (\ref{ga})
that the group of gauge transformations generated by the first class
constraints is nothing but $G_- = exp({\cal G}_-)$ and that this group
acts on ${\cal G}$ by group conjugation $x \rightarrow g x g^{-1}, g \in G_-$.
Now let $x$ be an element of $C$. In general it has the form
\be
\label{x}
x = t_+ +  \sum_{\bar{\al}} x^{\bar{\al}}t_{\bar{\al}}.
\ee
where $x^{\bar{\alpha}}$ are complex numbers.
It can be shown \cite{BT} that the gauge freedom $G_-$ can be
completely fixed by bringing $x$ to the form
\be
\label{gaugefix}
t_+ + \sum_{ t_{\bar{\al}} \in {\cal G}_{lw} } W^{\bar{\al}}t_{\bar{\al}},
\ee
where ${\cal G}_{lw} = \{ x \in {\cal G} \mid [t_-,x] = 0 \}$. Put
differently, for every $x$ of the form (\ref{x}), there exists a
unique $g \in G_-$ such that $g x g^{-1}$ is of the form (\ref{gaugefix}).
Since the element (\ref{gaugefix}) is the same for any element
within a certain
gauge orbit, the quantities $W^{\al}$ must be
gauge invariant. This means that they commute with the
constraints $\{\ph^{\al},W^{\bar{\al}}\} = 0$. By construction the
quantities $W^{\bar{\al}}$ from a complete set of generators of the
finite W-algebra in question. The only thing left is therefore to
calculate the Poisson relations
\be
\label{Wred}
\{ W^{\bar{\al}},W^{\bar{\bt}} \}^* = \{ W^{\bar{\al}},W^{\bar{\bt}} \}.
\ee
The equality in equation (\ref{Wred}) can be understood as follows.
In principle the bracket $\{.,.\}^*$ is simply $\{.,.\}$ where it is
understood that we take $\ph^{\al}$ {\em after} we have calculated the
Poisson bracket. However, as ${\cal G}_0 \oplus {\cal G}_-$ is a
subalgebra of ${\cal G}$ and since the quantities $W^{\bar{\al}}$
will be polynomials only in $J^{\bar{\al}}$, as follows from (\ref{x}),
we find that the Poisson bracket of $W^{\bar{\al}}$ and $W^{\bar{\bt}}$
will not involve any $J^{\al}$. This justifies (\ref{Wred}). It is
now time for some examples.

\subsection{Examples}
As a simple example take ${\cal G} = sl_2$. Also take as a basis
of $sl(2)$ the matrices $t_1,t_2$ and $t_3$ such that
\bea
J^at_a = \left( \begin{array}{cc} J^2 & J^1 \\ J^3 & - J^2 \end{array} \right).
\eea
The Kirillov Poisson structure then reads
\be
\{ J^2,J^1 \} = - J^1, \hspace{2cm} \{ J^2,J^3 \} = J^3,
\hspace{2cm} \{ J^1,J^3 \} = -2 J^2.
\ee
Obviously there is only one non-trivial $sl(2)$ embedding into $sl(2)$,
namely the algebra itself. Now, $sl(2)$ splits up into three
subalgebras of grade +1, 0  and -1,
\be
sl(2) = {\cal G}^{(+1)} \oplus {\cal G}^{(0)} \oplus {\cal G}^{(-1)}
\ee
given by the span of $t_3,t_2$ and $t_1$ respectively. Note that all grades
are integral, which means that ${\cal G}_n \equiv {\cal G}^{(n)}$. The
constraint becomes $\ph = J^1 - \chi(t_1) = J^1 - 1$,
since $t_1 = t_+$. $C^{\infty}(C)$ is given by the smooth
functions of $J^2$ and $J^3$. Next we look for gauge invariant
functions, i.e. functions which Poisson commute with the
constraint (after imposing  $\ph = 0$). We do this by the method
outlined above. Let $g$ be an arbitrary element of $G_-$, i.e.
\bea
g = exp \left( \begin{array}{cc} 0 & 0 \\ a & 0 \end{array} \right)
= \left( \begin{array}{cc} 1 & 0 \\ a & 1 \end{array} \right).
\eea
We can find the gauge invariant function $W$ by solving
\be
\label{gaugefixing}
g \left( t_+ + J^2t_2 + J^3t_3 \right) g^{-1} \equiv t_+ +W t_3
\ee
for $g$ and $W$. In matrix form this equation reads
\bea
\left( \begin{array}{cc} 1 & 0 \\ a & 1 \end{array} \right)
\left( \begin{array}{cc} J^2 & 1 \\ J^3 & -J^2 \end{array} \right)
\left( \begin{array}{cc} 1 & 0 \\ -a & 1 \end{array} \right)
\equiv  \left( \begin{array}{cc} 0 & 1 \\ W & 0 \end{array} \right).
\eea
This equation is satisfied if $a = J^2$. We then find that
\be
W \equiv J^3 + J^2J^2.
\ee
Note that indeed $\{ \ph,W \} = \{J^1,J^3 + J^2J^2 \} = -2J^2 + 2J^1J^2$,
such that after imposing $\ph = 0$ we have $\{ \ph,W \} = 0$.
As ${\cal W}$ has only one generator, the obtained W-algebra is trivial. %

In order to illustrate the construction in a slightly less trivial case
consider the so-called non-principal $sl_2$ embedding into $sl_3$.
For $sl_3$ we choose the basis
\be
J^at_a =\left(
\begin{array}{ccc}
J^4+\frac{1}{2}J^5 & J^2 & J^1 \\
J^6 & -2J^4 & J^3 \\
J^8 & J^7 & J^4-\frac{1}{2}J^5
\end{array}
\right).
\ee

The $sl(2)$ embedding is given by $t_0 = t_5,t_+ = t_1$
and $t_- = t_8$. ${\cal G}_{lw}$ is the span of $t_8,t_7,t_6$
and $t_4$. The grading of $sl(3)$ with respect to this $sl(2)$ subalgebra is
\be
sl(3) = {\cal G}^{(-1)} \oplus {\cal G}^{(-\frac{1}{2})}
\oplus {\cal G}^{(0)} \oplus {\cal G}^{(\frac{1}{2})} \oplus {\cal G}^{(1)},
\ee
where
\bea
{\cal G}^{(-1)} &=& {\rm span}\{t_8\}; \hspace{1cm} {\cal G}^{(0)} =
{\rm span}\{t_4,t_5\}; \hspace{1cm}
{\cal G}^{(- \frac{1}{2})} = {\rm span}\{t_6,t_7\} \\ \nn
{\cal G}^{(1)} &=& {\rm span}\{t_1\}; \hspace{1cm} {\cal G}^{(\frac{1}{2})} =
{\rm span}\{t_2,t_3\}.
\eea

Obviously this grading is {\em not} integral. The element $\dl$ can be
found in the appendix and reads in this case
$\dl = \frac{1}{3}{\rm diag}(1,1,-2)$.
The grading with respect to $\dl$ is
\be
sl(3) = {\cal G}_{-1} \oplus {\cal G}_0 \oplus {\cal G}_1,
\ee
where
\be
{\cal G}_{-1} = {\rm span}\{t_7,t_8\}; \hspace{1cm} {\cal G}_0 =
{\rm span}\{t_4,t_5,t_2,t_6\}; \hspace{1cm} {\cal G}_1 = {\rm span}\{t_1,t_3\}.
\ee
Obviously
\be
{\cal G}_L^{(\frac{1}{2})} = {\rm span}\{t_2\};
\hspace{2cm} {\cal G}_R^{(\frac{1}{2})} = {\rm span}\{t_3\}.
\ee
The one dimensional representation is given by
\be
\chi(t_1) = 1; \hspace{2cm} \chi(t_3) = 0.
\ee
According to the standard procedure the constraints are
\be
\ph^1 = J^1 - 1 = 0; \hspace{2cm} \ph^3 = J^3 = 0 .
\ee
Again we want to find the gauge invariant functions in the
variables $J^2,J^4,J^5,...,J^8$. The group of gauge transformations $G_-$ is
given by
\bea
G_- = \left( \begin{array}{ccc} 1 & 0 & 0 \\ 0 & 1 & 0 \\ a & b & 1
\end{array} \right).
\eea
As before we solve the equation
\bea
\left( \begin{array}{ccc} 1 & 0 & 0 \\ 0 & 1 & 0 \\ a & b & 1
\end{array} \right)
\left( \begin{array}{ccc} J^4 + \frac{1}{2}J^5 & J^2 & 1 \\
J^6 & -2 J^4 & 0 \\ J^8 & J^7 & J^4 - \frac{1}{2} J^5 \end{array} \right)
\left( \begin{array}{ccc} 1 & 0 & 0 \\ 0 & 1 & 0 \\ - a & - b & 1
\end{array} \right)
 =
\left( \begin{array}{ccc} W^4 & 0 & 1 \\ W^6 & -2W^4 & 0 \\ W^8 & W^7 & W^4
\end{array} \right) \nn
\eea
for $a,b$ and $W^i$. The result reads $a = \frac{1}{2} J^5$, $b = J^2$ and
\bea
W^4 &=& J^4; \hspace{2cm} W^7 = J^7 + \frac{1}{2}J^2J^5 - 3J^2J^4 \nn \\
W^6 &=& J^6; \hspace{2cm} W^8 = J^8 + \frac{1}{4}J^5J^5 + J^2J^6, \label{Wgen}
\eea
which can all easily be shown to be gauge invariant.
These are now the generators of the finite W-algebra.

Defining
\bea
j_+ &=& W^7; \hspace{1cm}
j_- = \frac{4}{3}W^6; \hspace{1cm}
j_0 = -4W^4 \nn \\
C &=& - \frac{2}{3} C_2 = - \frac{4}{3}(W^8 + 3W^4W^4).\label{W32gen}
\eea
and calculating the Poisson brackets between these quantities we find
\bea
\label{Wsl3}
\{j_0,j_{\pm}\}^* &=& \pm 2j_{\pm} \nn \\
\{j_+,j_-\}^* &=& j_0^2 + C,
\eea
Note that this algebra is identical
to the symmetry algebra of the two dimensional
anisotropic oscillator with frequency ratio 2:1.

We shall now consider another example which turns out to be of
unexpected physical significance. Take ${\cal G} = sl(4)$ and
choose the basis $\{t_a\}$ such that
\bea
J^at_a = \left( \begin{array}{cccc} \frac{1}{2}J^7 + J^8 + J^9 & J^5 +
J^6 & J^2 +J^3 & J^1 \\ J^{10} + J^{11} &  \frac{1}{2}J^7 - J^8 -
J^9 & J^4 & J^2 - J^3 \\ J^{12} + J^{13} & J^{14} & - \frac{1}{2}J^7 +
J^8 - J^9 & J^5 - J^6 \\ J^{15} & J^{12} - J^{13} & J^{10} - J^{11} & -
\frac{1}{2} J^7 - J^8 + J^9 \end{array} \right).
\eea
The $sl(2)$ embedding we consider is $t_0 = t_7,\; t_+ = t_2,\; t_- = t_{12}$.
The grading is given by
\be
\label{gradingsl4}
{\cal G} = {\cal G}^{(-1)} \oplus {\cal G}^{(0)} \oplus {\cal G}^{(1)},
\ee
where
\bea
{\cal G}^{(-1)} &=& {\rm span}\{t_{12},t_{13},t_{14},t_{15}\}; \hspace{1cm}
{\cal G}^{(0)} = {\rm span}\{t_{5},t_{6},t_{7},t_{8},t_9
t_{10},t_{11}\} \\ \nn
{\cal G}^{(1)} &=& {\rm span}\{t_{1},t_{2},t_{3},t_{4}\}; \hspace{1cm}
{\cal G}_{lw} = {\cal G}^{(-1)} \oplus {\rm span}\{t_{5},t_{8},t_{10}\}.
\eea
 From (\ref{gradingsl4}) we see that the grading is integral, so there is
no problem with second class constraints. The one-dimensional
representation $\chi: {\cal G}^{(1)} \rightarrow {\bf R}$ is
given by $\chi(t_2) = 1$ and $\chi(t_1) = \chi(t_3) = \chi(t_4) = 0$.
The constraints therefore read $J^1 = J^3 = J^4 = 0$ and $J^2 = 1$.
Performing the calculation of ${\cal W}$ as before, we find that the
finite W-algebra associated to this $sl(2)$ embedding can be written
as
\bea {}
\{L^a,L^b\} &=& f^{ab}_c L^c; \hspace{1cm}
\{L^a,R^b\} = f^{ab}_c R^c \nn \\ {}
\{R^a,R^b\} &=& (-2 H - C_2) f^{ab}_c L^c,
\eea
where $a = 1,2,3$ and $f_c^{ab}$ are the structure constants of $sl(2)$.
Note that this algebra is essentially the Runge-Lenz algebra
which is the symmetry algebra of a particle moving on $S^3$ in a
Coulomb potential, as we have seen in the previous chapter.
\subsection{Classical Miura transformation}
\label{miura_sect}
A careful examination of equations (\ref{x}) and (\ref{gaugefix})
reveals that the gauge
invariant generators  $W^{\bar{\alpha}}$ in general
have a very specific form. Let
$t_{\bar{\alpha}}$ be a lowest weight vector, i.e. $[t_-,t_{\bar{\alpha}}]=0$,
then
\be
W^{\bar{\alpha}}=W^{\bar{\alpha}}_-+W^{\bar{\alpha}}_0,
\ee
where $W^{\bar{\alpha}}_0$ contains all terms that only contain
$J^{\bar{\beta}}$ with $t_{\bar{\beta}} \in {\cal G}_0$, and
$W^{\bar{\alpha}}_-$ is the sum of the remaining terms.
Furthermore it turns out that $W^{\bar{\alpha}}_0 \neq 0$. From this
last fact we can derive a very important result. As the Poisson
bracket of two elements of degree zero is again of degree zero,
we find that we must have
\be
\{W^{\bar{\alpha}},W^{\bar{\beta}}\}_0=
\{W^{\bar{\alpha}}_0,W^{\bar{\alpha}}_0\},
\ee
where the subscript $0$  in $\{.,.\}$ means that we throw away everything
except the grade zero piece of (whatever comes out of) the Poisson bracket.
What this equation says is
that the $\{W^{\bar{\alpha}}_0\}$ form an algebra which is isomorphic
to the  algebra satisfied by the $\{W^{\bar{\alpha}}\}$. However, note
that $W_0^{\bar{\alpha}}$ only contains $J^{\bar{\beta}}$ of degree
zero which means that what we have done is to embed the finite $W$-algebra
into the Kirillov Poisson algebra of the semi-simple Lie
algebra ${\cal G}_0$ (denoted by
$K({\cal G}_0)$). The map
\be
{\cal W} \longrightarrow K({\cal G}_0)
\ee
is called the `classical (finite) Miura transformation'.

Let us consider an example. In the case of $sl(2)$ we have seen
that $W=J^3+J^0J^0\equiv W_-+W_0$, where $W_0=J^0J^0$.
This example was trivial due to the fact that it had only
one generator. More generally, however, in the case of the so
called `principal $sl(2)$ embeddings' into
$sl(n)$ we get abelian finite $W$-algebras with $n-1$ generators.
The reason for this is that for principal embeddings ${\cal G}_0$
is equal to the Cartan subalgebra, which is an abelian algebra.
The $W$-algebra is therefore also abelian. The non-trivial cases
arise for the non-principal embeddings.

Take for example again the case of ${\cal G}=sl(3)$ with the non-principal
embedding. The generators $W^{\bar{\alpha}}$ of this algebra were given in
(\ref{Wgen}). As ${\cal G}_0$ was the span of $t_4,t_5,t_2$ and $t_6$
we find that the grade 0 pieces of the generators are given by:
\bea
W^4_0 & = & J^4 \nn \\
W^6_0 & = & J^6 \nn \\
W^7_0 & = & \frac{1}{2}J^2J^5-3J^2J^4 \nn \\
W^8_0 & = & \frac{1}{4}J^5J^5+J^2J^6.
\eea
Now, defining
\bea
h & = & -\frac{3}{2}J^4-\frac{1}{4}J^5 \nn \\
e & = & J^2 \nn \\
f & = & J^6 \nn \\
s & = & \frac{1}{2}J^4-\frac{1}{4}J^5,
\eea
which satisfy the Poisson relations of $sl(2) \oplus u(1)$
\bea
\{h,e\} & = & e \nn \\
\{h,f\} & = & -f \nn \\
\{e,f\} & = & 2h \label{sl2u1},
\eea
we find, using equation (\ref{W32gen}), the following expressions
of the generators $j_0,j_+,j_-$ and $C$ in terms of $h,e,f$ and $s$:
\bea
j_0 & = & 2h-2s \nn \\
j_+ & = & e(h-3s) \nn \\
j_- & = & \frac{4}{3}f \nn \\
C & = & -\frac{4}{3}(h^2+ef+3s^2) \label{Miu}.
\eea
Using the relations (\ref{sl2u1}), one can easily verify that
these expressions indeed satisfy the algebra (\ref{Wsl3}). Note that
due to the fact that the expressions in the right hand side
of (\ref{Miu}) are quadratic in the cases of $j_+$ and $C$ it is not
possible
to express $h,e,f$ and $s$ similarly in terms of $j_0,j_\pm,C$, i.e.
to invert the Miura transformation. The reason for this
is that the Miura transformation is in general an homomorphism rather
than an isomorphism.

In the case of the Runge-Lenz algebra, which we constructed above
as a reduction of $sl(4)$, ${\cal G}_0$ is given by $sl(2)
\oplus sl(2) \oplus u(1)$. The explicit expressions of $L^a$ and $R^a$
in terms of the generators of ${\cal G}_0=sl(2)\oplus sl(2) \oplus u(1)$
can be found in \cite{thesis}.

\subsection{General form of $sl(2)$ related finite $W$-algebras}
Consider the plane in $\cal{G}$ made up of elements of the form
\be
x=t_++\sum_{t_{\bar{\alpha}}\in {\cal G}_{lw}}
x^{\bar{\alpha}}t_{\bar{\alpha}},
\label{vorm}
\ee
where $x^{\bar{\alpha}}$ are complex numbers.
That is, we put all $J^a$ to zero except of course $J^+$,
corresponding to $t_+$, which we put to 1 and  $J^{\bar{\alpha}}$
with $t_{\bar{\alpha}}\in
{\cal G}_{lw}$, which we leave unconstrained. It can be shown \cite{BT} that
this set of constraints (which is larger than the set of constraints
that lead to the constraint surface $C$) is completely second class,
that is, none of the above mentioned constraints are first class.
This means that the original Poisson bracket on $\cal{G}$, the
Kirillov Poisson bracket, induces a `Dirac
bracket' on the set of elements of the form (\ref{vorm}).
Recall that for a set of second class constraints $\phi^i$ the
Dirac bracket between two functions $f$ and $g$ is defined by
\be
\{ f,g\}_D=\{f,g\}-\{f,\phi^i\}\Delta_{ij}\{\phi^j,g\},
\ee
where $\Delta_{ij}$ is the inverse of the
matrix $\Delta^{ij}=\{\phi^i,\phi^j\}$.

Now, since the set of elements of the form (\ref{vorm}) is nothing but
$C/\cal{G}_-$, i.e. $C$ with all the gauge freedom removed we find that
the Dirac bracket algebra of the functions $J^{\bar{\alpha}}$ (for
$t_{\bar{\alpha}} \in {\cal G}_{lw}$) must be isomorphic to the
finite $W$-algebra associated to the $sl(2)$ embedding in question.

It is possible \cite{BT} to derive an elegant general formula for the
Dirac bracket algebra on ${\cal G}_{lw}$, and thus of the
finite $W$-algebra. Let $w\in {\cal G}_{lw}$ and let $Q_1$ and $Q_2$ be
smooth functions on ${\cal G}_{lw}$. Define $\mbox{grad}_wQ
\in {\cal G}_{hw}\equiv \{ x\in {\cal G}\mid [t_+,x]=0\}$ by
\be
(w',\mbox{grad}_wQ)=\frac{d}{d\epsilon}Q(w+\epsilon w')\mid_{\epsilon =0},
\ee
for all  $w \in {\cal G}_{lw}$. This uniquely defines  $\mbox{grad}_wQ$,
since ${\cal G}_{lw}$ and ${\cal G}_{hw}$ are non-degenerately paired
by the Cartan-Killing form $(.,.)$. The finite $W$-algebra associated
to the $sl(2)$ embedding in question is now (isomorphic to)
\be
\{Q_1,Q_2\}(w)=(w,[\mbox{grad}_wQ_1,\frac{1}{1+L\circ
 {\rm ad}_w},\mbox{grad}_wQ_2]),
\label{genkir}
\ee
where ${\rm ad}_w(.)=[w,.]$. The operator $L$ is essentially the inverse
of ${\rm ad}_{t_+}$. There is nevertheless a
problem, since $\mbox{Ker}({\rm ad}_{t_+})\neq 0$.
It {\em is} however a 1-1 map from $\mbox{Im}(
{\rm ad}_{t_-})$ to the image of
${\rm ad}_{t_+}$. One now takes $L$ to be the inverse of ${\rm ad}_{t_+}$
in this domain of definition and then extends it to all of $\cal{G}$ by 0.

Note that formula (\ref{genkir}) reduces to the ordinary
Kirillov Poisson bracket for the trivial $sl(2)$ embedding $t_0=t_-=t_+=0$.
For non-trivial $sl(2)$ embeddings however (\ref{genkir}) is a
highly non-trivial and in general non-linear Poisson structure as we have seen.

\subsection{$W$-coadjoint orbits}
In this section we are going to construct the coadjoint orbits  of finite
$W$-algebras associated to $sl_2$ embeddings. First we briefly  discuss
some general aspects of symplectic and coadjoint orbits and then we show
how one can construct finite $W$ symplectic orbits from the coadjoint orbits
of the Lie algebra to which they are associated.

Suppose we are given a Poisson manifold, that is a manifold $M$
together with a Poisson bracket $\{.,.\}$ on the space of smooth
functions on $M$. In general such a Poisson structure is not
associated to a symplectic form on $M$ since the Poisson bracket
can be degenerate, i.e. there may exist functions that Poisson
commute with all other functions. The Poisson bracket $\{.,.\}$
does however induce a symplectic form on certain submanifolds
of $M$. In order to see this consider the set of Hamiltonian
vector fields
\be
X_f(\cdot )=\{f,.\} \;\;\;\; \mbox{ for } f\in C^{\infty}(M).
\ee
Note that the set of functions on $M$ that Poisson commute with
all other functions are in the kernel of the map $f\mapsto X_f$.
 From this follows that in every tangent space
the span of all Hamiltonian vector fields is only a subspace of the
whole tangent space. We say that a Poisson structure is regular,
if the span of the set of Hamiltonian vector fields has the same
dimension in every tangent space.
Obviously a regular Poisson structure
defines a tangent system on the manifold $M$
and from the well known relation
\be
[X_f,X_g]=X_{\{f,g\}}
\ee
it then immediately follows that this system is integrable (in the
sense of Frobenius). Therefore $M$ foliates into a disjoint union of
integral manifolds of the Hamiltonian vector fields. Obviously the
restriction of the Poisson bracket to one of these integral manifolds
is nondegenerate and therefore associated to a symplectic form.
They are therefore called symplectic leaves.

The symplectic leaves play an important role in the representation
theory of Poisson algebras. A representation of a Poisson algebra
is a symplectic manifold $S$ together with a map $\pi$ from the
Poisson algebra to the space of smooth functions on $S$ that
is linear and  preserves both the multiplicative and Poisson
structure of the Poisson algebra. Also the Hamiltonian vector field
associated to $\pi (f) \in C^{\infty}(S)$ must be complete (i.e. defined
everywhere).
A representation is called {\em irreducible}, if span of the set
$\{X_{\pi (f)}(s)\mid f\in C^{\infty}(M)\}$ is equal to the tangent
space of $S$ in $s$ for all $s\in S$.
The role
of the symplectic leaves is clarified by the following theorem
\cite{NPL}. If a representation of the Poisson algebra $C^{\infty}(M)$
is irreducible,
then $S$ is symplectomorphic to a covering space of a symplectic leaf
of $M$.

 From the above it follows that it is rather important to
construct the symplectic leaves associated to finite $W$-algebras.
Looking at eq.(\ref{genkir}) it is clear that constructing these
symplectic orbits from scratch could be rather difficult. Luckily
we can use the fact that the symplectic orbits of the Kirillov
Poisson structure are known (by the famous Kostant-Souriau theorem
they are nothing but the coadjoint orbits and their covering spaces)
to construct these. The answer turns out to be extremely simple
and is given in the following theorem:\\[3mm]
\noindent
{\bf Theorem}
{\it Let ${\cal O}$ be a coadjoint orbit of the Lie algebra $g$, then
the intersection of ${\cal O}$ and $g_{lw}$ is a symplectic orbit
of the finite $W$-algebra $W(g;t_0)$.}\\[3mm]
\noindent
Proof: Let again $C=\{x\in g \mid \phi^{\alpha}(x)=0\}$. The
Hamiltonian vector fields $X^{\alpha}=\{\phi^{\alpha},.\}$ form
an involutive system tangent to  $C$ and therefore $C$ foliates.
$g_{lw}$ has one point in common with every leaf and denote the
canonical projection from $C$ to $g_{lw}$, which projects an element
$x\in C$ to the unique point $x'\in g_{lw}$, by $\pi$. This map
induces a map $\pi_*$ from the tangent bundle $TC$ of $C$ to
the tangent bundle $Tg_{lw}$ of $g_{lw}$. Let $f$ be a `gauge invariant'
function on $C$, i.e. $\{f,\phi_{\alpha}\}|_{C}=0$ then obviously
$X_f=\{f,.\}$ is a section of $TC$ (i.e. as a vector field it is
tangent to $C$ in every point of $C$). It need not be an element
of $Tg_{lw}$ however, but using the gauge invariance we can
project it back onto the gauge slice. This projection of $X_f$
on the the gauge slice is given by $\pi_*(X_f)\in Tg_{lw}$. By
construction the Dirac bracket now has the property that
$\{f,.\}^*=\pi_*(X_f)$. What we now need to show is that $\pi_*(X_f)$
is tangent to the coadjoint orbit $\cal O$. Well, obviously we have
$\pi_*(X_f)=X_f+Y$, where $Y$ is a tangent to the gauge orbit. Since
the spaces tangent to the gauge orbits are spanned by the Hamiltonian
vector fields $X^{\alpha}$, we find that $Y$ is a tangent vector of
$\cal O$. Because $X_f$ is by definition a tangent vector of $\cal O$ we see
that $\pi_*(X_f)$ is also tangent to the coadjoint orbit $\cal O$.
What we can conclude is that the Hamiltonian vector fields
on $g_{lw}$ w.r.t. the
reduced Poisson bracket are all tangent to the coadjoint orbit and
therefore the symplectic orbits are submanifolds of the coadjoint
orbits. The theorem now follows immediately \vspace{5mm}.

Let us now recall some basic facts on coadjoint orbits of simple
Lie algebras. Let $f$ be an adinvariant function on $g$, i.e.
$f(axa^{-1})=f(x)$ for all $a\in G$ and $x\in g$. Then we have
\ba
0 & = & \frac{d}{dt}f(e^{tx}ye^{-tx})|_{t=0}=\frac{d}{dt}
f(y+t[x,y])|_{t=0}
\nonumber \\
& = & \left( \mbox{grad}_yf,[x,y]\right) = \left( [\mbox{grad}_yf,x],y
\right),
\ea
which means that
\be
\left( [\mbox{grad}_yf,\mbox{grad}_yh],y\right) = \{f,h\}(y)=0.
\ee
We conclude that any adinvariant function Poisson commutes with all
other functions. Conversely any function that Poisson commutes with all
others is adinvariant, because $\{f,h\}=0$ for all $h$ implies that
$X_h(f)=0$ for all $h$, where $X_h$ is the Hamiltonian vector field
associated to $h$, which means that the derivative of $f$ in all
directions tangent to a symplectic orbit are zero.

Consider now the Casimir functions $\{C_i\}_{i=1}^{\mbox{rank}(g)}$
of the Lie algebra $g$. Certainly these Poisson commute with all
other functions and it can in fact be shown that the (co)adjoint (since
we have identified the Lie algebra with its dual the adjoint and
coadjoint orbits coincide)
orbits of maximal dimension of the Lie algebra are given by their
constant sets
\be
C_{\vec{\mu}}=\{x\in g \mid C_i(x)=\mu_i\, ;\; \mu_i \in \ce \, ;\;
i=1,\ldots ,\mbox{rank}(g)\}.
\ee
Their dimension is therefore $\mbox{dim}_{\ce} (g)-\mbox{rank}(g)$.

Coadjoint orbits are very important to the representation theory
of (semi)simple Lie groups. In order to see why let us recall the
Borel-Weil-Bott (BWB) theorem \cite{Bott}. Let $G$ be a compact
semisimple Lie group with maximal torus $T$ and let $R$ be a finite
dimensional irreducible representation of $G$. The space of highest
weight vectors of $R$ is 1 dimensional (call it $V$) and furnishes
a representation of $T$. The product space $G\times V$ is a
trivial line bundle over $G$ and its quotient by the action of $T$
(where $(a,v)\sim (at,t^{-1}v)$ for $a\in G$ and $t\in T$) is a
holomorphic line bundle over $G/T$ (which is a complex manifold). Now,
$L$ admits a $G$ action $(a,v)\rightarrow (a'a,v)$, where $a,a' \in G$
and $v\in V$, so the space of holomorphic sections of $L$ is naturally
a $G$ representation. The BWB theorem now states that this representation
is isomorphic to $R$. The BWB construction has however the restriction
that it applies only to compact groups. There does exist however a
generalization of the BWB construction called the coadjoint method
of Kirillov. In this method one generalizes $G/T$ to  certain
homogeneous spaces $G/H$ which can be realized as coadjoint orbits.
As we have mentioned above, a coadjoint orbit carries a natural
($G$ invariant) symplectic form $\omega$ inherited from the Kirillov
Poisson structure and can therefore be seen as a phase space of some
classical mechanical system. One then attempts to quantize this symplectic
manifold using the methods of geometric quantization \cite{geomquant}.
For this one is supposed to construct a holomorphic line bundle $L$
over the coadjoint orbit such that the first Chern class of $L$ is
equal to the (cohomology class of)
$\omega$. If such a line bundle exists the coadjoint orbit
is called quantizable. For a quantizable orbit the bundle $L$ can then
be shown to admit a hermitian metric whose curvature is equal to $\omega$.
The space of sections of $L$ thus obtains a Hilbert space structure
and is interpreted as the physical Hilbert space of the quantum system
associated to the coadjoint orbit. It carries a unitary representation
of the group $G$ and in fact one attempts to construct all unitary
representations in this way. Obviously this construction is a
generalization of the BWB method. Note that the fact that generic groups
have unitary irreducible representations only for a discrete set of
highest weights is translated into the fact that only a discrete
set of orbits is quantizable.

 From the above it is clear that the symplectic (or $W$-coadjoint) orbits
of finite $W$-algebras are extremely important especially in the
study of global aspects of $W$-algebras. In order to see this note
that coadjoint orbits are homogeneous spaces of the Lie group $G$
in question, and it is therefore tempting to interpret the symplectic
orbits of $W$-algebras as  $W$-{\em homogeneous spaces}. These
naturally carry much information on the global aspects of $W$-transformations.
In the notation used above
the symplectic orbits of the finite $W$-algebra $W(g;t_0)$ are given
by
\be
{\cal O}_{\vec{\mu}}=C_{\vec{\mu}} \bigcap g_{lw},
\ee
which are therefore (generically) of dimension $\mbox{dim}_{\ce} (g_{lw})-
\mbox{rank}(g)$.

Let us give some  examples. As we have seen, the finite $W$-algebras
associated to principal $sl_2$ embeddings are trivial in the sense
that they are Poisson abelian. This means that all Hamiltonian
vector fields are zero and that all symplectic orbits are points.
Again the simplest nontrivial case is the algebra $W_3^{(2)}$.
The 2 independent Casimir functions of $sl_3$ are
\be
C_1=\frac{1}{2} \mbox{Tr}(J^2) \;\;\; \mbox{ and } C_2 =
\frac{1}{3} \mbox{Tr}(J^3),
\ee
where $J=J^at_a$. The spaces ${\cal O}_{\vec{\mu}}$ can be
constructed by taking for $J$ the constrained and gauge fixed matrix
\be
J= \left(
\begin{array}{ccc}
j & 0 & 1 \\
G_+ & -2j & 0 \\
T & G_- & j
\end{array}
\right).
\ee
One then easily finds
\ba
C_1 & = & 3j^2+T \nonumber \\
C_2 & = & -2j^3 + G_+G_- +2jT.
\ea
Obviously $\mbox{dim}(g_{lw})=4$ and introducing the variables
$z_1=j; \; z_2=G_+ ;\; z_3=G_-;\; z_4=T$ we find that
\be
{\cal O}_{\mu_1,\mu_2}=\{\vec{z}\in \ce^4 \mid 3z_1^2+z_4=\mu_1
\; \mbox{ and } \; 2z_1z_4+z_2z_3-2z_1^3=\mu_2 \}.
\ee
The topological nature of these symplectic orbits becomes clearer if we
insert $z_4=\mu_1-3z_1^2$ into the second equation. We find that
topologically ${\cal O}_{\mu_1,\mu_2}$ is equivalent to the 2 dimensional
surface in $\ce^3$ determined by the equation
\be
z_2z_3=\mu_2-2\mu_1z_1+8z_1^3.
\ee
Compare this to the coadjoint orbits of $sl(2,\ce )$ which are given  by
\cite{Wittencoad}
\be
z_2z_3=z^2_1+\mu .
\ee
It is well known that the coadjoint orbits of $sl(2,{\bf R})$ and
$su(2)$, which are real forms of $sl(2,\ce )$, can be found by considering
the intersection of these complex coadjoint orbits with
the appropriate real subspaces of $sl(2,\ce )$.

In this section we have seen that it is easy to find the coadjoint orbits
of finite $W$-algebras by using the fact that they are reductions of
Lie algebras. Constructing these orbits would have been much more
complicated, if not practically impossible, if this information had not
been used.

\subsection{Semi-simple subalgebras of finite $W$-algebras}
\label{subalg_sect}
As we have seen most finite $W$-algebras are non-linear, i.e. the commutation
relations (Poisson brackets)
close on polynomials, not just linear combinations, of the generators.
It is however not uncommon for a finite $W$-algebra to have a
linear subalgebra. We have actually already seen an example of this in
the case of the Runge-Lenz algebra which was a reduction of $sl(4)$.
The generators $L^a$ in this algebra formed an $sl(2)$ subalgebra of
the Runge-Lenz algebra (which as a whole was non-linear). In this section
we will discuss how linear subalgebras arise and how one can predict them.

For this we need some basic facts on the theory of $sl(2)$ embeddings.
As is well known \cite{Dynkin}
$sl(2)$ embeddings into $sl(n)$ are completely characterized by the
way the fundamental (or defining) representation of $sl(n)$ branches
into irreducible $sl(2)$ representations. Furthermore, all
conceivable branchings are possible. So the fundamental ($n$-dimensional)
representation of $sl(n)$ can branch into a direct sum of
$n_1,n_2,n_3,...$ dimensional $sl(2)$ representations as long as
we have $n = n_1 + n_2 + n_3 + \cdots$. The number of
inequivalent $sl(2)$ embeddings is therefore equal to the number of
partitions of the number $n$. Conversely, any partition
determines an $sl(2)$ embedding up to inner automorphisms.

We are now ready to explain when semi-simple subalgebras of
finite $W$-algebras arise \cite{BT}. Suppose a certain $sl(2)$
representation occurs $m$ times in the branching of the
fundamental representation of $sl(n)$ then there will be an $sl(m)$
subalgebra in the resulting finite $W$-algebra. For example, in the
case of the Runge-lenz algebra one can easily check, using the
explicit form of the generators $t_0,t_+$ and $t_-$, that the 4
dimensional fundamental representation of $sl(4)$ decomposes into a
direct sum of two 2-dimensional representations of $sl(2)$. The
partition of 4 corresponding to this branching is $4=2+2$.
The fact that the same representation occurs twice immediately
leads us to expect that there will be an $sl(2)$ subalgebra in the
resulting finite W-algebra. Indeed we know this to be the case.
Another, rather trivial example is the case of the trivial $sl(2)$
embedding $t_0=t_{\pm}=0$. This case corresponds to the
partition $n=1+1+\ldots +1$, i.e. the singlet of $sl(2)$ is $n$ fold
degenerate. We therefore expect an $sl(n)$ subalgebra. This is
of course the original Lie algebra itself.

A much more interesting question is whether there exist non-linear
extensions of the gauge group $SU(3)\times SU(2) \times U(1)$ of the
standard model,
i.e. finite $W$-algebras that contain its Lie algebra as a subalgebra.
The answer turns out to be affirmative. The smallest $W$-algebra
that contains such a subalgebra must be a reduction of at least
$sl(7)$ because we need in the branching of the fundamental
representation  at least a double and a triple degeneracy.
The smallest $n$ for which it is possible to do this is 7.
The branching that does the trick is $7=1+1+1+2+2$.
Due to the fact that the singlet "1" has a threefold degeneracy
and the doublet "2" a double degeneracy we conclude that
the finite $W$-algebra will contain an $sl(3)$
and an $sl(2)$ subalgebra. The $u(1)$ subalgebra will be
there automatically due to the fact that the second order Casimir descends to
the finite $W$-algebra and becomes a generator of it.
The step from $sl(n)$ to $su(n)$ is then
made by taking the appropriate real form of the
$W$-algebra, but we will come back to this later.

We conclude that it is not difficult to construct non-linear extensions of
the gauge group of the standard model. The next step would of course be
to construct an actual gauge theory with this non-linear gauge algebra.
In doing so one might be helped by the fact that it is
a reduction of, and can be embedded into (by the Miura
transformation), a Lie algebra. An obvious procedure would therefore be
to `break' the symmetry of the linear theory to a non-linear subsymmetry
by adding terms to the action which are only invariant under the
transformations generated by the non-linear gauge algebra.
This is the principle on which for example Toda theories base their
non-linear $W$-symmetry (we will come back to this later).

The $W$-algebra associated to the branching $7=2+2+1+1+1$ can be embedded
into (the universal enveloping algebra of) $su(5)\oplus su(2)$. One
method of writing down a gauge theory based on this $W$-algebra would therefore
be to first consider an $su(5)\oplus su(2)$ gauge theory and then to add
to its action terms invariant under the transformations generated by the
$W$-algebra only. A further discussion of finite $W$ gauge theories will
be postponed until chapter~6.

\chapter{Quantum Finite $W$-algebras}
In principle a complete
quantum description of finite $W$-algebras involves two things:
First we need to
quantize the classical algebras constructed in the
previous chapter by associating
to them appropriate non-commutative associative algebras.
We come to this in a moment.
The second step is to construct the irreducible unitary
representations of these quantum algebras on Hilbert spaces.

Unfortunately the general
problem of quantizing finite $W$-algebras is not solved.
However, in certain special cases,
like the $W$-algebras associated to $sl(2)$ embeddings, the
quantization {\em is} known
\cite{BT}. In this chapter we discuss these results and
add some new ones on finite
$W$-algebras that are obtained by reduction w.r.t. Cartan subalgebras.
In the next chapter we
discuss unitary highest weight representations of
finite $W$-algebras and describe
some conjectures on Kac determinants and character formulas.

In quantum mechanics, quantization amounts to replacing Poisson
brackets by commutators, sometimes denoted by
\be \{ \cdot , \cdot \} \rightarrow \frac{i}{\hbar} [ \cdot , \cdot ]
\ee
In a mathematically more sophisticated language, this amounts to
replacing a Poisson algebra, which is commutative and associative,
$({\cal A}_0,\{.,.\} )$ by an associative but
non-commutative algebra ${\cal A}$ depending on a parameter $\hbar$,
such that
\begin{itemize}
\item ${\cal A}/\hbar {\cal A}\simeq {\cal A}_0$
\item if $\pi$ denotes the natural map $\pi : {\cal A} \rightarrow {\cal
A}/\hbar {\cal A}\simeq {\cal A}_0$, then
\be
\{\pi(X),\pi(Y)\}=
\pi((XY-YX)/\hbar).
\ee
\end{itemize}
The first condition simply expresses the fact that in the limit $\hbar
\rightarrow 0$ the original algebra is recovered from the new one, while
the second one tells you how to extract the Poisson brackets of the
old algebra from the commutators of the new algebra.

In most cases one has a set of generators
for ${\cal A}_0$, and ${\cal A}$ is completely fixed by giving
the commutation relations of these generators.

For example, let ${\cal A}_0$ be the Kirillov Poisson algebra $K({\G})$
associated to ${\G}$, then a quantization of
this Poisson algebra is the algebra
${\cal A}$
generated by the $J^a$ and $\hbar$, subject to the commutation relations
$[J^a,J^b]=\hbar f^{ab}_{\,\,\, c} J^c$ (note that here the $J^a$ are
no longer functions on ${\G}$ but the quantum objects associated
to them). Obviously, the Jacobi
identities are satisfied. Specializing to
$\hbar=1$, this algebra is precisely the universal enveloping
algebra ${\cal U}{\G}$ of  ${\G}$.

To find quantizations of finite $W$-algebras, one can first
reduce the $sl_n$ Kirillov Poisson algebra, and then try to
quantize the resulting algebras that we studied in the previous
sections. On the other hand, on can also first quantize and then
constrain. We will follow the latter approach, and thus study
the reductions of the quantum Kirillov algebra
\be \label{eq:qkir}
[J^a,J^b]=\hbar f^{ab}_{\,\,\, c} J^c.
\ee
We want to impose the same constraints on this algebra as we
imposed previously on the Kirillov Poisson algebra, to obtain
the quantum versions of the finite $W$-algebras $({\cal G},
{\cal L},\chi)$. Imposing constraints on quantum algebras can
be done using the
BRST formalism \cite{KoSt}, which is what we
will use in the next sections. It is important to realize that
we do by no means prove uniqueness of the quantization of the
finite $W$-algebra. There may exist other, non-equivalent quantizations.
The one that one finds by applying the BRST formalism starts with
a specific quantization of the Kirillov algebra, namely (\ref{eq:qkir}),
and leads to one specific quantization of the finite $W$-algebra.
Clearly, this is the most natural one from the Lie-algebraic point
of view, but it is an interesting open problem to find out whether
or not other quantizations of finite $W$-algebras exist. One might
for example examine what happens when one starts with a quantum group
rather than the quantum Kirillov algebra. This might lead to
$q$-deformations of finite $W$-algebras, which can -if they exist-
be seen as two-parameter quantizations of classical finite $W$-algebras.

\sectiona{BRST quantization of finite $W$-algebras}
Recall that the constraints we want to impose read
\be
J^{\alpha}-\chi(J^{\alpha})=0.
\ee
To make this into good quantum constraints, $\chi$ must be promoted
to a representation of the Lie algebra ${\cal L}$
into some associative algebra, which is a quantization of the
Poisson algebra in which ${\cal L}$ was represented classically
by means of $\chi$. This associative algebra will be denoted by
${\cal U}_{\chi}$. In the cases where $\chi=0$ or $\chi$ is a
one-dimensional character, this is trivial. For all other cases
we will simply assume this has been accomplished in some or another
way. Furthermore we
will take $\hbar=1$ for simplicity; the explicit $\hbar$
dependence can be determined afterwards.

To set up the BRST framework
we need to introduce anticommuting
ghosts and antighosts $c_{\alpha}$ and $b^{\alpha}$, associated
to the constraints that we want to impose \cite{KoSt}. They satisfy
$b^{\alpha}c_{\beta}+c_{\beta}b^{\alpha}=
\delta^{\alpha}_{\beta}$ and generate
the Clifford algebra $Cl({\cal L}\oplus {\cal L}^*)$.
The quantum Kirillov algebra is
just the  universal enveloping algebra ${\cal U}{\G}$,
and the total
space on which the BRST operator acts is $\Omega={\cal
U}{\G}\otimes {\cal U}_{\chi} \otimes Cl({\cal L}\oplus {\cal L}^*)$.
A ${\bf Z}$
grading on $\Omega$ is defined by
$\deg(J^a)=\deg({\cal U}_{\chi})=0$, $\deg(c_{\alpha})=+1$
and $\deg(b^{\alpha})=-1$,
and we can decompose $\Omega=\oplus_k \Omega^k $ accordingly.
The BRST differential on $\Omega$ is given by $d(X)=[Q,X]$,
where $Q$ is the BRST charge
\be \label{def:q}
Q=(J^{\alpha}-\chi(J^{\alpha}))c_{\alpha}-\hf
f^{\alpha\beta}_{\,\,\, \gamma} b^{\gamma}c_{\alpha}c_{\beta},
\ee
and $[.,.]$
denotes the graded commutator (as it always will from now on)
\be
[A,B]=AB- (-1)^{\mbox{deg}(A).\mbox{deg}(B)}BA.
\ee
Note that $\mbox{deg}(Q)=1$.
First-class constraints generate gauge transformations, and $Q$ acts
on generators precisely as gauge transformations, with the parameters
replaced by the anticommuting ghosts $c_{\alpha}$. Roughly, the
anti-ghosts $b^{\alpha}$ are needed to impose the first class constraints,
whereas the ghosts $c_{\alpha}$ are needed to perform the gauge fixing
on the constrained phase space.

The standard BRST complex associated to the first-class constraints
is
\be
\cdots \stackrel{d}{\rightarrow} \Omega^{k-1}
\stackrel{d}{\rightarrow} \Omega^{k}
\stackrel{d}{\rightarrow} \Omega^{k+1}
\stackrel{d}{\rightarrow}\cdots
\ee
Of interest are the cohomology groups of this complex, $H^k(\Omega;d)$,
defined by
\be
H^k(\Omega;d)=\frac{{ \rm ker}\, d : \Omega^{k-1} \rightarrow \Omega^k }{
{ \rm im}\, d : \Omega^{k} \rightarrow \Omega^{k+1}}.
\ee
By definition, the zeroth cohomology group is the quantization of the finite
$W$-algebra. It is straightforward to verify that the zeroth cohomology
is indeed a closed, associative algebra. To show that it is a quantization of
the classical finite $W$-algebra is less straightforward, but can also be done.
If the gauge group $\hat{H}$ generated by the first class constraints
acts properly on the constrained phase space $C$, one expects that all higher
cohomologies vanish, as they are generically related to singularities in the
quotient $C/\hat{H}$. In particular, if $\chi$ is a one-dimensional character
related to an $sl_2$ embedding, the higher cohomologies will vanish, but if
it is zero, they will not. Later on we will see examples of this phenomenon.
In the mathematics
literature the cohomology of the BRST complex is called the Hecke
algebra ${\cal H}\tripel$ associated to
$\tripel$. General Hecke algebras have not been computed, apart from those
where $\chi$ is the one-dimensional character related to the principal
$sl_2$ embedding in ${\cal G}$. In that case it was
shown by Kostant \cite{kost2} that the only non-vanishing
cohomology is $H^0(\Omega;d)$ and that it is isomorphic to the
center of the universal enveloping algebra. Recall that the
center of the ${\cal U}{\G}$ is generated by the set of independent
Casimirs of ${\G}$. This set is closely related to
the generators of standard infinite $W_n$-algebras; in that case
there is one $W$-field for each Casimir which
form a highly non-trivial algebra \cite{BBSS}. We see that for
finite $W$-algebras the same generators survive, but that
they form a trivial abelian algebra.

Since in the cases where $\chi$ is a character related to an $sl_2$
embedding the BRST cohomology can be completely calculated, we will
first restrict our attention to this situation. In particular, we will
verify Kostants result, but also see that for non
principal $sl_2$ embeddings quantum finite $W$-algebras are non-trivial.
Later we will come back
to the more general situation, in particular to $\chi=0$. Here,
the general answer is not known, and related to some interesting and
difficult open problems in mathematics. Therefore, we will restrict
our attention to a few examples to sketch the general idea. But before
that we first consider the cases related to $sl_2$ embeddings.

\sectiona{Quantum finite $W$-algebras from $sl_2$ embeddings}

As said previously, we want to compute the cohomology of $(\Omega;d)$.
Unfortunately this is a difficult problem to approach directly. We
therefore use a well known trick of cohomology theory, namely we split
the complex $(\Omega ;d)$ into a double complex and calculate the
cohomology via a spectral sequence argument. We will just sketch the
idea, a more detailed treatment can be found in $\cite{deBoerTjin}$.
Crucial is
that the operator $d$ can be
decomposed into two anti-commuting pieces. Write $Q=Q_0+Q_1$, with
\ba \label{def:q01}
Q_1 & = & J^{\alpha}c_{\alpha}-\hf
f^{\alpha\beta}_{\,\,\, \gamma} b^{\gamma}c_{\alpha}c_{\beta},
 \nn \\
Q_0 & = & -\chi(J^{\alpha})c_{\alpha},
\ea
and define $d_0(X)=[Q_0,X]$, $d_1(X)=[Q_1,X]$. Then one can
verify by explicit computation that
$d_0^2=d_0d_1=d_1d_0=d_1^2=0$. Associated to this decomposition
is a bigrading of $\Omega=\oplus_{k,l} \Omega^{k,l}$ defined by
\ba \label{def:bigrade}
\deg(J^a)=(-k,k), & & \mbox{ if }t_a\in {\G}_{k}  \nn \\
\deg(c_{\alpha})=(k,1-k), & & \mbox{ if }t_{\alpha} \in {\G}_{k}  \nn \\
\deg(b^{\alpha})=(-k,k-1), & & \mbox{ if }t_{\alpha} \in {\G}_{k},
\ea
with respect to which $d_0$ has degree $(1,0)$ and $d_1$ has
degree $(0,1)$. Thus $(\Omega^{k,l};d_0;d_1)$ has the structure
of a double complex. Explicitly, the action of $d_0$ and $d_1$
is given by
\ba \label{eq:q01}
d_1(J^a) & = & f^{\alpha a}_{\,\,\,b}J^bc_{\alpha},
 \nn \\
d_1(c_{\alpha}) & = & -\hf f^{\beta\gamma}_{\,\,\, \alpha}c_{\beta}
c_{\gamma},  \nn \\
d_1(b^{\alpha}) & = & J^{\alpha} + f^{\alpha \beta}_{\,\,\,
\gamma}b^{\gamma}c_{\beta}, \nn \\
d_0(J^a)=d_0(c_{\alpha}) & = & 0,  \nn \\
d_0(b^{\alpha}) & = & -\chi(J^{\alpha}).
\ea
To simplify the algebra, it is advantageous to introduce
\be \label{def:hj}
\hj^a=J^a+f^{a\beta}_{\,\,\, \gamma}b^{\gamma} c_{\beta}.
\ee
Our motivation to introduce these new elements $\hj^a$ is
twofold: first, similar expressions were encountered
in a study of the effective action for $W_3$ gravity \cite{BG},
where it turned out that the BRST cohomology for the infinite $W_3$
algebra case could conveniently be expressed in terms of
$\hj$'s; second, such expressions were introduced for the $J^a$'s
that live on the Cartan subalgebra of $g$ in \cite{FeFr}, and
simplified their analysis considerably. In terms of $\hj$ we
have
\ba \label{eq:q01hat}
d_1(\hj^a) & = & f^{\alpha a}_{\,\,\,\bar{\gamma}}
\hj^{\bar{\gamma}} c_{\alpha},  \nn \\
d_1(c_{\alpha}) & = & -\hf f^{\beta\gamma}_{\,\,\, \alpha}c_{\beta}
c_{\gamma},  \nn \\
d_1(b^{\alpha}) & = & \hj^{\alpha},  \nn \\
d_0(\hj^a) & = & -f^{a \beta}_{\,\,\, \gamma} \chi(J^{\gamma})
c_{\beta}, \\ \nn
d_0(c_{\alpha}) & = & 0,  \nn \\
d_0(b^{\alpha}) & = & -\chi(J^{\alpha}).
\ea

Now that we have a double complex, we can apply
the techniques of spectral sequences \cite{botttu}
to it, in order to compute the cohomology of $(\Omega;d)$.
The idea behind spectral sequences is to find a a series of
complexes $(\Omega_j;D_j)$, $j\geq 0$, such that
$\Omega_0=\Omega$, such that $\Omega_{j+1}=H^{\ast}(\Omega_j,D_j)$
and such that $\Omega_{\infty}=H^{\ast}(\Omega;d)$.
In practise this is only a convenient technique if the spectral
sequence degenerates at some point, which means that $\Omega_j$
does not change anymore for $j$ larger than some $j_0$.
If a complex is actually a double complex, like in our case, two
natural spectral sequences exist. One of these has $D_0=d_0$ and
$D_1=d_1$, the other one has $D_0=d_1$ and $D_1=d_0$.

The first people to propose using the theory of spectral sequences
in the setting of $W$-algebras were
Feigin and Frenkel \cite{FeFr}.  In fact, in this way they computed the
BRST cohomology in the infinite-dimensional case
but only for the special example of the principal $sl_2$
embedding (which are known to lead to the $W_N$ algebras).
However, their calculation has the drawback that it is
very difficult to generalize to arbitrary embeddings and that it
constructs the cohomology in an indirect way (via commutants of
screening operators). Their spectral sequence was based on a double
complex, with in the notation introduced above, $D_0=d_1$ and $D_1=d_0$.
In \cite{BT} it was first proposed to
calculate the BRST cohomology using the other spectral sequence with
$D_0=d_0$ and $D_1=d_1$.
As it turns out this has drastic simplifying consequences for the
calculation of the cohomology.

\subsection{The BRST cohomology}
\label{coho_sect}
The following theorem \cite{BT} gives the BRST cohomology on the level of
vector spaces.\\[3mm]
\noindent
{\bf Theorem} {\it As before let ${\G}_{lw}\subset {\G}$
be the kernel of the map
$\ad{t_-}:{\G}\rightarrow {\G}$. Then the BRST cohomology is
given by the following isomorphisms of vector spaces
\be \label{coho}
H^{k}(\Omega;d)\simeq ({\cal U}{\G}_{lw})\delta_{k,0}.
\ee }
The computation of the BRST cohomology is simplified
considerably due to the introduction of the new set of generators
$\hj^a$. The simplification arises due to the fact that
$H^{\ast}(\Omega;d)\simeq H^{\ast}(\Omega_{red};d)$, where $\Omega_{red}$ is
the subalgebra of $\Omega$ generated by
$\hj^{\bar{\alpha}}$ and $c_{\alpha}$.
We will not prove this here, but note that it is nothing but the statement
that the anti-ghosts $b^{\alpha}$ impose the constraints. The subalgebra
$\Omega_{red}$ is a quantum version of the constraint phase
space we have seen previously.
The reduced complex $(\Omega_{red};d)$ is described by the
following set of relations:
\ba \label{eq:q01hat2}
d_1(\hj^{\bar{\alpha}}) & = & f^{\alpha \bar{\alpha}}_{\,\,\,
\bar{\gamma}} \hj^{\bar{\gamma}} c_{\alpha},  \nn \\
d_1(c_{\alpha}) & = & -\hf f^{\beta\gamma}_{\,\,\, \alpha}c_{\beta}
c_{\gamma},  \nn \\
d_0(\hj^{\bar{\alpha}}) & = & -f^{{\bar{\alpha}}
\beta}_{\,\,\, \gamma} \chi(J^{\gamma})
c_{\beta},  \nn \\
d_0(c_{\alpha}) & = & 0, \nn \\ \nn
[\hj^{\bar{\alpha}},\hj^{\bar{\beta}}]
& = & f^{\bar{\alpha}\bar{\beta}}_{
\,\,\,\bar{\gamma}} \hj^{\bar{\gamma}},  \\ \nn
[\hj^{\bar{\alpha}},c_{\beta}] & = &
-f^{\bar{\alpha}\gamma}_{\,\,\,\beta} c_{\gamma},  \\ \mbox{}
[c_{\alpha},c_{\beta}] & = & 0.
\ea
At this stage we apply the spectral sequence technique. The inconvenient choice
$D_0=d_1$ and $D_1=d_0$ would lead us to compute the cohomology of $d_1$
on $\Omega_{red}$. This turns out to be very hard, and in addition
the spectral sequence will not degenerate ever. In other words, we would
have to compute cohomologies infinitely many times in order to find the
final answer. The choice $D_0=d_0$ and $D_1=d_1$ is much more convenient,
it turns out that the spectral sequence already degenerates after the
first step, i.e.  after taking
the $D_0$ cohomology, so that $\Omega_{\infty}=\Omega_1$.
Therefore, all that remains is to compute the
$d_0$ cohomology of $\Omega_{red}$.
To get an idea of the $d_0$ cohomology, we rewrite
 $d_0(\hj^{\bar{\alpha}})$ as
 \ba
 \label{rewri}
 d_0(\hj^{\bar{\alpha}}) & = &
 -\tr([\chi(J^{\gamma})t_{\gamma},t^{\bar{\alpha}}] t^{\beta}
 c_{\beta} )  \nn \\
 & =  & -\tr([t_+,t^{\bar{\alpha}}]t^{\beta}c_{\beta} ).
 \ea
 From this it is clear that $d_0(\hj^{\bar{\alpha}})=0$ for
 $t^{\bar{\alpha}}\in {\G}_{hw}$. Furthermore, since $t_{\bar{\alpha}}\in
 {\G}_0\oplus {\G}_-$ and $\dim({\G}_{lw})=\dim({\G}_0)$, it
 follows that for each $\beta$ there is a linear combination
$a(\beta)_{\bar{\alpha}}\hj^{\bar{\alpha}}$ with
 $d_0(a(\beta)_{\bar{\alpha }}\hj^{\bar{\alpha}})=c_{\beta}$.
 This can be used to prove that
 \be \label{E1}
 H^k(\Omega_{red};d_1) =
 \bigotimes_{t_{\bar{\alpha}}\in {\G}_{lw}}
\ce[\hj^{\bar{\alpha}}]\delta_{k,0} =
 ({\cal U}{\G}_{lw})\delta_{k,0}.
 \ee
 Because the spectral sequence stops here, the total cohomology
of $(\Omega;d)$ is the same as (\ref{E1}), and this proves the theorem.
We see that the ghosts $c_{\alpha}$ have been 'used' to keep only
the generators in ${\G}_{lw}$. This is the quantum counterpart of
the gauge fixing procedure we saw previously.

As expected, there is only cohomology of degree
zero, and furthermore, the elements of ${\G}_{lw}$ are in one-to-one
correspondence with the components of ${\G}$ that made up the
lowest weight gauge in section 1. Therefore
$H^{\ast}(\Omega;d)$ really is a quantization of the
finite $W$-algebra. What remains to be done is to compute the algebraic
structure of $H^{\ast}(\Omega;d)$.
What we have computed in the previous theorem is an isomorphism of
vector spaces rather than an isomorphism of algebras.
The only thing that
(\ref{coho}) tells us is that the product of two elements $a$
and $b$ of bidegree $(-p,p)$ and $(-q,q)$ is given by the
product structure on ${\cal U}{\G}_{lw}$, modulo terms of bidegree
$(-r,r)$ with $r<p+q$. To find these lower terms we need
explicit representatives of the generators of
$H^0(\Omega;d)$ in $\Omega$. Such representatives can be
constructed using the so-called tic-tac-toe construction
\cite{botttu}, another important ingredient of the theory of
spectral sequences: take some $\phi_0\in {\G}_{lw}$, of bidegree $(-p,p)$.
Then $d_0(\phi_0)$ is of bidegree $(1-p,p)$. Since
$d_1d_0(\phi_0)=-d_0d_1(\phi_0)=0$, and there is no $d_1$ cohomology of
bidegree $(1-p,p)$, $d_0(\phi_0)=d_1(\phi_1)$ for some $\phi_1$ of
bidegree $(1-p,p-1)$. Now repeat the same steps for $\phi_1$,
giving a $\phi_2$ of bidegree $(2-p,p-2)$, such that
$d_0(\phi_1)=d_1(\phi_2)$. Note that $d_1d_0(\phi_1)=
-d_0d_1(\phi_1)=-d_0^2(\phi)=0$. In this way we find a sequence
of elements $\phi_l$ of bidegree $(l-p,p-l)$. The process stops
at $l=p$. Let
\be \label{1x1x17}
W(\phi)=\sum_{l=0}^{p} (-1)^l \phi_l.
\ee
Then
$dW(\phi)=0$, and $W(\phi)$ is a representative of $\phi_0$ in
$H^0(\Omega;d)$. The algebraic structure of $H^0(\Omega;d)$ is
then determined by calculating the commutation
relations of $W(\phi)$ in $\Omega$, where $\phi_0$ runs over a
basis of ${\G}_{lw}$. This is the quantum finite $W$-algebra.
The non-linearity comes from the fact that the $\phi_l$ are
polynomials of order $l+1$ in the hatted generators $\hj$.

Let us now give an example of the construction described above.

\subsection{Example}
\label{w32_sect}
Consider again the nonprincipal $sl_2$ embedding into $sl_3$
associated to the following partition
of the number 3: $3=2+1$. We constructed the classical $W$-algebra associated
to this embedding earlier. We shall now quantize this Poisson algebra
by the methods developed above.
Take the following basis of $sl_3$:
\be \label{basis21}
r_at_a=\mats{\frac{r_4}{6}-\frac{r_5}{2}}{r_2}{r_1}{r_6}{-
       \frac{r_4}{3}}{r_3}{r_8}{r_7}{\frac{r_4}{6}+\frac{r_5}{2}}.
\ee
Remember that
(in the present notation)
the $sl_2$ embedding is given by $t_+=t_1$, $t_0=-t_5$ and
$t_-=t_8$. The nilpotent subalgebra ${\G}_+$ is spanned by
$\{t_1,t_3\}$, ${\G}_0$ by $\{t_2,t_4,t_5,t_6\}$ and ${\G}_-$ by
$\{t_7,t_8\}$. The $d_1$ cohomology of $\Omega_{red}$ is
generated by $\{\hj^4,\hj^7,\hj^6,\hj^8\}$, and using the
tic-tac-toe construction one finds representatives for these
generators in $H^0(\Omega_{red};d)$:
\ba \label{reps21}
W(\hj^4) & = & \hj^4,  \nn \\
W(\hj^6) & = & \hj^6,  \nn \\
W(\hj^7) & = & \hj^7-\hf\hj^2\hj^5-\hf\hj^4\hj^2+\hf\hj^2,  \nn \\
W(\hj^8) & = & \hj^8+\deel{1}{4}\hj^5\hj^5+\hj^2\hj^6-\hj^5.
\ea
Let us illustrate the tic-tac-toe construction by working out the form
of $W(\hj^7)$ in somewhat more detail. Starting with $\hj^7$, we
find
\be d_1(\hj^7)=\hj^2 c_1 - \frac{1}{2} \hj^4 c_3 - \frac{1}{2} \hj^5 c_3.
\ee
To find a $\phi_1$ such that $d_0(\phi_1)$ equals this, one can either
write down the most general form of $\phi_1$ with arbitrary coefficients,
\ba
\phi_1 & = & a_1 \hj^2\hj^2 + a_2 \hj^2\hj^5 + a_3 \hj^5\hj^5
 + a_4 \hj^4\hj^2 + a_5 \hj^2\hj^6 \nonu
& & + a_6 \hj^5\hj^4 + a_7 \hj^5\hj^6 + a_8 \hj^2 + a_9 \hj^5,
\ea
or make a more clever guess using the form of $d_0(\hj^5)=2c_1$ and
$d_0(\hj^2)=-c_3$. In our case this immediately tells us that
$a_1=a_3=a_5=a_6=a_7=0$, and that $a_2=\frac{1}{2}$ and $a_4=\frac{1}{2}$.
The value of $a_8$ and $a_9$ has to fixed by explicit computation,
and one finds
\be \phi_1=\hf\hj^2\hj^5+\hf\hj^4\hj^2-\hf\hj^2,
\ee
leading to the form of $W(\hj^7)$ in (\ref{reps21}).

Let us introduce another set of generators
\ba \label{newgens21}
C & = & -\deel{4}{3} W(\hj^8)-\deel{1}{9}W(\hj^4)W(\hj^4)-1,
\nonu
j_0 & = & -\deel{2}{3}W(\hj^4)+1,  \nn \\
j_+ & = &  W(\hj^7),  \nn \\
j_- & = & \deel{4}{3} W(\hj^6).
\ea
The commutation relations between these generators are given by
\ba \label{alg21}
[j_0,j_+] & = & 2j_+,  \nn \\ \nn
[j_0,j_-] & = & -2j_-,  \\ \nn
[j_+,j_-] & = & j_0^2+C,   \\ \mbox{}
[C,j_+]=[C,j_-]=[C,j_0] & = & 0.
\ea
These are precisely the same as the relations for the
finite $W_3^{(2)}$ algebra given in \cite{Finite} and (\ref{W_3^(2)}).
Notice that in this case the quantum relations are identical to the
classical ones.
The explicit
$\hbar$ dependence can be recovered simply by multiplying the
right hand sides of (\ref{alg21}) by $\hbar$.

In the appendix we also discuss the explicit quantization of all
the finite $W$-algebras that can be obtained from $sl_4$.
There one does encounter certain quantum effects, i.e the quantum
relations will contain terms of order $\hbar^2$ or higher.

\subsection{Quantum Miura transformation}
\label{qmiura_sect}
In section~\ref{miura_sect} we saw that there exists a realization of
classical finite $W$-algebras in terms of the Kirillov Poisson algebra
${\cal K}({\cal G}_0)$. A similar Miura transformation can be constructed
in the quantum case, using the explicit form of the generators as
described at the end of section~\ref{coho_sect}. Denote by $W(\phi)^{0,0}$
the part of $W(\phi)$ of bidegree $(0,0)$, so it is $(-1)^p \phi_p$
in the notation of (\ref{1x1x17}). The quantum Miura transformation
is simply given by the map $W(\phi)\rightarrow W(\phi)^{0,0}$. As
was shown in \cite{deBoerTjin} this map is an isomorphism of
algebras. The generators $\hat{J}^{\bar{\alpha}}$, with
$\bar{\alpha}$ restricted to ${\G}_0$, form an algebra that
is isomorphic to ${\cal U}{\G}_0$, the universal enveloping algebra
of ${\G}_0$. As we will see later, these facts play an important role
in the representation theory of finite $W$-algebras. Let us illustrate
the quantum Miura transformation for the example we considered in the
previous section. If we introduce $s=(\hat{J}^4+3\hat{J}^5)/4$,
$h=(\hat{J}^5-\hat{J}^4)/4$, $f=2\hat{J}^6$ and $e=\hat{J}^2/2$,
then $h,e,f$ form an $sl_2$ Lie algebra with $[h,e]=e$,
$[h,f]=-f$ and $[e,f]=2h$ while $s$ commutes with everything.
The quantum Miura transformation one finds
from (\ref{newgens21}) by restricting these expressions to their bidegree
$(0,0)$ part reads, in terms of $s,e,f,h$,
\ba
C & = & -\frac{4}{3}(h^2+\frac{1}{2} ef + \frac{1}{2} fe)-
\frac{4}{9} s^2 + \frac{4}{3} s - 1,  \nn \\
H & = & 2h-\frac{2}{3} s+1,   \nn \\
E & = & -2(s-h-1)e,  \nn \\
F  & = &  \frac{2}{3} f.
\ea

\sectiona{Quantum finite $W$-algebras not from $sl_2$ embeddings}

As promised, we now turn to the more difficult case of finite $W$-algebras
that can not be obtained from $sl_2$ embeddings. Important is the case
where we take
$\chi=0$, which we will study first. In general these BRST cohomologies
are very
difficult to calculate. One exception to the rule is the case where one chooses
$\cal L$ to be the Cartan subalgebra (we considered the classical
counterpart of this
case in the previous chapter). We explicitly quantize this algebra
in the case of $sl(3)$.

\subsection{$\chi=0$}

The BRST charge in this case reads
\be
Q=J^{\alpha}c_{\alpha} - \frac{1}{2} f^{\alpha\beta}_{\,\,\, \gamma}
b^{\gamma} c_{\alpha} c_{\beta}.
\ee
The action of the differential $d$ is given by
\ba \label{eqzero:q01}
d(J^a) & = & f^{\alpha a}_{\,\,\,b}J^bc_{\alpha},
 \nn \\
d(c_{\alpha}) & = & -\hf f^{\beta\gamma}_{\,\,\, \alpha}c_{\beta}
c_{\gamma},  \nn \\
d(b^{\alpha}) & = & J^{\alpha} + f^{\alpha \beta}_{\,\,\,
\gamma}b^{\gamma}c_{\beta}.
\ea
To simplify the algebra, it is again advantageous to introduce
\be \label{defzero:hj}
\hj^a=J^a+f^{a\beta}_{\,\,\, \gamma}b^{\gamma} c_{\beta}.
\ee
In terms of $\hj$ we
have
\ba \label{eqzero:q01hat}
d(\hj^a) & = & f^{\alpha a}_{\,\,\,\bar{\gamma}}
\hj^{\bar{\gamma}} c_{\alpha},  \nn \\
d(c_{\alpha}) & = & -\hf f^{\beta\gamma}_{\,\,\, \alpha}c_{\beta}
c_{\gamma},  \nn \\
d(b^{\alpha}) & = & \hj^{\alpha}.
\ea
Here $\bar{\gamma}$ runs over a basis of the orthocomplement
${\cal L}^{\perp}$ of ${\cal L}$, ${\cal G}={\cal L}^{\perp}
\oplus {\cal L}$. Notice that in contrast to the cases obtained from
$sl_2$ embeddings, ${\cal L}^{\perp}$ need not be a closed Lie algebra
itself. This makes life extra complicated.

Although we have made a simplification by going to the
hatted generators $\hj^a$, they satisfy more complicated commutation
relations than the original generators $J^a$. They read
\ba \label{extrarel}
[\hj^a,\hj^b] & = & f^{ab}_{\,\,\,c} \hj^c -b^{\alpha} c_{\beta}
(f^{\beta b}_{\,\,\, \bar{\gamma}}
 f^{\bar{\gamma}a}_{\,\,\, \alpha}+
f^{a \beta }_{\,\,\, \bar{\gamma}}
 f^{\bar{\gamma}b}_{\,\,\, \alpha}), \nonu
[c_{\alpha},\hj^a] & = & f^{a \beta}_{\,\,\, \alpha} c_{\beta}, \nonu
[b^{\beta},\hj^a] & = & f^{\beta a}_{\alpha} b^{\alpha}.
\ea

Let us define ${\cal F}^p$ as the vector space
spanned by all products of at most $p$ generators $\hj^a$ and arbitrary
many ghosts and anti-ghosts. Clearly, the differential $d$ preserves this
space. Furthermore, the relations (\ref{extrarel}) show that
${\cal F}^p \,{\cal F}^q\subset {\cal F}^{p+q}$ and that the induced
algebraic structure on the quotient ${\cal F}^p/{\cal F}^{p-1}$ is
precisely that of a set of commuting variables $\hj^a$ and anti-commuting
variables $b^{\alpha},c_{\alpha}$. These are just the classical relations
between the variables. The subspaces ${\cal F}^p$ define a so-called
filtration of $\Omega$, and associated to such a filtration is another
spectral sequence \cite{specseq}. What we have shown here is that the
computation of the
first term in this particular sequence is equivalent to the
computation of the cohomology
of $d$ assuming classical commutation relations between the $\hj^a$ and
the ghosts. It would be great if the spectral sequence would degenerate
after this first step, but although we expect that this happens in
many cases, we do not have a general proof. The problem is that certain
higher cohomologies will survive, and to show that the spectral sequence
degenerates one has to do an explicit computation of all the higher
operators $D_i$. To avoid this problem we will from now on assume that
the spectral sequence indeed degenerates after the first step, but
keep in mind that in all explicit examples this has to be verified explicitly.

The problem has now basically been reduced to a 'classical cohomology'
computation. The following observation is that we can (like in the
$sl_2$ case) go to a reduced vector space consisting of polynomials
in $c_{\alpha}$ and $\hj^{\bar{\gamma}}$
(using the K\"unneth theorem, see also \cite{BT}). Let $V_{{\cal L}^{\perp}}$
denote the commutative algebra freely generated by the $\hj^{\bar{\gamma}}$.
$V_{{\cal L}^{\perp}}$ is nothing but the constrained phase space we have
seen previously.
We want to compute the cohomology of $d$ on $V_{{\cal L}^{\perp}}\otimes
\wedge {\cal L}^{\ast}$,
the last term $\wedge {\cal L}^{\ast}$ representing the ghosts $c_{\alpha}$.
This particular cohomology problem is a well-known one in mathematics.
Since ${\cal L}^{\perp}$ is an ${\cal L}^{\ast}$ module in the obvious way,
so is $V_{{\cal L}^{\perp}}$, and the cohomologies we are computing are
nothing but the
Lie algebra cohomologies $H^n({\cal L}^{\ast},V_{{\cal L}^{\perp}})$.
Alternatively, they
can be described in terms of the ${\rm Ext}$ groups
${\rm Ext}^n_{{\cal U}{\cal L}^{\ast}}(\ce,V_{{\cal L}^{\perp}})$,
which might be
computable if we had a resolution of $V_{{\cal L}^{\perp}}$ in terms of
injective
${\cal U}{\cal L}^{\ast}$ modules, but we do not know how to write down
such a resolution. In any case, if the spectral sequence degenerates
after one step and we would know these Lie algebra cohomologies, to obtain the
final full result for the original problem still requires an additional
amount of work. First, we have to use a tic-tac-toe type of construction
to find the explicit generators, and secondly we have to compute their
commutation relations. In the latter we have to be careful to identify
terms which differ by a BRST exact expression. Altogether the result of
this procedure will be that the BRST cohomology is a quantum deformation
of the classical Lie algebra cohomology, granted that the spectral
sequence degenerates after the first step.

Extra subtleties further arise if one wants to write down a generating
basis for the quantum finite $W$-algebra. The number of generators may
be much smaller than the number of generators of the corresponding
classical Poisson algebra. This is due to the fact that the Poisson
algebra is commutative, and the quantum algebra is not. Take for
example ${\cal G}=sl_2$, and ${\cal L}=0$. In this trivial case the classical
algebra is the Kirillov Poisson algebra based on $sl_2$, with generators
$J^-,J^0,J^+$. The quantum algebra is the quantum
Kirillov algebra (\ref{eq:qkir}), but this one is generated by
just $J^+$ and $J^-$, by
virtue of the relation $J^0=\frac{1}{\hbar}[J^+,J^-]$. This relation becomes
singular in the limit $\hbar \rightarrow 0$, explaining the apparent
discrepancy between the two. Similar phenomena take place in conformal
field theory, see \cite{BoFeHo}.

Since the zeroth cohomology $H^0({\cal L}^{\ast},V_{{\cal L}^{\perp}})$
is simply the space $(V_{{\cal L}^{\perp}})^{{\cal L}^{\ast}}$
of ${\cal L}^{\ast}$ invariants of $V_{{\cal L}^{\perp}}$, and the
quantum $W$-algebra is
by definition the zeroth cohomology, we find that the quantum finite $W$
algebra, assuming the spectral sequence degenerates, is as a vector space
isomorphic to $(V_{{\cal L}^{\perp}})^{{\cal L}^{\ast}}$. To give a
basis of this space with
its classical multiplication rule is a problem of so-called invariant theory.
The answer to this question is known only for special $V_{{\cal L}^{\perp}}$
and ${\cal L}^{\ast}$.
Again, this resembles closely the situation in conformal field theory as
explained in \cite{BoFeHo}. Thus, even classically we cannot in general
give a minimal set of generators of the algebra $({\cal G},{\cal L},0)$.

To illustrate these rather abstract statements we will now give a
few examples.

\subsection{The algebra $({\cal G},{\cal G}^+,0)$}
\label{gplus_alg}

Suppose that we have chosen ${\cal L}$ as in the case of an $sl_2$
embedding, but with $\chi=0$. This problem is interesting for
a variety of reasons. Firstly, in the case of the principal $sl_2$
embedding it is a interesting open problem in mathematics.
And secondly, it is the first term in the
'difficult' spectral sequence that we encountered before in the
case where $\chi$ was a one-dimensional character, and might
shed some light on the $W$-algebras obtained from $sl_2$ embeddings.
More generally, we can look at any ${\cal G}^+$ which is the positive
eigenspace of some grading element $\delta$ of the Cartan subalgebra.
The nice feature of this case is that ${\cal L}^{\perp}={\cal G}^-
\oplus {\cal G}^0$ is a closed subalgebra. That implies that we can
rigorously reduce the cohomology problem to one for the
subalgebra generated by $\hj^{\bar{\gamma}}$ and the ghosts $c_{\alpha}$.
The classical zeroth cohomology can easily be worked out, it is
just the Kirillov Poisson algebra based on ${\cal G}^0$.
According to the spectral sequence argument above, the quantum algebra
can as a vector space be at most isomorphic to this. On the other
hand, by explicit computation one verifies that the quantum Kirillov
algebra based on ${\cal G}^0$ is contained in the quantum algebra.
Combining these two facts proves
\be \label{hhh00}
H^0({\cal G},{\cal G}^+,0)={\cal U}{\cal G}^0.
\ee
The really difficult problem is to compute the higher cohomologies here.
For this, we only know the answer for $sl_2$, if ${\cal G}^+$ is
generated by $J^+$. Then
\be H^1(sl_2,{\cal G}^+,0)=c \,\ce[J^-]. \ee

\subsection{The algebra $({\cal G},{\cal T},0)$}
\label{torusalg}

In this example we choose ${\cal L}={\cal T}$ to be the maximal
torus of ${\cal G}$. Now, ${\cal L}^{\perp}$ is no longer a
closed subalgebra.
However, we are in the fortunate circumstances
that $\hj^{a}=J^{a}$. This means in particular that
the quantum algebra is generated by expressions
\be J_{\alpha_1} J_{\alpha_2} \cdots J_{\alpha_n}, \ee
where $J_{\alpha_i}$ is the generator $J^{\bar{\gamma}}$ associated to
the (positive or negative) root $\alpha_i$, and $\sum \alpha_i=0$.
To find a minimal basis is a separate problem. For example, for $sl_3$
a set of generators is
\ba
A_1 & = & J_{\alpha_1} J_{-\alpha_1} \nonu
A_2 & = & J_{\alpha_2} J_{-\alpha_2} \nonu
A_3 & = & J_{\alpha_1+\alpha_2} J_{-\alpha_1-\alpha_2} \nonu
B & = & ( J_{-\alpha_1} J_{-\alpha_2} J_{\alpha_1+\alpha_2} -
J_{\alpha_1} J_{\alpha_2} J_{-\alpha_1-\alpha_2} )/2 \nonu
C & = & ( J_{-\alpha_1} J_{-\alpha_2} J_{\alpha_1+\alpha_2} +
J_{\alpha_1} J_{\alpha_2} J_{-\alpha_1-\alpha_2} )/2.
\ea
The BRST charge in this case is
\be Q=c_0 H_0 + c_1 H_1, \ee
where $H_0=[J_{\alpha_1},J_{-\alpha_1}]$ and $H_1=[J_{\alpha_2},
J_{-\alpha_2}]$.
This is an example where the commutators do not close in themselves,
but only modulo BRST exact expressions.
For example, when working out $[A_2,C]$ one encounters the
expression $[J_{\alpha_2},J_{-\alpha_2}]$, which lives in the
Cartan subalgebra. Elements of the Cartan subalgebra can be
moved to either the left or the right of an expression. Only if they stand
to the left or right will the resulting expression be
BRST exact (it is the
$d$ of an anti-ghost times something) and can be put equal to zero.
In this way we find $[A_2,C]$ is purely BRST exact, so the commutator
vanishes. The final result for the quantum algebra can be conveniently
expressed as follows. Introduce
\be \tilde{B}=B+\frac{\hbar}{2}(A_1-A_2+A_3). \ee
Then the brackets are
\ba
[A_{i},A_{i+1}] & = & 2\hbar\tilde{B} \nonu
[A_{i},\tilde{B}] & = & \hbar(A_{i+1} A_i -A_i A_{i-1}),
\ea
which is indeed a quantum deformation of the classical result.
The generators satisfy the following relation
\be
C^2-\tilde{B}^2  =  \frac{1}{2} (A_1 A_2 A_3 +A_3 A_2 A_1)
+ \frac{\hbar}{4}(A_1-A_2+ A_3)^2.
\ee

It is amusing to note that in this particular case with ${\cal L}$
equal to a maximal torus there is an exact result for the zeroth
cohomology,
\be H^0({\cal G},{\cal T},0)=({\cal U}{\G})^{\cal T}/{\cal T}.
\ee
In the right hand side, we first perform a gauge fixing by
looking at the centralizer of ${\cal T}$. In this resulting
algebra, ${\cal T}$ generates a left ideal and we can divide out by this
ideal, corresponding to imposing the constraints. Thus we have
interchanged the usual order of first imposing constraints and
then gauge fixing.

\subsection{Further finite $W$-algebras}
\label{sect434}

The last case we have not considered so far is when $\chi$ is some
representation of ${\cal L}$ in a non-trivial algebra. An example
is is to take ${\cal G}=sl_3$, ${\cal L}$ equal to the
upper triangular matrices, and a representation of $\chi$ in
the one-dimensional Heisenberg algebra generated by $p$ and $q$
with $[p,q]=1$. Explicitly, one can take $\chi(J_{\alpha_1})=
p$, $\chi(J_{\alpha_2})=q$, and $\chi(J_{\alpha_1+\alpha_2})=1$.
In this case one presumably recovers $W_3^{(2)}$, the $p$ and
$q$ corresponding to a set of auxiliary variables that can also
be introduced in the infinite-dimensional case \cite{B}.
In fact, from the
infinite-dimensional case \cite{SeTr} we know that the analysis
for higher dimensional representations $\chi$ may become more
complicated in that the spectral sequence does not degenerate after
the first step any more. For certain higher dimensional $\chi$ that are
inspired by $sl_2$ embeddings, one can nevertheless
still compute the cohomology exactly. For all other type
of representations $\chi$ little is known, and this remains a region
consisting of a vast number of new unexplored finite $W$-algebras.
If ${\cal U}_{\chi}$ is isomorphic to a copy of ${\cal U}{\cal L}$,
and $\chi$ is the isomorphism between the two, then the cohomology
reduces essentially to ${\cal UG}^{\cal L}$, the space of ${\cal L}$
invariants in ${\cal G}$. For some cases, this space of invariants
has been studied in \cite{lyon}. In addition, we want to mention that
there exists an alternative method to quantize $W$-algebras\cite{NPL}.
Using this method one finds a faithfull $*$-representation of the
maximally non-compact real form (see next chapter) of a given finite
$W$-algebra.

For applications to physics we are interested in the
representation theory of quantum finite $W$-algebras. This is the
subject of the next chapter.

\chapter{The Representation Theory}
We now turn to the representation theory for finite $W$-algebras.
Physically the
representations that are most interesting are
the unitary irreducible representations
on Hilbert spaces. However, as finite $W$-algebras are non-linear it is not
completely obvious what unitary representations are.
Usually a representation of a
group is called unitary if all group elements are represented by
unitary operators
on some Hilbert space. For finite $W$-algebras there is no such
thing as an abstract group
which one can get by exponentiating the algebra. Nevertheless it is
possible to
define the concept of unitarity for finite $W$-algebras.
The reason for this is
that if the generators of a finite $W$-algebra are
represented by (finite) matrices
one can exponentiate them without running into
trouble because $\exp(A)$ converges
for any matrix $A$. Therefore even though there is no formal
grouplike object associated
to a finite $W$-algebra we {\em can} exponentiate its elements in any given
representation. Now, not every element of a complex
finite $W$-algebra can exponentiate
to a unitary matrix, as we will explain in a moment
(this is also true for Lie algebras).
It is therefore necessary to consider real forms of
finite $W$-algebras before one
addresses the problem of constructing unitary representations.

Deriving finite $W$-algebras from compact (real) Lie algebras
such as $su(n)$ is complicated
by  the fact that these Lie algebras do not admit Gaussian decompositions.
The question therefore arises whether the constructions
developed up to now are physically
interesting at all, as non-linear algebras involving compact Lie algebras
seem not to be
included. In fact they {\em are} included. The
point is that one should consider the Lie
algebras discussed above as {\em complex} Lie algebras,
that is $sl(n)$ means $sl(n;\ce )$.
Applying the reduction procedures one obtains
complex finite $W$-algebras. Only then
should questions about compactness and unitarity be addressed by
studying the real forms
of the complex $W$-algebra. As we shall see, a given finite
$W$-algebra admits many real forms not all of which
admit unitary representations.

The next problem we address is defining the
concept of a highest weight representation.
For this one needs a Poincare-Birkhoff-Witt (PBW)
like theorem for finite $W$-algebras.
Using this we discuss some conjectures for the
Kac determinants and character formulas
for finite $W$ highest weight modules. These conjectures were
proposed in \cite{Peter&Koos}, although a discussion of the PBW
theorem, of real forms of finite $W$-algebras and some other
details are lacking in \cite{Peter&Koos}.
\sectiona{Real forms and unitary representations}
Let ${\cal W}^{\ce}$ denote a complex finite $W$-algebra (where we explicitly
specified the fact that the algebra is a module over the complex numbers)
obtained from a complex
Lie algebra by the construction discussed in the previous sections.
A representation
$\rho: {\cal W}^{\ce} \rightarrow {\rm lin}({\cal H})$, where $({\cal H},
\langle.,. \rangle)$ is a (complex Hilbert space), can never be such that
$\exp(\rho(W))$ is a unitary operator on ${\cal H}$ for all $W \in {\cal W}$.
One can see this as follows. For $\exp(\rho(W))$ to be unitary, we must have
\be
\exp(\rho(W))^{\dagger} = \exp(\rho(W))^{-1},
\ee
from which follows
\be
\label{unirho}
\rho(W)^{\dagger} = - \rho(W).
\ee
Suppose (\ref{unirho}) holds for $W\in {\cal W}$.
Now consider $W' = i W$.
We have $\rho(W')^{\dagger} = (i \rho(W))^{\dagger} =
-i \rho(W)^{\dagger} = \rho(W')$. From this it follows
that $\exp(\rho(W'))$ is {\em not} unitary.
We see that $W$ and $iW$ can never
simultaneously give rise to a unitary operator.

What one needs to do is go to `real forms' of ${\cal W}^{\ce}$.
It is clear from the argument above, that if $\exp(\rho(W))$ is
 unitary, then $\exp(\rho(\al W))$ is also unitary,
 {\em provided $\al$ is real}. Real subspaces of
${\cal W}$ therefore have a chance of being represented unitarily.
These real vector spaces should, however, be invariant
 with respect to taking commutators. This requirement
 leads one to consider the so called anti-involutions of
order 2 of ${\cal W}$.

A two-anti-involution of a complex
finite $W$-algebra ${\cal W}^{\ce}$ is a
map $\om: {\cal W}^{\ce} \rightarrow {\cal W}^{\ce}$ such that
\bea
&1)& \om^2 = 1  \nn \\
&2)& \om(\al w_1 + \bt w_2) = \bar{\al} \omega (w_1) +
\bar{\bt} \omega (w_2)  \nn \\
&3)& \om(w_1w_2) = \om(w_2) \om(w_1),
\eea
where $\al$ and $\bt$ are complex numbers
and $w_1,w_2 \in {\cal W}^{\ce}$. From 1) it follows that $\om$ has
two eigenvalues: $\pm 1$. Consider the negative eigenspace of $\om$
\be
{\cal W}_-^{\ce} = \{ W \in {\cal W} \mid \om(W) = - W \}.
\ee
This space is actually a real subspace of ${\cal W}^{\ce}$,
as one can see as follows: Let $W \in {\cal W}$
and $\al \in \ce$, then $\om(\al W) = - \bar{\al} W$.
If $\al W$ is to be an element of ${\cal W}^{\ce}_-$,
we must have $\bar{\al} = \al$. We conclude that $\al W$ is
only an element of ${\cal W}_-^{\ce}$ if $\al$ is real.

${\cal W}_-^{\ce}$ is also closed under commutation, because
\be
\label{comm}
\om([w_1,w_2]) \equiv [\om(w_2),\om(w_1)]
\ee
by property 3). Therefore, if $w_1,w_2 \in {\cal W}_-^{\ce}$, we find
\be
\om([w_1,w_2]) = - [w_1,w_2],
\ee
which means that $[w_1,w_2]$ is an element
of ${\cal W}_-^{\ce}$ if $w_1$ and $w_2$ are. ${\cal W}_-^{\ce}$
is therefore actually a subalgebra of ${\cal W}_-^{\ce}$.
We have thus found a closed real subalgebra of ${\cal W}_-^{\ce}$
that stands a chance of admitting unitary representations.
 From now on we denote the space ${\cal W}_-^{\ce}$
by ${\cal W}_{\om}^{\re}$, where we have made the $\om$ dependence
explicit and in analogy with the Lie algebra case we
call it a `real form' of ${\cal W}^{\ce}$.

A unitary representation of the real form ${\cal W}_{\om}^{\re}$ is
now defined as a representation of ${\cal W}_{\om}^{\re}$ in
some (complex) Hilbert space ${\cal H}$ (with inner
product $\langle.,. \rangle$), such that for all
elements $W \in {\cal W}_{\om}^{\re}$
\be
\exp(\rho(W))^{\dagger} = \exp(\rho(W))^{-1},
\ee
or equivalently
\be
\rho(W)^{\dagger} = \rho(\om(W)) \equiv - \rho(W).
\ee

In general not all real forms of a given ${\cal W}^{\ce}$
algebra will admit unitary representations and in this sense the
situation is similar to that in group theory.

Let us look at some examples. Consider again the (complex) finite $W$-algebra
\be
\label{W32}
[j_0,j_{\pm}] = \pm 2j_{\pm}; \hspace{2cm} [j_+,j_-] = j_0^2 + C.
\ee
This algebra has two independent anti-involutions $\om_1$ and $\om_2$, given by
\bea \label{1x4x10}
\om_1(j_0) &=& j_0; \hspace{2cm} \om_2(j_0) = -j_0  \nn \\
\om_1(j_{\pm}) &=& j_{\mp}; \hspace{2cm} \om_2(j_{\pm}) = \pm j_{\pm}  \nn \\
\om_1(C) &=& C; \hspace{2cm} \om_2(C) = C.
\eea
The negative eigenvalue eigenspace of $\om_1$ is spanned by
\be
\frac{{(-i)}^{n - 1}}{2} ( \sg_{k_0} ... \sg_{k_n} - \sg_{k_n} ... \sg_{k_0} )
\ee
and
\be
\frac{{(-i)}^n}{2} ( \sg_{k_0}  ... \sg_{k_n} + \sg_{k_n} ... \sg_{k_0}),
\ee
with for $0\leq p\leq n$ $k_p \in \{1,\ldots,4\}$, and $n = 0,1,2\ldots$, where
\be \label{1x4x13}
\sg_1 = i j_0; \hspace{1cm} \sg_2 = j_+ - j_-;
\hspace{1cm} \sg_3 = i(j_+ + j_-), \hspace{1cm} \sg_4 = i C.
\ee

The non-zero commutation relations between the generators $\sg_k$ read
\be \label{1x4x14}
[\sg_1,\sg_2] = 2 \sg_3; \hspace{1cm} [\sg_2,\sg_3] =
2 \sg_4 - 2 i \sg_1^2; \hspace{1cm} [\sg_3,\sg_1] = 2 \sg_2.
\ee
This is the real form of (\ref{W32}) with respect to $\om_1$.
The unitary irreducible representations of this algebra were
constructed in \cite{Finite}.
The two dimensional representations read
\bea
\sg_1 &=& \left( \begin{array}{cc} i(x + 1) & 0 \\ 0 &
i(x - 1) \end{array} \right) \hspace{2cm}
\sg_2 = \left( \begin{array}{cc} 0 & \sqrt{2x} \\ -
\sqrt{2x} & 0 \end{array} \right)  \nn \\
\sg_3 &=& \left( \begin{array}{cc} 0 & i \sqrt{2x}\\ i
\sqrt{2x} & 0 \end{array} \right) \hspace{2cm}
\sg_4 = \left( \begin{array}{cc} - i(1 + x^2) & 0 \\ 0 &
-i(1 + x^2) \end{array} \right),
\eea
where $x > 0$ is real. Note that $\exp(\al_a\sg_a)$ is unitary
for all real numbers $\al_a$.

Now consider $\om_2$. In this case
the negative eigenvalue eigenspace is spanned by
\be
\frac{{(-i)}^{n - 1}}{2} ( S_{k_0} ... S_{k_n} - S_{k_n} ... S_{k_0})
\ee
and
\be
\frac{{(-i)}^n}{2} ( S_{k_0}  ... S_{k_n} + S_{k_n} ... S_{k_0}),
\ee
with for $0\leq p\leq n$ $k_p \in \{3,+,-,0\}$, and $n = 0,1,2\ldots$, where
\be \label{1x4x18}
S_3 = j_0; \hspace{1cm} S_+ = i j_+;
\hspace{1cm} S_- = j_-; \hspace{1cm} S_0 = i C.
\ee
The non-zero commutation relations between the generators read
\be \label{1x4x19}
[S_3,S_{\pm}] = \pm 2 S_{\pm}; \hspace{1cm} [S_+,S_-] = i S_3^2 + S_0.
\ee
This real form can {\em not} have a non-trivial finite-dimensional
unitary representation. In order to see this, assume that $v_0$ is
a eigenvector of $S_3$ (where we omit explicit reference to the
representation $\rho$). As $S_3^{\dagger} = - S_3$ in a unitary
representation, all eigenvalues of $S_3$ are purely
imaginary, i.e. $S_3 v_0 = (i \lm) v_0$, where $\lm \in \re$.
Now, consider the vector $S_{\pm}v_0$. It is also an
eigenvector of $S_3$, as follows from
\be
S_3 S_{\pm}v_0 = (S_{\pm} S_3 \pm 2 S_{\pm}) v_0 = (i \lm \pm 2) S_{\pm} v_0,
\ee
but with eigenvalue $i \lm + 2$. This contradicts the fact
that $S_3$ only has purely imaginary eigenvalues,
therefore $S_{\pm} v_0$ must be zero. This means that the
representation is at most one dimensional.

\subsection{Real forms of general finite $W$-algebras}

In the previous section we have demonstrated that real forms of a
complex finite $W$-algebra can be obtained from anti-involutions
$\omega$. Since our construction of finite $W$-algebras is based
on the computation of a certain BRST cohomology, the question
arises whether real forms of the Lie algebra lying above can
be used to construct ones of the finite $W$-algebra. In the example
we just considered one can however easily convince oneself that
none of the two anti-involutions comes from any of the three
real forms of $sl_3$ (corresponding to $sl(3,\re),su(3)$ and
$su(1,2)$). To understand what causes this, notice that if the
BRST operator is given by $d(X)=[Q,X]$, then
an anti-involution $\omega$ of ${\cal G}$ descends to
an anti-involution of $H^{\ast}(d)$ iff $H^{\ast}(d)=H^{\ast}(\omega(d))$,
where $\omega(d)(X)=[\omega(Q),X]$. In particular, a sufficient
condition for this to happen is that $\omega(Q)$ is
proportional to $Q$. This shows

An anti-involution on ${\cal G}$ gives rise to one on the finite ${\cal W}$
algebra $({\cal G},{\cal L},\chi)$, if $\omega({\cal L})={\cal L}$ and
$[\omega,\chi]=0$.

For example, this shows that the anti-involution $\omega(X)=-X$ of
${\cal G}$\footnote{We assume here and in the sequel that a basis of ${\cal G}$
with real structure constants has been chosen, typically the one
corresponding to the maximal non-compact real form of ${\cal G}$, which for
${\cal G}=sl(n,\ce)$ is $sl(n,\re)$}
always gives rise to an anti-involution of $({\cal G},
{\cal H},0)$. It is unknown whether or not any other anti-involutions
of these algebras exist, and since we will not discuss their
representation theory in detail, we will for the remainder of
this section focus on those $W$-algebras that can be obtained
from $sl_2$ embeddings. Let us denote the generators of $sl_2$
by $\{t_-,t_0,t_+\}$, and the corresponding images in ${\cal G}$
by $J^-,J^0,J^+$. Furthermore, we will assume that the one-dimensional
representation $\chi$ is given by $\chi(J^+)=\xi$, where now
$\xi$ is an arbitrary complex number. The Lie algebra ${\cal G}$ can
be decomposed in terms of the half-integral eigenvalues of
${\rm ad}(J^0)$ as ${\cal G}=\oplus_n {\cal G}^{(n)}$
(see also (\ref{grading2})). Naively,
one would think that the choice of $\xi$ is irrelevant, since
there is an automorphism of ${\cal G}$ where every element of
${\cal G}^{(n)}$ is rescaled by $\lambda^n$ for some nonzero
complex number $\lambda$, and one can use this to put $\xi=1$.
However, this rescaling does not scale the generators of the
$W$-algebra in a simple way, since the explicit form of these
generators in terms of the anti-commuting variables $b^{\alpha},
c_{\alpha}$ and the $J^{a}$ contains terms with different ${\rm ad}(J^0)$
eigenvalues. Therefore, varying $\xi$ gives a one-parameter family
of finite $W$-algebras that are not necessarily isomorphic.
An interesting situation arises when $\xi$ is purely imaginary. In that
case the anti-involution $\omega(X)=-X$ of ${\cal G}$ satisfies
$[\omega,\chi]=0$ and gives rise to an anti-involution on the
corresponding finite $W$-algebra. In particular, the second
anti-involution $\omega_2$ in (\ref{1x4x10}) can be obtained in this way,
since for the embedding that describes the algebra $W_3^{(2)}$ it turns out
that the resulting algebras are isomorphic for all $\xi$. More
explicitly, for arbitrary $\xi$ the algebra (\ref{W32}) becomes
\be \label{W32xi}
[j_0,j_{\pm}] = \pm 2j_{\pm}; \hspace{2cm} [j_+,j_-] = \xi^{-1} j_0^2 + C.
\ee
One sees that indeed for purely imaginary $\xi$ an anti-involution
is obtained by sending each generator $X$ to $-X$. A subsequent
basis transformation $C \rightarrow \xi^{-1} C$ and $j_{+} \rightarrow
\xi^{-1} j_{+}$ yields exactly the anti-involution $\omega_2$ in
(\ref{1x4x10}). Notice that taking $\xi$ imaginary brings one
automatically in the basis described in equations (\ref{1x4x18})
and (\ref{1x4x19}).

In general, we will call the finite $W$-algebras that one gets for
imaginary $\xi$, with their corresponding anti-involution, the maximal
non-compact real form of $({\cal G},{\cal L},\chi)$. In analogy with
the corresponding situation for Lie algebras,
these algebras do not have any interesting
finite-dimensional unitary representations.
The argument is the same as the one we presented for the case of $W_3^{(2)}$
at the end of the previous section.

To have a more interesting representation theory, we would like to have
the analogue of the compact real form for finite $W$-algebras, like
the anti-involution $\omega_1$ in (\ref{1x4x10}). The construction of
this real form is somewhat more involved, since it cannot be induced
from one on ${\cal G}$. Let us denote the parabolic subalgebra
$\oplus_{n\leq 0} {\cal G}^{(n)}$ of ${\cal G}$ by ${\cal P}$, the
centralizer of $sl_2$ in ${\cal G}$ by ${\cal K}$, the restriction
of the Cartan subalgebra to ${\cal K}$ by ${\cal H}_{\cal K}$ and
the orthocomplement of ${\cal H}_K$ in ${\cal H}$ by ${\cal H}_S$
(so ${\cal H}={\cal H}_S \oplus {\cal H}_K$). The centralizer of
${\cal K}$ in ${\cal G}$ is a direct sum of some subalgebra ${\cal G}_S$
(with Cartan subalgebra ${\cal H}_S$) and
the $u(1)$ factors of ${\cal K}$. This algebra ${\cal G}_S$ plays
an important role in the discussion of the representation theory
\cite{Peter&Koos}. In most cases, ${\cal G}_S$ is the subalgebra of
${\cal G}$ in which the $sl_2$ is principally embedded. ${\cal K}$
is precisely the semi-simple subalgebra of the finite $W$-algebra
described in section~\ref{subalg_sect}.

In section~\ref{coho_sect} we saw that the computation of the
BRST cohomology could be reduced to one for a reduced complex, where
only the $J^a$ which are not in ${\cal L}$ appear. If there are no
degree $1/2$ subspaces in ${\cal G}$, so that ${\cal G}^{\pm (1/2)}=0$ and
the grading provided by $t_0$ is integral, then exactly the $J^a$ that
correspond to the parabolic algebra ${\cal P}$ appear. If the degree
$1/2$ subspaces are nonzero, then there exists an other, equivalent
formulation of the BRST cohomology involving auxiliary fields, and
one can go to a modified reduced complex, where still only generators
of ${\cal P}$ appear. Rather then trying to reduce an anti-involution
of the whole Lie algebra ${\cal G}$, we can also try to start with one
of the reduced complex and reduce that to the finite $W$-algebra.
It is straightforward to analyze what conditions such an anti-involution
has to satisfy and one finds that any anti-involution $\omega$ of
the parabolic algebra ${\cal P}$ gives rise to one
of the finite $W$-algebra $({\cal G},{\cal L},\chi)$ if $\omega(J^{-})=
(\xi/\bar{\xi}) J^{-}$.

To prove the existence of anti-involutions is not so easy in general.
For example, the existence of the Cartan involution for ordinary Lie
algebras becomes only transparent in certain distinguished bases of
the Lie algebra, like the Cartan-Weyl basis. We do not have similar
distinguished bases of ${\cal P}$ at our disposal, so that we can only
conjecture the following

There exists an anti-involution on ${\cal P}$ which restricts to the
Cartan involution on ${\cal K}$ and to minus the identity on ${\cal H}_S$.
We will call the corresponding real form of the finite $W$-algebra the
compact real form.

In all known examples such an anti-involution exists. For example, in the
example in the previous section the anti-involution of ${\cal P}$ corresponding
to the anti-involution $\omega_1$ in (\ref{1x4x10}) is
\be
\omega_1 \mats{a-b}{0}{0}{c}{-2 a}{0}{e}{d}{a+b} =
 \mats{a+b}{0}{0}{d}{-2 a}{0}{e}{c}{a-b}.
\ee
Another example is the finite $W$-algebra corresponding to the $sl_2$
embedding in $sl_4$ under which the fundamental representation of
$sl_4$ decomposes as $2 \oplus 2$. In that case
\be
\omega \mat{A}{0}{C}{B} = \mat{B^t}{0}{C^t}{A^t},
\ee
where $A,B,C$ are two by two matrices.

It is an open problem to classify all possible real forms of  finite
$W$-algebras, and to prove the existence of a compact real form in general.

\sectiona{Highest weight representations}

In this section we work out the construction of highest
weight representations of finite $W$-algebras, and the conjectured
form of the Kac determinant and the Kazhdan-Lusztig conjecture for such
representations\cite{Peter&Koos}.
The latter relates the characters of highest
weight representations to those of irreducible representations.

The definition of highest weight representations of Lie algebras uses
the decomposition of the Lie algebra into generators corresponding
to the positive roots, the negative roots and the Cartan subalgebra. A
highest weight module is built by acting on a particular vector with
arbitrary negative root generators. This vector is by definition
annihilated by all the positive root generators, and has specific
eigenvalues with respect to the generators of the Cartan subalgebra.
If the Hamiltonian of a physical system would be part of the
Cartan subalgebra, a physical motivation to look at such representations
would be that their energy is bounded from below.

To imitate this construction, we need a similar type of decomposition of
the finite $W$-algebra. The part of the Cartan subalgebra which survives
in the finite $W$-algebra is precisely ${\cal H}_{\cal K}$, and
every generator of the finite $W$-algebra has well defined eigenvalues
with respect to the adjoint action of the generators of ${\cal H}_{\cal K}$.
Thus we can associate to every generator of the $W$-algebra a root in
the root space ${\cal H}_{\cal K}^{\ast}$, which is the dual of
${\cal H}_{\cal K}$. To find out what are the positive and negative roots,
we assume there is an automorphism of the Lie algebra ${\cal G}$ which
maps ${\cal G}_S$ into a subalgebra generated by $\{E_{\pm \alpha_i},
H_{\alpha_i} \}_{i\in S}$, where $S$ is a subset of the set of
simple roots of ${\cal G}$. From now on we will assume we are in a
basis of ${\cal G}$ where this is the case. Any root $\alpha$ of
${\cal G}$ can be orthogonally projected onto
an element (called $\pi(\alpha)$) of the root space
${\cal H}^{\ast}_{\cal K}$.
The image of the positive roots of ${\cal G}$
under this projection (except those which project to zero) is what
we will call the positive roots of ${\cal H}_{\cal K}$. This then defines
the following decomposition of the generators of the finite $W$-algebra
\be {\cal W}= {\cal W}_- \oplus {\cal W}_0 \oplus {\cal W}_+.
\label{triang} \ee
The same decomposition can be written down not just for the generators,
but also for the universal enveloping algebra of the
$W$-algebra. Typically, ${\cal W}_0$ contains
${\cal H}_{\cal K}$, a certain number of central elements that commute
with everything and do not give new roots (one of such generators is
the remnant of the quadratic Casimir of the original Lie algebra), and
other elements whose adjoint action cannot be diagonalized.

The properties of such a root-space type decomposition of the generators
of the finite $W$-algebra are different from those of Lie algebras.
Here, there can be root spaces of dimension larger than one, and
one does not have subalgebras isomorphic to $sl_2$ corresponding to
any root. In addition, in the case of Lie algebras $[E_{\alpha},E_{\beta}]$
vanishes if $\alpha+\beta$ is not a root or zero.
Finite $W$-algebras on the other hand
satisfy non-linear relations, so even if
$\alpha+\beta$ is not a root, one can still have relations of the type
$[E_{\alpha+\gamma},E_{\beta-\gamma}]\sim E_{\alpha} E_{\beta}$.

An important property that finite $W$-algebras share with Lie algebras
is that the Poincar\'e-Birkhoff-Witt theorem still holds. Choose a basis
of the finite $W$-algebra so that the generators are in one-to-one
correspondence with elements of ${\rm ker}({\rm ad}_{t_-})$ and
have a well defined non-negative half-integer eigenvalue with
respect to $-{\rm ad}_{t_0}$. For example, choose the generators in
one-to-one correspondence with the lowest weights of the irreducible
$sl_2$ representations in which ${\cal G}$ decomposes under the action
of the embedded $sl_2$. If this basis is $\{W_i\}_{i\in I}$,
we can order the index set $I$ in some arbitrary fashion,
and the Poincar\'e-Birkhoff-Witt theorem states
that a basis for the finite $W$-algebra (considered as a vector space)
is given by all expressions of the type
\be
 W_{i_1} W_{i_2} \cdots W_{i_n},
\ee
where $i_1\leq i_2 \cdots \leq i_n$. To prove this, we first note that
from the explicit form of the generators, as obtained from the
tic-tac-toe construction, one can see that the commutator of
two generators of $-{\rm ad}_{t_0}$ eigenvalue $\lambda_1$ and $\lambda_2$
does not contain generators or products of generators of $-{\rm ad}_{t_0}$
eigenvalue larger than $\lambda_1+\lambda_2$. More precisely,
the linear terms have eigenvalues $\leq \lambda_1+\lambda_2$,
the quadratic terms have eigenvalues $\leq \lambda_1+\lambda_2-1$,
the third order terms have eigenvalues $\leq \lambda_1+\lambda_2-2$, etc.
Armed with this observation the theorem can now be proven using
double induction on the number of generators and their
$-{\rm ad}_{t_0}$ eigenvalue.

A particular useful ordering of the generators of the finite $W$-algebra
is to put all generators in ${\cal W}_-$ to the left,
those corresponding to ${\cal W}_0$ in the middle and those
corresponding to ${\cal W}_+$ to the right.
Then one can define a highest-weight
representation in the usual fashion, by acting with the $W$-algebra
on a particular vector $|\lambda\rangle$, which (i) is annihilated by
the generators in ${\cal W}_+$ and (ii) has certain eigenvalues with
respect to the generators in ${\cal W}_0$. The Poincar\'e-Birkhoff-Witt
theorem then tells us that the resulting module is spanned by the
vectors
\be W_{i_1} W_{i_2} \cdots W_{i_n}|\lambda\rangle, \ee
where $i_1\leq i_2 \cdots \leq i_n$, and the $W_i$ are elements of
${\cal W}_-$.  This agrees precisely with the standard definition.

At this stage we should discuss how one constructs highest weight
representations of finite $W$-algebras in terms of representations
of ${\cal G}$. One possibility is to use the BRST procedure once
more. The finite $W$-algebra was the BRST cohomology of
a certain complex constructed out of ${\cal G}$, and it is
possible to extend this complex to one where the Lie algebra
is replaced by a representation of the Lie algebra. The cohomology
of this complex then automatically gives representations of the finite
$W$-algebra. This procedure has some disadvantages, however.
Firstly, it is rather cumbersome if one has to compute a new cohomology for
every representation, and secondly, finite-dimensional representations
of ${\cal G}$ typically yield just the trivial representation of the
finite $W$-algebra. For a discussion, see \cite{thesis}.

A better procedure is to take explicit representatives for the
generators of the finite $W$-algebra, and to use those to
obtain finite $W$-representations as subspaces of representations
of the universal enveloping algebra of ${\cal G}$\footnote{
Strictly speaking one also needs a representation of the
$b^{\alpha},c_{\alpha}$ Clifford algebra, for which one
can take the module obtained by acting on a vector $|0\rangle$
that is annihilated by all the $c_{\alpha}$. In that case one
never has to deal with the Clifford algebra explicitly, and
we will ignore it from now on.}. This only works if the generators
form an exact closed algebra, not one up to BRST exact terms\footnote{
It would for instance be nice if one could express the generator
of the finite $W$-algebra corresponding to the quadratic Casimir
simply as $\eta_{ab} J^a J^b$, but then the finite $W$-algebra
no longer closes and it is in general not clear whether the
other generators can be modified in order to close the algebra.}. The only
realization which has this property is the one that one gets from
the tic-tac-toe construction in section~\ref{coho_sect}. This is not
yet the complete story, however. One would also like to get
highest weight representations of the finite $W$-algebra from
highest weight representations of ${\cal G}$. However, from
(\ref{newgens21}) one sees that $C$ contains a lowering operator
of ${\cal G}$, namely $J^8$. Hence $C$ is not automatically
diagonal on a highest weight of ${\cal G}$. To achieve also this
we have to use the quantum Miura transformation, which tells
us that we can consistently restrict the generators to their
bidegree $(0,0)$ part. If we do this, then $C$ will be diagonal
on a highest weight of ${\cal G}$, and we end up with a highest
weight representation of the finite $W$-algebra. Thus, to summarize,
we can construct highest weight representations of finite $W$-algebras via
\be \begin{array}{c}
\mbox{{\rm highest weight representations of }}{\cal G} \\
\downarrow \mbox{{\rm restriction}} \\
\mbox{{\rm highest weight representations of }}{\cal G}_0 \\
\downarrow \mbox{{\rm Miura}} \\
\mbox{{\rm highest weight representations of }}{\cal W}
\end{array}
\ee

Assume from now on that we are
in the basis of ${\cal G}_S$ where ${\cal G}_S$ is
generated by $\{E_{\pm \alpha_i},
H_{\alpha_i} \}_{i\in S}$. The above construction of
highest weight representations of the finite $W$-algebra
leads to a non-linear parametrization
of the eigenvalues of the generators of ${\cal W}_0$
in terms of the weights of ${\cal G}$. This is a correct
counting of degrees of freedom: ${\cal W}_0$ consists of generators
obtained from ${\cal G}_S$ in addition to those of ${\cal H}_{\cal K}$.
Since the $sl_2$ is principally embedded in ${\cal G}_S$, we get
${\rm rank}({\cal G}_S)$ generators in ${\cal W}_0$ from this,
so that ${\rm dim}({\cal W}_0)={\rm rank}({\cal G}_S)+
{\rm dim}({\cal H}_{\cal K})={\rm dim}({\cal H}_S)+
{\rm dim}({\cal H}_{\cal K})={\rm rank}({\cal G})$. The ${\cal W}_0$
eigenvalues corresponding to ${\cal H}_{\cal K}$ are linear
in terms of the ${\cal G}$ weight $\Lambda$. The ${\cal W}_0$
eigenvalues corresponding to the generators obtained from ${\cal G}_S$
will typically be non-linear, and be reminiscent of the Casimirs of
${\cal G}_S$. Indeed, the conjectured form of the Kac determinant,
to be discussed in a moment, shows the ${\cal W}_0$ eigenvalues
correspond to the $W_S$ invariants that can be constructed out of
$\Lambda$; $W_S$ is the Weyl group of ${\cal G}_S$, and its action on
the weights $\Lambda$ will be discussed later.

As we said previously, every generator of the finite
$W$-algebra has well defined eigenvalues with respect to the adjoint
action of ${\cal H}_{\cal K}$, and this defines for every $W$-generator
an root in ${\cal H}^{\ast}_{\cal K}$. The set of roots corresponding
to the generators of ${\cal W}_+$ will be denoted by $\Delta_+^{\cal W}$.
We will write $\Delta_+^{\cal W}
=\{(\bar{\alpha}_i,\mu_i)\}_{i \in I}$, where $\bar{\alpha}_i \in
{\cal H}_{\cal K}^{\ast}$, and $\mu_i$
labels the potential degeneracy of such a root. The corresponding
generators of ${\cal W}_+$ will be denoted by $W_{\bar{\alpha}}^{\mu}$,
and those of ${\cal W}_-$ by  $W_{-\bar{\alpha}}^{\mu}$. Furthermore
we will by $\bar{\Delta}^S_+$ denote the set of positive roots of
${\cal G}$ minus the set of positive roots of ${\cal G}_S$, and
denote by $\pi$ the canonical (orthogonal) projection from the root
space of ${\cal G}$ to that of ${\cal K}$. On ${\cal H}^{\ast}_{\cal K}$
we can define the Kostant partition function $P(\bar{\alpha})$,
which is the number of inequivalent series of nonnegative integers
$\{t_i\}_{i \in I}$ such that $\bar{\alpha}=
\sum_i t_i \bar{\alpha}_i$. Now consider the highest weight module
generated by $|\Lambda\rangle$. It consists of the vectors
\be \label{module_basis}
W^{\mu_1}_{-\bar{\alpha}_1} W^{\mu_2}_{-\bar{\alpha}_2}
\cdots W^{\mu_n}_{\bar{\alpha}_n}|\Lambda\rangle
\equiv W^{\{\mu\}}_{\{-\bar{\alpha}\}}|\Lambda\rangle , \ee
with $(\bar{\alpha_1},\mu_1) \geq (\bar{\alpha_2},\mu_2) \geq
\cdots \geq (\bar{\alpha_n},\mu_n)$ with respect to some ordering
of the roots of $\Delta_+^{\cal W}$. The states with
$\sum_i \bar{\alpha_i}=\bar{\beta}$ span some finite-dimensional
vector space $V_{\bar{\beta}}$. If $\omega$ denotes the
anti-involution corresponding to the compact real form of ${\cal W}$,
and we define $|\Lambda\rangle^{\dagger}=\langle \Lambda |$ and
$\langle \Lambda | \Lambda \rangle=1$, then unitarity
of the highest weight representation implies that
the ${\rm dim}(V_{\bar{\beta}}) \times {\rm dim}(V_{\bar{\beta}})$
dimensional matrix
\be M_{\bar{\beta}}(\Lambda)_{\{\mu',\bar{\alpha}'\},\{\mu,
\bar{\alpha}\}}= \langle\Lambda | \omega
(W^{\{\mu'\}}_{\{-\bar{\alpha}'\}} )
W^{\{\mu\}}_{\{-\bar{\alpha}\}} |\Lambda \rangle \ee
is hermitian and has only positive eigenvalues. Reducibility of the
highest weight representation is indicated by zero eigenvalues of
this matrix, and both unitarity and reducibility can be studied by
looking at the so-called Kac determinant of this matrix, which we
denote by $M_{\bar{\beta}}(\Lambda)$. We conjecture the following
form of the Kac determinant, similar to the one in \cite{Peter&Koos}
\be \label{kacdet}
 M_{\bar{\beta}}(\Lambda)= K(\bar{\beta},\Lambda)
\prod_{k>0} \prod_{\alpha \in
\bar{\Delta}^S_+} (
\langle \Lambda + \rho_{{\cal W}}, \alpha \rangle - \frac{k}{2}
\langle \alpha , \alpha \rangle )^{P(\bar{\beta}-k\pi(\alpha))},
\ee
where
$K(\bar{\beta},\Lambda)$ is a positive constant and
and $\langle , \rangle$ is the usual positive definite invariant inner
product on ${\cal H}^{\ast}$. The vector $\rho_{{\cal W}}$ that
appears in (\ref{kacdet}) depends on the details of the Miura
transformation and the $sl_2$ embedding. To construct it, we
look at the generator of the finite $W$ algebra which is written
in the tic-tac-toe form of section~\ref{coho_sect}, and which
differs from the quadratic Casimir $\eta_{ab}J^a J^b$ of ${\cal G}$
by a BRST-exact quantity. The bidegree $(0,0)$ of this generator,
which appears in the Miura transformation, has a certain eigenvalue
on the highest weight state $|\Lambda\rangle$.
The eigenvalue in terms of $|\Lambda\rangle$ is of the form
\be
a \langle \Lambda + \rho_{{\cal W}}, \Lambda + \rho_{{\cal W}}
\rangle + b
\ee,
for some constants $a,b$, and this defines $\rho_{{\cal W}}$.
One of the properties of $\rho_{{\cal W}}$ is that the
${\cal W}_0$ eigenvalues are invariant under the following
action of $W_S$
\be \label{aux2}
w\cdot \Lambda=w(\Lambda + \rho_{\cal W}) - \rho_{\cal W}.
\ee
This is consistent with (\ref{kacdet}), since one easily verifies
that $ M_{\bar{\beta}}(w\cdot\Lambda) = M_{\bar{\beta}}(\Lambda)$
for $w\in W_S$. Thus the Kac determinant can be rewritten as a
polynomial function of the ${\cal W}_0$ eigenvalues.

To illustrate the use of the Kac
determinant we will now work it out in more detail for the finite
$W$-algebra described in section~\ref{w32_sect}.

\subsection{Example}
\label{w32_sect2}
In this example the algebra ${\cal G}_S$ is generated by $\{t_1,t_5,t_8\}$,
which is not generated by $\{H_{\alpha_i},E_{\pm\alpha_i}\}_{i\in S}$.
Therefore we first perform an automorphism of the $sl_3$ to make sure
that ${\cal G}_S$ is generated by $\{H_{\alpha_1},E_{\pm\alpha_1}\}$. This
automorphism is
\be X\rightarrow \mats{1}{0}{0}{0}{0}{1}{0}{1}{0} X
\mats{1}{0}{0}{0}{0}{1}{0}{1}{0} \ee
and the new basis of $sl_3$ is
\be \label{basis21a}
r_at_a=\mats{\frac{r_4}{6}-\frac{r_5}{2}}{r_1}{r_2}{r_8}{
\frac{r_4}{6}+\frac{r_5}{2}}{r_7}{r_6}{r_3}{-\frac{r_4}{3}}.
\ee
Strictly speaking $J$'s are elements of the dual Lie algebra ${\cal G}^{\ast}$.
We identify the latter with the Lie algebra via the pairing $(X,Y)=
{\rm tr}(XY)$. Furthermore, highest weight representations of ${\cal G}$
correspond to lowest weight representations of ${\cal G}^{\ast}$,
and it is the highest weight representations of ${\cal G}^{\ast}$ that
give the highest weight representations of the finite $W$-algebra.
To correct for this, we apply the Chevalley automorphism
$H\rightarrow -H, E_{\alpha} \leftrightarrow E_{-\alpha}$ to
${\cal G}$. This leads to the following identification between
the $J$'s and $sl_3$ matrices
\be r_aJ^a=\mats{-r_4+r_5}{r_1}{r_2}{r_8}{-r_4-r_5}{r_7}{r_6}{r_3}{2r_4}.
\ee

The semi-simple subalgebra ${\cal K}$ is a $u(1)$,
and is spanned by $t_4$ which is proportional to
$2H_{\alpha_2}+H_{\alpha_1}$. Therefore, the root space
${\cal H}_{\cal K}^{\ast}$ is spanned by $\alpha_2+\frac{1}{2}\alpha_1$.
Furthermore we find that $\bar{\Delta}^S_+=\{\alpha_2,\alpha_1+\alpha_2\}$,
and that $\pi(\alpha_2)=\pi(\alpha_1+\alpha_2)=\alpha_2+\frac{1}{2}\alpha_1$.
Hence the positive roots of the $W$-algebra correspond to positive
multiples of $2\alpha_1+\alpha_1$, and we see that
${\cal W}_0$ is generated by $C$ and $j_0$, that
${\cal W}_+$ is generated
 by $j_+$ with root $\alpha_2+\frac{1}{2}\alpha_1$ and that
${\cal W}_-$ is generated
by $j_-$ with root $-(\alpha_2+\frac{1}{2}\alpha_1)$. The anti-involution
corresponding to the compact real form is the same as $\omega_1$ in
(\ref{1x4x10}).

The explicit form of the generators as given in (\ref{reps21}) and
(\ref{newgens21}) can now be used to deduce what the eigenvalues
of $j_0$ and $C$ on a $sl_3$ highest weight state $|\Lambda>$ are.
If we parametrize
the weight $\Lambda$ as $q_1 \lambda_1 + q_2 \lambda_2$,
where $\lambda_{1,2}$ are the fundamental weights of $sl_3$, we
find for the $j_0$ and $C$ eigenvalues $h,c$ respectively
\ba \label{ch-values}
h & = & 1+ \frac{2q_1+4q_2}{3},   \nn \\
c & = & -\frac{4}{9}(q_1^2+q_2^2+q_1 q_2)-\frac{4}{3} q_2 -1.
\ea
Alternatively, $c$ can be written as
$-\frac{2}{3} \langle \Lambda + \alpha_2 , \Lambda + \alpha_2 \rangle+
\frac{1}{3}$, and we deduce that $\rho_{{\cal W}}=\alpha_2$.
Therefore, element $w$ of the Weyl group acts on a weight $\Lambda$ as
\be
w \cdot \Lambda=w(\Lambda + \rho_{{\cal W}}) - \rho_{{\cal W}}.
\ee
The Weyl group is $\ze_2$ and is generated by the reflection $s_{\alpha_1}$
in the line perpendicular to $\alpha_1$. It acts on $q_1,q_2$ as
\be \label{weyl_act}
s_{\alpha_1}(q_1) = -q_1+2       ; \hspace{2cm}
s_{\alpha_1}(q_2) = q_1 + q_2 + 1,
\ee
and leaves $h,c$ invariant.

For the Kac determinant at $\bar{\beta}=p (\alpha_2+\frac{1}{2}\alpha_1)$
we find
\be
M_p(q_1,q_2) = K_p(q_1,q_2) \prod_{k=1}^{p} (q_2+2-k)(q_1+q_2+1-k),
\ee
which can in agreement with our expectations be rewritten in terms of
$c,h$ as
\be
M_p(q_1,q_2) = K_p(q_1,q_2) (\frac{3}{4})^p \prod_{k=1}^{p}
(c+(h+1-k)^2 + \frac{1}{3} (k^2-1) ).
\ee
This result can be compared with the results in \cite{Finite} by
working out explicit commutation relations. Then one finds out
that $K_p(q_1,q_2)=p! (\frac{4}{3})^p$.

The Kac determinant can now be used to obtain information about
unitary and reducible representations of the finite $W$-algebra.

Unitarity of a representation implies that the Kac determinant is positive
for every $\bar{\beta}$. In this example the representation is
spanned by $\{j_-^l|\Lambda\rangle\}_{l\geq 0}$ and the Kac determinant
gives us the norm
\be M_p(q_1,q_2)=\langle \Lambda | j_+^p j_-^p | \Lambda \rangle, \ee
so that positivity of the Kac determinant is here equivalent to unitarity.
It follows that the highest weight representation is unitary if
(i) $q_2<{\rm min}(-q_1,-1)$ or (ii) there exists a nonnegative integer
$r$ such that ${\rm max}(r-2,r-1-q_1)<q_2<{\rm min}(r-1,r-q_1)$.
If the Kac determinant has a zero at some level $p$, then the norm
of $j_-^p|\Lambda\rangle$ is zero and this vector can be consistently
put equal to zero, leading to a finite-dimensional
representation of the finite $W$-algebra. The condition for this is
that either $q_2=p-2$ or $q_1+q_2=p-1$. This leaves one parameter free
which we parametrize by an arbitrary
real number $x$ as $q_1+2q_2=\frac{3}{2}(p+x-2)$, so that
$h=p+x-1$ and $c=(1-p^2)/3-x^2$.
When are these finite-dimensional representations unitary? From the
Kac determinant we see that if $q_2=p-2$ then $q_1+q_2+1-k$
must be positive for $0\leq k \leq p-1$, and if $q_1+q_2=p-1$
then  $q_2+2-k$ must be positive for $0\leq k \leq p-1$. Therefore
we find that (i) $q_2=p-2$ and $q_1>0$ or (ii) $q_1+q_2=p-1$ and
$q_2>p-3$. Both possibilities imply the same condition for
$x$, namely $\frac{3}{2}(p+x-2)>2p-4$, or $x>(p-2)/3$.
Finally, let us look at the maximal reducible representations.
These are representations where the Kac determinant has a maximal
number of zeroes, corresponding to finite-dimensional reducible
representations of the finite $W$-algebra. Such representations
exist if for some integer $l$ satisfying $0<l<p$ we have
(i) $q_2+2-p=0$ and $q_1+q_2+1-l=0$ or (ii) $q_1+q_2+1-p=0$ and $q_2+2-l=0$.
In terms of $x$ this implies $p+l-3=\frac{3}{2}(p+x-2)$, or
$x=(2l-p)/3$. For this value of $x$ the finite $W$-algebra has
a $p$-dimensional representation with an $l$-dimensional
subrepresentation.

\sectiona{Characters of finite $W$-algebras}

In this final part of the description of the representation theory of finite
$W$-algebras we will take a closer look at the characters of these
representations. In the description of the Kac determinant we
explained that every highest weight representation can be decomposed
into finite-dimensional vector spaces $V_{\bar{\beta}}$, and the
Kac determinant was defined as ${\rm det}\langle v_i | v_j \rangle$ where
the $|v_i\rangle$ formed a certain distinguished basis of
$V_{\bar{\beta}}$. The formal
character of a highest weight representation $R$
with highest weight $\Lambda$ of a finite $W$-algebra is now defined as follows
\be  \label{defchar}
\mbox{{\rm ch }} R = \sum_{\bar{\beta}} {\rm dim}(V_{\bar{\beta}})
e^{\pi(\Lambda)-\bar{\beta}},
\ee
where $\pi$ is the orthogonal projection of the weight space
of ${\cal G}$ on that of ${\cal K}$.
This is a formal expression in the sense that it contains the
ill-defined object ${\rm exp}(\bar{\beta})$. The best way to think
of these objects is that ultimately we want to view the character
as a function on ${\cal H}_{\cal K}$, and for $\lambda \in
{\cal H}_{\cal K}$ the function  ${\rm exp}(\bar{\beta})$
is defined as ${\rm exp}(\bar{\beta})(\lambda)=
{\rm exp}(\langle\bar{\beta},\lambda \rangle)$. In particular
these exponentials satisfy the usual property
${\rm exp}(\bar{\alpha}) {\rm exp}(\bar{\beta})={\rm exp}(\bar{\alpha}+
\bar{\beta})$. The combination $\pi(\Lambda)-\bar{\beta}$
which appears in the exponent in (\ref{defchar}) is exactly what gives
the eigenvalues of the generators in ${\cal H}_{\cal K}$ when
acting on the states in $V_{\bar{\beta}}$.

The Verma module $M(\Lambda)$
with highest weight $\Lambda$ is by definition the
highest weight module that is freely generated by $|\Lambda\rangle$,
so that in particular we keep all the states with zero norm.
Every highest weight module with highest weight $\Lambda$ is a
quotient of $M(\Lambda)$, and the quotient of $M(\Lambda)$ and
its maximal proper submodule $M'(\Lambda)$ is an irreducible
highest weight module $L(\Lambda)=M(\Lambda)/M'(\Lambda)$. All
irreducible highest weight modules are of the form $L(\Lambda)$ for
some $\Lambda$, and our goal in this section is to study
the characters of the modules $L(\Lambda)$ in terms of the
characters of the Verma modules $M(\Lambda)$.

The characters of the Verma modules $M(\Lambda)$ are easily obtained,
since they are freely generated by ${\cal W}_-$, and we can use
the Poincar\'e-Birkhoff-Witt theorem to write down a basis
for $M(\Lambda)$ (see (\ref{module_basis})). From this one
deduces that
\be \label{aux1}
{\rm ch} M(\Lambda) = \frac{e^{\pi(\Lambda)}}{\prod_{\bar{\alpha}
\in \Delta_+^{\cal W}} (1-e^{-\bar{\alpha}}) }.
\ee
Non-trivial submodules of $M(\Lambda)$ can arise if $M(\Lambda)$
contains a singular vector, which is a vector (not equal to $|\Lambda
\rangle$) annihilated by all generators in ${\cal W}_+$. If we denote
this vector by $|\Lambda'\rangle$, then the Verma module $M(\Lambda)$
contains the submodule $M(\Lambda')$. If the latter would happen to be
the maximal proper submodule of $M(\Lambda)$, we would find
that $L(\Lambda)=M(\Lambda)/M(\Lambda')$. Since the character of a
quotient of two modules is simply the difference of the two characters,
this would lead to the following identity for characters
\be \mbox{{\rm ch }} L(\Lambda) =
\mbox{{\rm ch }} M(\Lambda) -
\mbox{{\rm ch }} M(\Lambda').
\ee
More general situations can occur, of course. For example, if
$M(\Lambda)$ contains three singular vectors $|\Lambda_1\rangle$,
 $|\Lambda_2\rangle$ and  $|\Lambda_3\rangle$, and $|\Lambda_3\rangle$
is contained in both $M(\Lambda_1)$ and $M(\Lambda_2)$, then the maximal
proper submodule is generated by $M(\Lambda_1)$ and $M(\Lambda_2)$,
but it would not be correct to subtract their characters from $M(\Lambda)$
to get the character of $L(\Lambda)$, since we would have subtracted the
vectors in $M(\Lambda_3)$ twice. Rather, the correct formula is
\be \mbox{{\rm ch }} L(\Lambda) =
\mbox{{\rm ch }} M(\Lambda) -
\mbox{{\rm ch }} M(\Lambda_1) -
\mbox{{\rm ch }} M(\Lambda_2) +
\mbox{{\rm ch }} M(\Lambda_3).
\ee
A further complication is the possible existence of subsingular vectors.
These are vectors that become singular after modding out a module generated
by singular vectors. In that case one also needs to subtract the
submodules generated by these subsingular vectors. Altogether this leads
to an expression for the character of $L(\Lambda)$ as a linear
combination of those of $M(\Lambda)$ with integer coefficients.
Since we know the latter explicitly (\ref{aux1}), this immediately yields
the explicit characters of $L(\Lambda)$.

The conjectured character formula in \cite{Peter&Koos} deals with the
maximally degenerate representations of the finite $W$-algebra.
These are representations for which the Kac determinant has
a maximal number of vanishing factors. Vanishing factors in the
Kac determinant are closely related to the existence of singular
vectors (the latter have zero norm), and an alternative definition
of maximally degenerate representations is those representations with
a maximum number of singular vectors. For these representations the
character formulas will be the most complicated ones.
A factor $\langle \Lambda + \rho_{\cal W},\alpha \rangle - \frac{k}{2}
\langle \alpha, \alpha \rangle$ in the
Kac determinant vanishes (for ${\cal G}=sl_n$) if
$\langle \Lambda + \rho_{\cal W},\alpha \rangle$ is a positive
integer, since every root has length two. If $\rho$ denotes one
half of the sum of the positive roots, then $\langle \rho,\alpha_i
\rangle=1$ for all simple roots $\alpha_i$. Therefore, if
$\Lambda+\rho_{\cal W}=\lambda+\rho$, and $\lambda$ is a dominant
integral weight (a linear combination of the fundamental weights with
nonnegative integer coefficients), then $\langle \Lambda+\rho_{\cal W},
\alpha \rangle$ will be a positive integer for any root $\alpha$.
 From now on we will restrict our attention to such weights.

The Weyl group $W$ of ${\cal G}$ acts on $\Lambda$ via (\ref{aux2}).
The subgroup $W_S$ does not change the representation of the finite
$W$-algebra, so from that point of view we may identify weights
and their images under the action of $W_S$. Thus, the orbit of
the weight under the Weyl group gives a set of weights of the
finite $W$-algebra in one-to-one correspondence with the coset
$W_S \!\setminus\! W$, and the conjecture is that these are precisely all the
singular vectors of the representation of the finite $W$-algebra.

The Weyl group is generated by the reflections in the hyperplanes
perpendicular to the simple roots. Every element of the Weyl
group can be written as the product of such reflections, and
the minimal number of reflections in simple roots needed to
obtain an element $w$ of the Weyl group is called the length of that
element, $l(w)$. A particular ordering of the elements of the Weyl
group plays in important role in the character formula, the so-called
Bruhat ordering (see e.g. \cite{coxeter}). Write $w'\rightarrow w$
if $w=tw'$ and $l(w)>l(w')$, where $t$ is the reflection in the
hyperplane perpendicular to some root (not necessarily simple).
Then we say that $w'<w$, if there is a sequence $w' \rightarrow w_1
\rightarrow \cdots \rightarrow w_{m-1} \rightarrow w$. Now given
a Weyl group $W_S$ associated to some subset $S$ of the set of
simple roots, we define $W^S$ as the set of $w\in W$ such that
$l(sw)>l(w)$ for all reflections $s$ in any hyperplane perpendicular
to any root (not necessarily simple) of ${\cal G}_{S}$.
Any element $w\in W$ can be uniquely written as $uv$ with $u\in W_S$
and $v\in W^S$, and $W^S$ is a set of representatives for the
coset $W_S\!\setminus\! W$. We will identify $W_S\!\setminus\! W$ with
the set $W^S$ and
give it a partial ordering by restricting the Bruhat ordering on
$W$ to $W^S$.

We can now write down the Kazhdan-Lusztig conjecture for finite
$W$-algebras, as proposed in \cite{Peter&Koos}. It reads
\be \label{KL}
\mbox{{\rm ch }} L( \tau \cdot (w_{{\rm max}} \cdot \Lambda) )
= \sum_{\sigma \leq \tau \in
W_S \! \setminus \! W} (-1)^{l(\sigma)}
(-1)^{l(\tau)} \tilde{P}^S_{\sigma \tau}(1) \mbox{{\rm ch }}
M(\sigma \cdot (w_{{\rm max}} \cdot \Lambda) ).
\ee
Here, $w_{{\rm max}}$ is the longest element of $W$
(so that $w_{{\rm max}}.\Lambda+\rho_{\cal W}-\rho$ is an
anti-dominant weight), and
 $\tilde{P}^S_{\sigma,\tau}(x)$ are the so-called dual relative
Kazhdan-Lusztig polynomials associated to $W_S$, see \cite{kalu,gema}.
In \cite{kalu} this character identity is proven for ordinary
simple Lie algebras, when $S$ is the empty set.

One of the implications of this character formula is that the singular
vectors in $M(\Lambda)$ are in one-to-one correspondence with the
elements of $W_S\! \setminus\! W \sim W^S$, and that the Bruhat ordering tells
us precisely which singular vectors can be obtained from which
other singular vectors by acting on them with ${\cal W}_-$.

To conclude this section we illustrate this character formula with the
example studied in sections~\ref{w32_sect} and~\ref{w32_sect2}. The
Weyl group of $sl_3$ is $S_3$ and the Bruhat ordering on $S_3$
coincides with the ordering with respect to the length of the elements,
i.e. $w<w' \leftrightarrow l(w)<l(w')$. If we denote the reflections
in the lines perpendicular to the simple roots $\alpha_1$ and $\alpha_2$
by $s_1$ and $s_2$, then $W_S=\{1,s_1\}$ and $W^S=\{1,s_2,s_2s_1\}$.
We represent the highest
weight as in section~\ref{w32_sect2} as $\Lambda=q_1 \lambda_1
+ q_2 \lambda_2$. The maximal Weyl group element is $s_1s_2s_1$ and by
explicit computation one finds that
\ba
w_{{\rm max}} \cdot \Lambda & = & (-1-q_2)\lambda_1 + (-1-q_1)\lambda_2
 \nn \\
s_2 \cdot (w_{{\rm max}} \cdot \Lambda) & = &
(-q_1-q_2)\lambda_1 + (q_1-3)\lambda_2
 \nn \\
s_2 s_1 \cdot ( w_{{\rm max}} \cdot \Lambda )
& = & (2-q_1)\lambda_1 + (q_1+q_2-1)\lambda_2.
\ea
In view of (\ref{weyl_act}) representations with weight
$s_2 s_1 \cdot ( w_{{\rm max}} \cdot \Lambda )$  can be identified with
those with weight $s_1s_2 s_1 \cdot ( w_{{\rm max}} \cdot \Lambda )$
which is just $\Lambda=q_1\lambda_1+q_2\lambda_2$, and we will do
so in the equations that follow.
The values of the dual relative Kazhdan-Lusztig polynomials are given
by $\tilde{P}^S_{\sigma\tau}(1)=1$ for all $\sigma\leq\tau$, with
one exception. If $\sigma=1$ and $\tau=s_2s_1$ then it vanishes.
Substituting everything in the character formula yields three
equations
\ba
\mbox{{\rm ch }} L(-1-q_2,-1-q_1) & = &
\mbox{{\rm ch }} M(-1-q_2,-1-q_1)
 \nn \\
\mbox{{\rm ch }} L(-q_1-q_2,q_1-3) & = &
\mbox{{\rm ch }} M(-q_1-q_2,q_1-3) -
\mbox{{\rm ch }} M(-1-q_2,-1-q_1)
 \nn \\
\mbox{{\rm ch }} L(q_1,q_2) & = &
\mbox{{\rm ch }} M(q_1,q_2) -
\mbox{{\rm ch }} M(-q_1-q_2,q_1-3) .
\ea
These identities agree with the picture of the representation of this finite
$W$-algebra as sketched at the end of section~\ref{w32_sect2}.
An explicit expression for the character of $L(q_1,q_2)$ is now
easily obtained. The positive root in $\Delta_+^{{\cal W}}$ is
$\alpha_2+\frac{1}{2} \alpha_1=\frac{3}{2} \lambda_2$, and
$\pi(\Lambda)=(q_2+\frac{1}{2}q_1)\lambda_2$. Thus
\be
\mbox{{\rm ch }} L(q_1,q_2) = \frac{e^{(2q_2+q_1)\lambda_2/2}-
e^{(q_1-q_2-6)\lambda_2/2}}{1-e^{-3\lambda_2/2} }.
\ee
The formal limit $\lambda_2 \rightarrow 0$ of the right hand side
yields $q_2+2$ and this is the dimension of $L(q_1,q_2)$. If we
parametrize as in the end of section~\ref{w32_sect2} the maximal
reducible representations by $q_1=l-2$ and $q_2=p-l-1$, we find
that the dimension of $L(q_1,q_2)$ is $l$. This is indeed correct since
for these values of $q_1,q_2$ the $p$ dimensional finite representation
had an $l$ dimensional subrepresentation which is precisely $L(q_1,q_2)$.

\chapter{Constructing Theories with Finite $W$-Symmetries}

In Chapter 2 we have seen some finite $W$-algebras that appear in
simple physical systems. This happened more or less by coincidence,
a priori we had no reason to expect a finite $W$-algebra to show up
(although anisotropic harmonic oscillators have quite generally
extended non-linear symmetry algebras for rational frequency
ratios, see e.g. \cite{Grieken}). Here we want to pose the reverse
question, namely, can one given some finite $W$-algebra construct
theories that have this $W$-algebra as their symmetry algebra? An additional
interesting question is whether or not one can build
gauge theories based on finite $W$-algebras. If we would succeed in
constructing a gauge theory for a finite $W$-algebra that contains
$SU(3)\times SU(2) \times U(1)$, this might be a new candidate for
a Grand Unified Theory. Such finite $W$-algebras certainly exist,
as we briefly explained at the end of section 3.3.6. Unfortunately,
we have not succeeded in constructing a gauge theory for finite
$W$-algebras except in one dimension. In this section we want to
sketch some of the ideas and problems involved in the construction
of theories with finite $W$-symmetries. This is definitively not
a finished chapter in the theory of finite $W$-algebras, and it will
have more the character of a series of remarks than that of a finished
theory.

\sectiona{Problems with finite $W$-symmetries}
\label{obstru}

We start with some particular (classical) finite $W$-algebra given
by the Poisson brackets
\be \label{z1}
\{ W_{\alpha}, W_{\beta} \} = P_{\alpha\beta}(W_{\gamma}),
\ee
where the $P_{\alpha\beta}$ are certain polynomials in $W_{\gamma}$.
Suppose that we have some field theory with fields $\phi_i$ with
a finite $W$-symmetry. This means that there are transformation
rules $\delta \phi_i = \sum_{\alpha}
\epsilon_{\alpha} \delta_{\alpha}\phi_i$ that leave the action invariant,
where $\epsilon_{\alpha}$ is the constant parameter corresponding to the
generator $W_{\alpha}$. Associated to these transformation rules are
a set of conserved currents $j^{\mu}_{\alpha}$ that can be found
in the usual way through the Noether procedure. The corresponding conserved
charges, that generate the symmetries, are
\be \label{z2} Q_{\alpha} = \int d^{d-1}x j^0. \ee
The statement that the theory is invariant under the finite $W$-algebra
(\ref{z1}) means that the conserved charges have to satisfy precisely
(\ref{z1}), i.e.
\be \label{z3}
\{ Q_{\alpha}, Q_{\beta} \} = P_{\alpha\beta}(Q_{\gamma}).
\ee
It is hard to see how one could realize (\ref{z3}). The left hand side
contains two integrations $\int d^{d-1} x$, and one of these disappears
after integrating over the delta function that arises in the equal-time
Poisson brackets. Hence, the left
hand side contains exactly one integration $\int d^{d-1} x$. The
right hand side contains as many integrations $\int d^{d-1} x$ as
the degree of $P_{\alpha\beta}$. These two facts can only be made to
agree with each other if (i) the finite $W$-algebra is linear or (ii)
$d=1$. Case (i) is precisely what we are not interested in, since
that brings us back in the realm of ordinary Lie algebras, and we
will come back to $d=1$ later. Therefore, it seems that we have some
kind of no-go theorem in dimensions $d>1$. What else could we try to do?
First, the above argument assumes the symmetries are local. If one drops
this assumption, it might still be possible to do something. We have not
analyzed this possibility, partly because of the problems in dealing
with theories with non-local symmetries. A second possibility is to
change the definition of what we mean by a theory with finite $W$-symmetries,
and this will be the subject of the next section. The reader who has
some knowledge of the corresponding situation with 'infinite' $W$-algebras
in two dimensions may wonder how these escape the no-go theorem given
above. The reason is that the 'infinite' $W$-algebras have an infinite
number of generators, and this makes it possible to convert the
integrations that remain in the right hand side of (\ref{z3}) into
infinite sums of generators. Since we restrict attention to finite
$W$-algebras with a finite number of generators, this does not provide
a way out either.

\subsection{Another definition of finite $W$-symmetries}

In \cite{SSN} an attempt was made to write down a gauge theory
for a non-linear algebra which is a deformation of $su(2)$. This construction
involved a set of scalar fields, which are not needed in the gauge
theory of pure $su(2)$. This can be rephrased in the language of
the previous paragraph by saying that rather than looking for a theory which
has (\ref{z3}) as its symmetry algebra, we look at one which has
the following algebra of symmetries
\be \label{z4}
\{ Q_{\alpha}, Q_{\beta} \} = P_{\alpha\beta}^{\gamma}(T_{\delta})Q_{\gamma},
\ee
where the $T_{\delta}$ are extra scalars, one for each generator of the
finite $W$-algebra. The $T_{\delta}$ are added by hand. Now the left
and right hand side of (\ref{z4}) contain the same number of integrations
and the objection of the previous paragraph no longer applies. In some sense
on might say that a minimal finite $W$-multiplet necessarily involves
an extra set of scalar fields. We have not yet specified what the
polynomials $P_{\alpha\beta}^{\gamma}$ are. A first guess could be
to require
\be \label{z6}
P_{\alpha\beta}^{\gamma}(Q_{\delta})Q_{\gamma}=
P_{\alpha\beta}(Q_{\delta}),
\ee
 so that (\ref{z4}) becomes identical
to (\ref{z3}) upon identifying $Q_{\alpha}$ with $T_{\alpha}$.
It is, however, not clear that with this choice the Jacobi identities
are satisfied. To verify the Jacobi identities, we need also the
bracket
\be \label{z5} \{ Q_{\alpha}, T_{\beta} \} = S_{\alpha \beta}(T_{\gamma}).
\ee
Furthermore, we assume that the equal time Poisson bracket of
$T_{\alpha}$ with $T_{\beta}$ vanishes. Then the Jacobi identities give
two differential equations for $S$ and $P$. If we choose
$P_{\alpha\beta}^{\gamma}$ as in (\ref{z6}), we cannot give an
explicit solution for $S$ or even prove that a polynomial solution
always exists. A much more natural choice is
\ba \label{z7}
S_{\alpha\beta}(T_{\gamma})&  = &P_{\alpha\beta} (T_{\gamma}), \nn \\
P_{\alpha\beta}^{\gamma}(T_{\delta}) &=&
 \frac{\partial}{\partial T_{\gamma}} P_{\alpha\beta} (T_{\delta}),
\ea
since now the Jacobi identities $\{Q_{\alpha},\{Q_{\beta},Q_{\gamma} \}
\}+{\rm cycl}=0$ and $\{Q_{\alpha},\{Q_{\beta},T_{\gamma} \}
\}+{\rm cycl}=0$ follow directly from the Jacobi identities for the
original $W$-algebra. These conventions differ slightly from those
in \cite{SSN}, but agree after rescaling the deformation parameter
in \cite{SSN} by a factor of two.

Is it possible to realize the algebra (\ref{z4}), (\ref{z5}) in some field
theory? Although this would not be a finite $W$-invariant theory in the
narrow sense, it would at least be a finite $W$-inspired theory. If there
are no further fields in the theory apart from the scalars $T_{\alpha}$,
it is natural to look at sigma-models with Lagrangian density
\be \label{z8}
{\cal L} = -\frac{1}{2} G^{\alpha \beta}(T_{\gamma}) \partial^{\mu}
T_{\alpha} \partial_{\mu} T_{\beta} - V(T_{\gamma}).
\ee
Invariance under the transformations generated by $Q_{\gamma}$ yields
the equations
\ba  \label{z9}
\frac{1}{2} \partial^{\rho} G^{\alpha\beta} P_{\rho\gamma}+
\frac{1}{2} G^{\rho\beta} \partial^{\alpha} P_{\rho\gamma}+
\frac{1}{2} G^{\alpha\rho} \partial^{\beta} P_{\rho\gamma}& =& 0,  \nn \\
\partial^{\rho} V P_{\rho\gamma}& =& 0.
\ea
The equation for the potential simply states that $V$ is in the center
of the finite $W$-algebra. If the latter is generated by certain
polynomials $C_i(W_{\alpha})$, then $V$ is a function of these $C_i$.
The center of a finite $W$-algebra from an $sl_2$ embedding consists of
the Casimirs of the original Lie algebra, since these clearly commute
with the BRST operator. Of course, depending on the explicit representatives
one chooses to represent the $W$-algebra generators, the generators of
the center of the $W$-algebra will in general differ from the Casimirs
by certain BRST-exact terms.

The solution to the first equation in (\ref{z9}) is less obvious. If the
symmetry algebra is a Lie algebra then one can choose $G^{\alpha\beta}$
to be an invariant inner product on the Lie algebra. However, Lie algebras
are special in the sense that $Q_{\alpha}$ and $T_{\alpha}$ transform in
the same way under $W$-transformations. In general, $Q_{\alpha}$
transforms in the same way as $\partial_{\mu} T_{\alpha}$, which differs
from the way $T_{\alpha}$ transforms. We have examined two
examples of non-linear algebras in detail. One of them is the
finite $W$-algebra $W_3^{(2)}$ and the other one is the following
non-linear deformation of $sl_2$\footnote{Non-linear algebras of this type
are among the few examples of non-linear algebras that are similar
to finite $W$-algebras and that have been studied previously, see remark
at the end of section 2.2.2.}
\be
[j_0,j_{\pm}] = \pm j_{\pm} \;\;\;\;\; [j_+,j_-]=\phi '(j_0),
\ee
where $\phi(j_0)$ is neither a linear nor quadratic function of $j_0$. The
center of this algebra is (classically) generated by $C_1=2j_+j_-+\phi (j_0)$.
The finite $W_3^{(2)}$ algebra can be written in the same way as (6.1.10),
where now $\phi =\frac{1}{3}j_0^3+Cj_0$ and $C$ is an additional generator
that commutes with everything. The center of this algebra is generated
by $C_1-j_+j_-+\phi$ and $C_2=C$. In both cases we have explicitly solved
the differential equation for $G^{\alpha\beta}$ and we have found that the
only allowed kinetic terms are in either case of the form
\be
G^{ij}(C_k)\partial^{\mu}C_i\partial_{\mu}C_j.
\ee
These are not really interesting, since they only induce dynamics for
the gauge invariant combinations of the $T_{\alpha}$. It would be
interesting to know whether a similar result holds for all non-linear
algebras. Before we start to look at other possibilities, we will
now first examine what happens if finite $W$ gauge fields are included.

\subsection{Gauge fields for finite $W$-symmetries}

Gauge fields appear if we want to make finite $W$-symmetries local, i.e.
we allow the parameter of the gauge transformations to be space-time
dependent. To find the transformation rules for the gauge fields one
can for example propose a covariant derivative for $T_{\alpha}$,
$D_{\mu} T_{\alpha}=\partial_{\mu} T_{\alpha} - h^{\beta}_{\mu}
R_{\beta\alpha}(T)$, and require that $D_{\mu} T_{\alpha}$
satisfies the same transformation rule under local $W$-transformations
as $\partial_{\mu} T_{\alpha}$ under global ones. This leads
to the following result
\ba \label{z12}
D_{\mu} T_{\alpha} & = & \partial_{\mu} T_{\alpha} - h^{\beta}_{\mu}
 P_{\beta\alpha}(T)  \nn \\
\delta h^{\beta}_{\mu} & = &
\partial_{\mu} \epsilon^{\beta} -\epsilon^{\eta}
\partial^{\beta} P_{\eta\gamma}(T) h^{\gamma}_{\mu},
\ea
where $\epsilon^{\gamma}$ is the
local parameter for the finite $W$-transformation generated by $Q_{\gamma}$.
With these definitions one
finds indeed
\be \label{z13}
\delta (D_{\mu} T_{\alpha}) = \epsilon^{\eta} \partial^{\gamma}
P_{\eta\alpha}(T) D_{\mu} T_{\gamma}.
\ee
It is not at all clear, however, that (\ref{z13}) is the appropriate
way to define an object that transforms `covariantly' under finite
$W$-transformations. Firstly, it is not clear how to define another
covariant derivative $D^{(2)}_{\mu}$ so that $D^{(2)}_{\nu} D_{\mu} T_{\alpha}$
transforms also covariantly, as $\partial_{\nu} \partial_{\mu} T_{\alpha}$
transforms differently from $\partial_{\mu} T_{\alpha}$, already under
finite $W$-transformations.
Secondly, one could as well consider modified transformation rules for
$h^{\beta}_{\mu}$ such as
\be \label{z14}
\delta h^{\beta}_{\mu}  =
\partial_{\mu} \epsilon^{\beta} -\epsilon^{\eta}
\partial^{\beta} P_{\eta\gamma}(T) h^{\gamma}_{\mu} +
\epsilon^{\eta} (D_{\mu} T_{\gamma}) \Sigma^{\beta\gamma}_{\eta}(T),
\ee
to obtain for $D_{\mu} T_{\alpha}$ the following transformation rule
\be \label{z15}
\delta (D_{\mu} T_{\alpha}) = \epsilon^{\eta} (\partial^{\gamma}
P_{\eta\alpha}(T) +P_{\alpha\beta}(T) \Sigma^{\beta\gamma}_{\eta} )
 D_{\mu} T_{\gamma}.
\ee
If we can choose $\Sigma$ in such a way that $D_{\mu} T_{\alpha}$ transforms
in the same way as $T_{\alpha}$, then we can immediately write down an
invariant kinetic term for the $T_{\alpha}$, based on one of the elements
of the center of the finite $W$-algebra. We have, however, no clue
whether this can be done or whether it is sensible. Alternatively, one
could try to fix $\Sigma$ by requiring the gauge algebra to close on
the gauge fields. One can easily compute that
\be \label{z16}
[\delta_{\epsilon_1},\delta_{\epsilon_2}] h^{\beta}_{\mu} =
\delta_{\epsilon_3} h^{\beta}_{\mu} - \epsilon_1^{\eta}
\epsilon_2^{\sigma} (D_{\mu} T_{\lambda}) X^{\mu\lambda}_{\eta\sigma},
\ee
where $\epsilon_3^{\beta}=\partial^{\beta} P_{\gamma\eta}
\epsilon_1^{\eta} \epsilon_2^{\sigma}$ and
\ba \label{z17}
 X^{\mu\lambda}_{\eta\sigma} & = &
-\partial^{\beta} P_{\eta\gamma} \Sigma^{\gamma\lambda}_{\sigma}
+\partial^{\beta} P_{\sigma\gamma} \Sigma^{\gamma\lambda}_{\eta}
+\partial^{\lambda} P_{\sigma\gamma} \Sigma^{\beta\gamma}_{\eta}
-\partial^{\lambda} P_{\eta\gamma} \Sigma^{\beta\gamma}_{\sigma}
+P_{\gamma\theta}\Sigma_{\sigma}^{\theta\lambda}
\Sigma_{\eta}^{\beta\gamma}
 \nn \\  & &
-P_{\gamma\theta}\Sigma_{\eta}^{\theta\lambda}
\Sigma_{\sigma}^{\beta\gamma}
+P_{\sigma\gamma} \partial^{\gamma} \Sigma^{\beta\lambda}_{\eta}
-P_{\eta\gamma} \partial^{\gamma} \Sigma^{\beta\lambda}_{\sigma}
-\partial^{\gamma} P_{\sigma\eta} \Sigma^{\beta\lambda}_{\gamma}
-\partial^{\lambda} \partial^{\beta} P_{\sigma\eta}.
\ea
Clearly, the gauge algebra closes if $X^{\mu\lambda}_{\eta\sigma}=0$,
but the significance of this equation for $\Sigma$ remains to be seen.
Finally, one would like to write down an invariant action for
the gauge fields, preferably in terms of a generalized curvature,
which would be a non-linear version of Yang-Mills
theory. It would be interesting to analyze for each of the possible
transformation rules for the gauge fields whether or not such invariant
actions exist. An attempt in this direction was made in \cite{SSN},
where an invariant action was constructed for a non-linear deformation
of $su(2)$, to first order in the deformation parameter. However, it is not
clear whether their result can be extended systematically to an arbitrary
order in the deformation parameter.

\subsection{Other possibilities in $d>1$}

In this final section on dimensions larger than one, we will briefly indicate
some other possible approaches to the construction of invariant actions for
finite $W$-algebras.
\begin{itemize}
\item
In the construction sketched so far, we did not use the fact that our
finite $W$-algebras were obtained from a Lie algebra by imposing constraints.
It would be nice if we could somehow use this fact to our advantage.
Suppose we have at our disposal a theory which is
invariant under some global symmetry
algebra ${\cal G}$.
Associated to these ${\G}$ transformations are certain conserved charges whose
Poisson brackets form precisely ${\cal G}$. We can impose the constraints
that some of these charges have to be equal to either zero or one, but
since these constraints are non-local, it is not obvious what the
best way to impose them is. Adding the constraints with a Lagrange multiplier
to the action makes the action non-local and difficult to handle.
\item
Alternatively, if we would start with an action with a local symmetry based
on the algebra ${\cal G}$, we could try to impose constraints on the
gauge fields. Putting some components of the gauge field equal to one
breaks Lorentz invariance, since the gauge field transforms as a vector,
and not as a scalar. Putting components equal to zero is only compatible
with Lorentz invariance if we put the same components equal to zero
for all $A_{\mu}$, $\mu=0\ldots d-1$. This is such a strong requirement
that typically no interesting non-linear symmetry will be left.
A possible way around these obstructions is to perform some twisting,
so that $A_{\mu}$ transforms in a non-standard way under Lorentz
transformations. Whether or not this is possible remains to be explored,
but this is for example what one does in two dimensions to to get
infinite $W$-algebras.
\item
Previously, we have examined the representation theory of finite $W$-
algebras in some detail. Given some $n$-dimensional unitary
representation of a finite $W$-algebra, one can always take fields
$\phi_i$, $i=1\ldots n$ that transform in this representation, and
write down a finite $W$-invariant term $\phi_i^{\ast} \phi_i$.
These are clearly not very interesting, as they are invariant under
$U(n)$ and the finite $W$-algebra is realized as a subalgebra of
$U(n)$. Now one can write down terms $\phi_i^{\ast} M_{ij} \phi_j$
which are also invariant under $U(n)$ if $M$ transforms in the
adjoint representation. An interesting question is whether one can
come up with some constrained field $M_{ij}$ so that this term
is no longer $U(n)$ invariant, but still invariant under finite
$W$-transformations, so that it can be used to break $U(n)$ to a finite
$W$-algebra. Another way to phrase this question is whether
the finite $W$-orbit on $M_{ij}$ is the same as the $U(n)$ orbit,
or strictly smaller?
\item
Can one use the fact that many finite $W$-algebras look as if they
are deformations of some Lie algebra? There is a one-to-one
correspondence between generators of a finite $W$-algebra associated
to an $sl_2$ embedding, and the generators of the Lie algebra ${\cal G}_0$
(see section~\ref{miura_sect}), given by the requirement that they
belong to the same $sl_2$ orbit. We do not know whether a finite
$W$-algebra can always be seen as a deformation of ${\cal G}_0$, and
whether this would be useful for the construction of actions.
\item
The universal enveloping algebra of a finite $W$-algebra can be viewed as
an infinite-dimensional Lie algebra, subject to additional relations. Can
one construct a theory invariant under this infinite-dimensional Lie
algebra and systematically impose the extra relations?
\end{itemize}

Fortunately, the obstruction explained in section~\ref{obstru} does not
apply in one dimension. For the remainder, we will examine some
of the possibilities that exist in $d=1$.

\sectiona{Finite $W$-invariance in $d=1$}
\label{sect62}

In one dimension it is possible to realize finite $W$-algebras in terms
of conserved charges. However, this fact in itself is not yet sufficient
to construct a theory invariant under finite $W$-symmetries. For
that one needs a realization of finite $W$-algebras in terms of known
objects, such as creation and annihilation operators or generators of
a Lie algebra. Or alternatively, one can use the fact that
finite $W$-algebras were obtained by imposing constraints on the generators of
a Lie algebra. We will now briefly examine these possibilities in turn.

\subsection{Imposing constraints}

Suppose we have some action which is invariant under the global symmetry
algebra ${\cal G}$. Associated to these global symmetries is a set
of local conserved charges that obey $\{Q_{\alpha},Q_{\beta}\}=
f^{\gamma}_{\alpha\beta} Q_{\gamma}$. Then we can immediately write down
an action in which we impose the constraints necessary to get the
finite $W$-algebra $({\cal G},{\cal L},\chi)$ by adding to the action the
term
\be \label{z18}
\int dt \, {\rm Tr} (A (Q-\chi(Q)))+ \ldots,
\ee
where $A$ is an ${\cal L}$ valued gauge field, and
$Q=Q_{\alpha} T^{\alpha}$. The dots indicate possible terms that are
of higher order in $A$.
This action has a local gauge invariance generated
by the first class constraints $Q-\chi(Q)$, under which $A$ transforms as
a kind of gauge field. How $A$ transforms exactly depends on the details
of the theory.
If we perform a BRST gauge fixing of this gauge symmetry
than the action becomes the original action plus a free ghost action,
and the BRST operator that generates the BRST symmetries of this action
is precisely the one we used to analyze quantum finite $W$-algebras,
with $J$ replaced by $Q$. In particular, the Hilbert space of the
theory is given by the BRST cohomology and carries a representation of
the finite $W$-algebra. This is a genuine finite $W$-invariant theory,
but the $W$-invariance is only apparent on the level of the Hilbert
space.

In some cases one starts with theories whose symmetry algebra contains
two copies of ${\cal G}$. When this happens it is possible to impose
constraints on both algebras, and it can happen that one can explicitly
integrate out the Lagrange multipliers, thus yielding an action without
BRST symmetry but with a finite $W$-invariance. One particular example
is to start with the action of a point particle moving in a group manifold.
This will lead to the celebrated Toda theories, and we will describe this
example in some more detail.

One starts with the action for a free particle moving on the group
manifold $G$ (the Lie algebra of $G$ is ${\cal G}$).
The metric on $G$ is given by extending the
Cartan-Killing metric $(t_a,t_b)=\tr( t_at_b)$ on the Lie
algebra ${\cal G}$ all over $G$ in a left-right invariant way.
This leads to the
familiar action
\be \label{freeac}
S[g]=\frac{1}{2} \int dt \tr \left(
g^{-1}\frac{dg}{dt} g^{-1}\frac{dg}{dt} \right).
\ee
It satisfies the following identity
\be \label{polwie}
S[gh]=S[g]+S[h]+\int dt \tr \left(
g^{-1} \frac{dg}{dt} \frac{dh}{dt} h^{-1} \right),
\ee
from which one immediately deduces the equations of motion,
\be \label{eqnmot}
\frac{d}{dt}
\left( g^{-1} \frac{dg}{dt} \right)=
\frac{d}{dt}
\left(  \frac{dg}{dt} g^{-1} \right)=0.
\ee
The action (\ref{freeac}) is invariant under $g\rightarrow h_1 g
h_2$ for constant elements $h_1,h_2\in G$. This leads to the
conserved currents $J,\bar{J}$ given by
\be \label{conscurr}
J
=\frac{dg}{dt}g^{-1} \equiv J^at_a
\,\,\,\, \mbox{\rm and} \,\,\,\,
\bar{J}
=g^{-1}\frac{dg}{dt} \equiv \bar{J}^at_a.
\ee
The equations that express the conservation of these currents in
time coincide with the equations of motion of the system, so
fixing the values of these conserved quantities completely fixes
the orbit of the particle once its position on $t=0$ is
specified. In this sense the free particle on a group is a
completely integrable system. The conserved quantities form  a
Poisson algebra \cite{papad}
\be \label{pois}
\{J^a,J^b\}=f^{ab}_{\,\,\, c}J^c,
\ee
with similar equations for $\bar{J}$. This is precisely the
Kirillov Poisson bracket we used as a starting point for the
construction of finite $W$-algebras. These were obtained by
imposing constraints on the Poisson algebra (\ref{pois}), and we
want to do the same here to get systems with finite $W$
symmetry. Actually, we already have the first explicit example at our
disposal here. If we consider the trivial embedding of $sl_2$ in
$SL_n$, then the finite $W$-algebra is the Kirillov Poisson
algebra (\ref{pois}). The action (\ref{freeac}) is the
generalized Toda theory for the trivial embedding. The conserved
currents of this generalized Toda theory form a Poisson algebra
that is precisely the finite $W$-algebra associated to the
trivial embedding.

Finite $W$-algebras were obtained by imposing a set of first
class constraints $\pi_{{\cal L}}(J)=\chi(J)$,
where $\pi_{{\cal L}}$ is the projection on ${\cal L}$. Here we want
to impose the same constraints, together with
similar constraints on
$\bar{J}$,
\be \label{constr}
\pi_{{\cal L}^{\ast}}(\bar{J})=\chi^{\ast}(\bar{J})
\,\,\,\,\, \mbox{\rm and} \,\,\,\,\,
\pi_{{\cal L}}(J)=\chi(J).
\ee
There are two equivalent ways to deal with these constraints.
One can either reduce the equations of motion, or reduce the
action for the free particle.
Let us first reduce the
equations of motion, where we restrict our attention to
the finite $W$-algebras obtained from an $sl_2$ embedding.
 If $G_{\pm}$ denote the subgroups of $G$
with Lie algebra ${\cal L},{\cal L}^{\ast}$, and $G_0$ the subgroup with Lie
algebra ${\cal G}_0$, then almost every element $g$ of $G$ can be
decomposed as $g_-g_0g_+$, where $g_{\pm,0}$ are elements of the
corresponding subgroups, because $G$ admits a generalized Gauss
decomposition $G=G_-G_0G_+$\{Strictly speaking
$G_-G_0G_+$ is only dense in $G$, but we will ignore this subtlety
in the remainder\}. Inserting $g=g_-g_0g_+$ into (\ref{constr})
we find
\ba \label{solco}
g_0^{-1} t_+ g_0 & = & \frac{dg_+}{dt} g_+^{-1} \nonu
g_0 t_- g_0^{-1} & = & g_-^{-1} \frac{dg_-}{dt}.
\ea
In the derivation of these equations one uses that
$\pi_{{\cal L}}(g_-t_+g_-^{-1})=t_+$, and a similar equation with
$\pi_{{\cal L}^{\ast}}$ and $t_-$,
 which follow from the fact that $t_{\pm}$
have degree $\pm 1$. The constrained currents look like
\ba \label{jconstr}
J & = & g_-\left(t_+ + \frac{dg_0}{dt} g_0^{-1} +
g_0 t_- g_0^{-1} \right) g_-^{-1}, \nonu
\bar{J} & = & g_+^{-1}\left(t_- + g_0^{-1} \frac{dg_0}{dt} +
 g_0^{-1} t_- g_0 \right) g_+.
\ea
The equations of motion now become
\ba \label{eqnm2}
0=g_-^{-1} \frac{dJ}{dt} g_- & = & \frac{d}{dt}
\left( \frac{dg_0}{dt} g_0^{-1} \right) + [g_0t_-g_0^{-1},t_+],
\nonu
0=g_+ \frac{d\bar{J}}{dt} g_+^{-1} & = & \frac{d}{dt}
\left( g_0^{-1} \frac{dg_0}{dt} \right) +
[t_-,g_0^{-1}t_+g_0],.
\ea
which are generalized finite Toda equations as will be shown in a
moment.

Alternatively, one can reduce the action by writing
down the following gauged version of the action
\begin{eqnarray} \label{wzwg}
S[g,A_+,A_-] & = & \frac{1}{2} \int dt \tr \left(
g^{-1}\frac{dg}{dt} g^{-1}\frac{dg}{dt}
+A_+^2 +A_-^2 \right) \nonumber \\
&   & + \int dt \tr
\left( A_-(J-\chi(J))+A_+(\bar{J}-\chi^{\ast}(\bar{J}))+A_-gA_+
g^{-1} \right).
\end{eqnarray}
This action is invariant under the following transformations
\ba \label{invar}
g & \rightarrow & h_- g h_+,\nonu
A_- &  \rightarrow & h_-A_-h_-^{-1}-\frac{dh_-}{dt}h_-^{-1},
\nonu
A_+ &  \rightarrow & h_+^{-1}A_+h_+ -h_+^{-1}\frac{dh_+}{dt},
\ea
where $h_{\pm}$ are arbitrary elements of $G_{\pm}$.
We assume here that $\chi$ is either constant or zero. If
$\chi$ is some higher dimensional representation, one
needs in addition to (\ref{wzwg}) a Lagrangian describing these
additional degrees of freedom.
In the case where the finite $W$-algebra comes from an $sl_2$
embedding
we can use the gauge invariance to put $g_+=g_-=e$
(where $e$ is the unit element of the group $G$) in the Gauss
decomposition of $g$, thus we can take $g=g_0\in G_0$. Then from
the equations of motion for $A_{\pm}$ we find $A_+=g_0^{-1}t_+
g_0$ and $A_-=g_0t_- g_0^{-1}$
(The terms $A_+^2$ and $A_-^2$ are not present in this case).
 Substituting these back into the
action it reduces to
\be \label{finalact2}
S[g_0]  =  \frac{1}{2} \int dt \tr \left(
g_0^{-1}\frac{dg_0}{dt} g_0^{-1}\frac{dg_0}{dt} \right)
 - \int dt \tr
\left( g_0t_-g_0^{-1} t_+ \right).
\ee
The equations of motion for this action are indeed given by
(\ref{eqnm2}), showing the equivalence of the two approaches.

This generalized Toda action describes a particle moving on
$G_0$ in some background potential. Two commuting copies of the
finite $W$-algebra leave the action (\ref{finalact2}) invariant
and act on the space of solutions of the
equations of motion (\ref{eqnm2})\footnote{More precisely, the
symmetries of (\ref{finalact2}) form an algebra that is on-shell
isomorphic to a finite $W$-algebra}. An explicit proof of this will
be given in the next section. This action is only given
infinitesimally, because we do not know how to exponentiate
finite $W$-algebras. One can, however, sometimes
find subspaces of the space of solutions that constitute a
minimal orbit of the $W$-algebra, see for example \cite{palla}
where this was worked out for the infinite $W_3$ algebra.

For the principal embeddings of $sl_2$ in $sl_n$, the equations
of motion reduce to ordinary finite Toda equations of the type
\be \label{fintod}
\frac{d^2q_i}{dt^2}+\exp\left(\sum_{j=1}^{n-1}K_{ij}q_j\right)=0,
\ee
where $i=1,\ldots, n-1$, $K_{ij}$ is the Cartan matrix of
$sl_n$, and $g_0=\exp(q_iH_i)$.

The general solution of the the equations of motion
(\ref{eqnm2}) can be constructed as follows. Let
$h_0^{(1)},h_0^{(2)}$ be elements of $G_0$.
Let $X_0$ be an arbitrary element of $g_0$.
If $g_0(t)$ is defined by the Gauss decomposition
\be \label{allsol}
g_-(t)g_0(t)g_+(t) = h_0^{(1)} \exp t(X_0+(h_0^{(1)})^{-1}t_+
h_0^{(1)}+h_0^{(2)}t_-(h_0^{(2)})^{-1}) \,\,h_0^{(2)},
\ee
then $g_0(t)$ is the most general solution of (\ref{eqnm2}).
The easiest way
to study the action of the finite $W$-algebra on these solutions,
is to use the explicit transformation rules (\ref{zz05}).
This might provide a valuable tool in the study of the solutions
(\ref{allsol}).

In the case when $\chi=0$ in (\ref{wzwg}) it is not so straightforward
to pick a good gauge for $g$, unless ${\cal L}={\cal G}^+$, which corresponds
to the case described in section~\ref{gplus_alg}. Then a good gauge
choice is to pick $g$ in $G_0$, and integrating out $A_{\pm}$ is
trivial. The result is
 \be \label{finalact3}
S[g_0]  =  \frac{1}{2} \int dt \tr \left(
g_0^{-1}\frac{dg_0}{dt} g_0^{-1}\frac{dg_0}{dt} \right).
\ee
The symmetry of this theory is ${\cal G}_0\times {\cal G}_0$, which is
in perfect agreement with (\ref{hhh00}).

Finally, let us present an action which has the finite $W$-algebra
obtained by setting the Lie algebra generators in a Cartan subalgebra
equal to zero, as discussed in section (\ref{torusalg}). We take the
action (\ref{wzwg}), but since it is not easy to find a good gauge
choice for $g$, we put $A_+=0$, ie we impose only constraints on
$J$, not on $\bar{J}$. This is a special case of (\ref{z18}).
A good gauge choice is then for example $g=g_-g_+$, where $g_{\pm}\in G_{\pm}$
and $G=G_-TG_+$ is a standard Gauss decomposition of $G$. Integrating
out $A$ yields the following action
\be \label{z20}
S[g_-,g_+]= \int dt \tr
\left(g_-^{-1} \frac{dg_-}{dt}  \frac{dg_+}{dt}g_+^{-1}
\right)-\int dt \tr \left( \pi_{{\cal T}}
\left(g_-\frac{dg_+}{dt}g_+^{-1}g_-^{-1}\right)^2\right),
\ee
where $\pi_{\cal T}$ is the projection on the Lie algebra ${\cal T}$.

Any of the actions in this section can in principle be used as a starting
point for a theory with local finite $W$-symmetries in one dimension.
We are not going to discuss this here; basically one can use the same
techniques as one uses in two dimension to gauge infinite $W$-algebras,
see eg \cite{mythesis}.

\subsection{Realizations of finite $W$-algebras in terms of lie algebras}

The Miura transformation provides a realization of a finite $W$-algebra
that comes from an $sl_2$ embedding in terms of the generators of
${\cal G}_0$ (see section~\ref{miura_sect}). Can one, given such a
realization, find an action invariant under finite $W$-transformations?
One way is to first express the generators of ${\cal G}_0$ in terms
of oscillators (see \cite{thesis,Kostant}),
thereby providing a Fock realization of the finite
$W$-algebra. Subsequently, one can try to use the method in the next
section to find an invariant action. Alternatively, one can
try to use the realization in terms of ${\cal G}_0$. An obvious
guess is to look for actions of the type
\be \label{zz01}
S[g_0]  =  \frac{1}{2} \int dt \tr \left(
g_0^{-1}\frac{dg_0}{dt} g_0^{-1}\frac{dg_0}{dt} \right)
 - \int dt V(g_0).
\ee
The kinetic part of this action has a ${\cal G}_0\times {\cal G}_0$
invariance, and the problem is now to find a potential $V(g_0)$ that
reduces this invariance to a finite $W$-invariance. To be able to
say something more we first need to work out some properties of the
polynomials in terms of $J_0=\frac{dg_0}{dt} g_0^{-1}$ that the Miura
transformation gives us and that form a realization of the finite $W$-algebra.

The (classical) Miura transformation can be described as follows.
We start with the Kirillov Poisson algebra (\ref{pois}), and
decompose $J$ as $J_- +J_0 +J_+$, and in addition we will decompose
$J_-=J_{-1}+J_{-2}+\ldots$ according to (\ref{grading}).
After imposing the constraints we get $J_{{\rm constr}}=J_-+J_0+t_+$.
The finite $W$-algebra is the Poisson algebra of the polynomial.
$P(J_-,J_0)$ that are gauge invariant under the gauge transformations
generated by the first class constraints on the constrained phase
space, i.e. $P(J_-,J_0)$ satisfies
\be \label{xgauge_inv}
\tr \left( \left( \frac{\delta P(J_-,J_0)}{\delta J_-} +
 \frac{\delta P(J_-,J_0)}{\delta J_0} \right) [J_- + J_0 + t_+, \epsilon_-]
\right)=0,
\ee
where $\epsilon_-$ is an arbitrary parameter with values in ${\cal G}_-$.
We can rewrite (\ref{xgauge_inv}) as
\be \label{ygauge_inv}
\pi_{+} \left[  \frac{\delta P(J_-,J_0)}{\delta J_-} +
 \frac{\delta P(J_-,J_0)}{\delta J_0}  , J_- + J_0 + t_+
\right]=0,
\ee
with $\pi_+$ the projection on ${\cal G}_+$. The Miura transformation
does not give $P(J_-,J_0)$, but just $P(J_0)=P(J_-,J_0)|_{J_-=0}$.
If we insert $J_-=0$ in (\ref{ygauge_inv}) and then project it
onto ${\cal G}_{+1}$, we find
\be \label{pident}
\left[Q(J_0),J_0\right]+\left[ \frac{\delta P(J_0)}{\delta J_0},t_+
\right]=0,
\ee
with $Q(J_0)=(\delta P(J_-,J_0)/\delta J_{-1})|_{J_-=0}$.

Under a small $W$-transformation generated by $P(J_0)$, the potential
$V(g_0)$ transforms as
\be \label{del_V}
\delta V(g_0) = \epsilon \tr \left(
\frac{\delta P(J_0)}{\delta J_0} \frac{\delta V(g_0)}{\delta g_0 g_0^{-1}}
\right).
\ee
In order for the perturbed action to be $W$-invariant, we want this
variation to be total derivative. In order to be able to use the identity
(\ref{pident}) in (\ref{del_V}), we need to require that
\be \label{rule1}
 \frac{\delta V(g_0)}{\delta g_0 g_0^{-1}} = [t_+,R(g_0)].
\ee
This allows us to rewrite
\be \label{del_V2}
\delta V(g_0) = \epsilon \tr \left(
Q(J_0)[R(g_0),J_0]
\right).
\ee
If in addition to (\ref{rule1}) we require that
\be \label{rule2}
[R(g_0),J_0]=\frac{d}{dt} T(g_0)
\ee
for some functional $T(g_0)$,
then $\delta V(g_0)$ is a total time derivative modulo the equation of
motion $dJ_0/dt=0$ of the unperturbed part of (\ref{zz01}).
We can then modify the transformation rule of $g_0$ to cancel the
equation of motion terms in $\delta V(g_0)$, but this gives rise to
a new variation of the potential $V(g_0)$, and we have to check that this
is again a total time derivative. Clearly, this is a somewhat cumbersome
procedure and it is not clear that it terminates. One obvious solution to
(\ref{rule1}) and (\ref{rule2}) is the potential $V(g_0)=\tr(g_0 t_-
g_0^{-1} t_+)$, with $R(g_0)=g_0 t_- g_0^{-1}$ and $T(g_0)=-R(g_0)$.
We do not know whether other simple solutions to (\ref{rule1}) and
(\ref{rule2}) exist. Rather than verifying step by step that (\ref{finalact2})
is invariant under finite $W$-transformations, we will give a direct proof
of this. This proof, given below, will show that the procedure sketched
in this section is in general not very efficient for finding
invariant actions, but may
provide some clues for finding other and better techniques.

The Toda action was obtained by imposing constraints on both $J$ and
$\bar{J}$. The corresponding constraints first brought
$J$ in the form $J=J_-+J_0+t_+$, and subsequently in the form
(\ref{jconstr}), where $g_-$ is a function of $g_0$ determined by
(\ref{solco}). If we substitute this special constrained form of $J$
in the gauge invariant polynomials $P(J_0,J_-)$, then the $g_-$
dependence disappears, since we are precisely interested in $G_-$
invariant polynomials, and we find the following polynomials
in terms of $g_0$ and its time derivatives:
\be \label{consqua}
P(g_0)\equiv P(J_0,J_-)|_{J_0=\frac{dg_0}{dt} g_0^{-1}, J_{-1}=
g_0t_-g_0^{-1}, J_{-2}=0,J_{-3}=0,\ldots}
\ee
We claim that these are precisely the conserved quantities of the Toda theory.
To prove this, we take (\ref{ygauge_inv}) and deduce from it that we must
in particular have
\be \label{y2gauge_inv}
\tr \left( J_{-1} \left[  \frac{\delta P(J_-,J_0)}{\delta J_-} +
 \frac{\delta P(J_-,J_0)}{\delta J_0}  , J_- + J_0 + t_+
\right] \right) =0.
\ee
If we put $J_{-2}=J_{-3}=\ldots=0$ in this equation, and denote
\be
P(J_{-1},J_0)=P(J_-,J_0)|_{J_{-2}=0,J_{-3}=0,\ldots}
\ee
etc., we find
\ba
\tr\left( J_{-1} \left(
\left[ \frac{\delta P(J_{-1},J_0)}{\delta J_{-1}}       ,J_0 \right] +
\left[ \left. \frac{\delta P(J_{-2},J_{-1},J_0)}{\delta J_{-2}}
 \right|_{J_{-2}=0} ,J_{-1} \right] +  \right. \right. & & \nn \\
\left. \left.
 \left[ \frac{\delta P(J_{-1},J_0)}{\delta J_0}       ,t_+ \right]
\right) \right)=0. & &
\ea
The middle term in this equation drops out, and the remainder can be
rewritten as
\be \label{zz03a}
\tr\left(
\frac{\delta P(J_{-1},J_0)}{\delta J_{-1}} [J_0,J_{-1}] +
\frac{\delta P(J_{-1},J_0)}{\delta J_0} [J_0,t_+]
\right)=0.
\ee
Now if we put $J_0=\frac{dg_0}{dt} g_0^{-1}$ and $J_{-1}=g_0 t_- g_0^{-1}$,
then
\ba
\frac{dJ_0}{dt} & = & [t_+,J_{-1}]  \nn \\
\frac{dJ_{-1}}{dt} & = & [J_0,J_{-1}].
\ea
The first equation is the Toda equation of motion, and the second one is
a straightforward algebraic identity. Using these, (\ref{zz03a}) can be
rewritten as
\be \label{zz03}
\tr\left(
\frac{\delta P(J_{-1},J_0)}{\delta J_{-1}} \frac{dJ_{-1}}{dt} +
\frac{\delta P(J_{-1},J_0)}{\delta J_0} \frac{dJ_0}{dt}
\right)=0,
\ee
and this is nothing but $\frac{d}{dt} P(g_0)$, proving that
$P(g_0)$ is a conserved quantity in the Toda theory.

The transformation rules that leave (\ref{finalact2}) invariant can now
immediately be deduced. They read
\be  \label{zz05}
\delta g_0 = \epsilon \left( \left. \frac{\delta P(J_{-1},J_0)}{\delta J_{0}}
\right|_{J_0=\frac{dg_0}{dt}g_0^{-1},J_{-1}=g_0t_-g_0^{-1} } \right) g_0.
\ee
To verify that these transformations form indeed a finite $W$-algebra,
we compute the variation of another conserved quantity $Q(g_0)=
Q(J_{-1},J_0)|_{J_0=\frac{dg_0}{dt}g_0^{-1},J_{-1}=g_0t_-g_0^{-1} }$ under
the finite $W$-transformations generated by $P(g_0)$
\be \label{zz04}
\delta Q = \tr \left(
 \left( \frac{d}{dt} \left(
\frac{\delta P}{\delta J_0} \right) +
\left[ \frac{\delta P}{\delta J_0}, J_0 \right] \right)
\frac{\delta Q}{\delta J_0} +
\left[\frac{\delta P}{\delta J_0},J_{-1} \right]
\frac{\delta Q}{\delta J_{-1}} \right).
\ee
Using the exact identity (\ref{zz03a}) and the Toda equations of motion,
we derive that
\be
\frac{d}{dt}\left( \frac{\delta P}{\delta J_0} \right) =
 \left[ \frac{\delta P}{\delta J_{-1}}, J_{-1} \right]
\ee
modulo equations of motion. Inserted into (\ref{zz04}) this yields
\be
\delta Q = \tr \left(  \left[ \frac{\delta P}{\delta J_{-1}}, J_{-1} \right]
  \frac{\delta Q}{\delta J_0} +
\left[ \frac{\delta P}{\delta J_0}, J_0 \right] \frac{\delta Q}{\delta J_0}
+\left[\frac{\delta P}{\delta J_0},J_{-1} \right]
\frac{\delta Q}{\delta J_{-1}} \right).
\ee
This corresponds exactly to the brackets of the finite $W$-algebra, with
$J_{-2}=\cdots=0$. Therefore the symmetry transformations
(\ref{zz05}) form a finite $W$-algebra modulo field equations,
i.e. the algebra is on-shell isomorphic to a finite $W$-algebra.

\subsection{Realizations of finite $W$-algebras in terms of oscillators}

Given a realization of a finite $W$-algebra in terms of a set of oscillators,
or more general in terms of the generators of an algebra ${\cal A}$, one
can try to compute the centralizer of the $W$-algebra in ${\cal A}$, and
take any of the generators of the centralizer as the generator of time
translations in some physical system. For this to make sense one
wants the generator to be Hermitian, so that it can be identified with
the Hamiltonian. Any action whose Hamiltonian is identical to this one
is an action invariant under $W$-transformations, provided the commutation
relations found by canonical quantization agree with those given by the
algebraic structure of ${\cal A}$. For any realization, the center of the
finite $W$-algebra is always a subalgebra of the centralizer in ${\cal A}$,
and these are the first candidates to look at.

As an example, we can take the oscillator realization of $W_3^{(2)}$ in
section~(\ref{aniso}). It is given by
\ba
j_+ & = & \frac{1}{\sqrt{3}} a^2 b^{\dagger},  \nn \\
j_- & = & \frac{-1}{\sqrt{3}} a (b^{\dagger})^2,  \nn \\
j_0 & = & \frac{2}{3}b^{\dagger} b - \frac{2}{3}
a^{\dagger} a,  \nn \\
C & = & -(\frac{1}{3}  a^{\dagger} a + \frac{2}{3} b^{\dagger} b )^2
 -(\frac{1}{3}  a^{\dagger} a + \frac{2}{3} b^{\dagger} b ).
\ea
The center of $W_3^{(2)}$ contains $C$, and the Hamiltonian of the
anisotropic harmonic oscillator is given by $\frac{3}{4}\sqrt{1-4C}$.
Although this is non-polynomial in terms of $C$, it is polynomial in
terms of the oscillators and belongs to the centralizer of the finite
$W$-algebra.  The other generator of the center of $W_3^{(2)}$ is
$C_3=j_0^3+2j_0+3j_0C+3j_+j_-+3j_-j_+$,
 which is equal to $h/3-16h^3/27$ with
$h$ the Hamiltonian. This illustrates the fact that the centralizer
of $W_3^{(2)}$ in the oscillator algebra is generated by $h$\footnote{If
one allows non-polynomial expressions in terms of the oscillators, this
is no longer true, as can be seen from the example $e^{n\pi i a^{\dagger} a}$,
which commutes with all generators of $W_3^{(2)}$.}. It would be interesting
to know whether the example of the anisotropic harmonic oscillator can
somehow be obtained as a reduction of a system with $sl(3)$ symmetry.
This would open the door for the construction of many more examples
of quantum mechanical systems with finite $W$-symmetry.

To conclude, let us mention one more possibility. So far we tried to
find systems with a finite $W$-algebra as its symmetry algebra. We
could also be less restrictive, and demand that it is just a spectrum
generating algebra. For our example of $W_3^{(2)}$, this would mean
that we can also take $j_0$ as our Hamiltonian, since $j_+$ and $j_-$
map $j_0$ eigenstates to $j_0$ eigenstates. Incidentally, an explicit
example where this is the case
is known \cite{VeSh}. Consider a sequence of Schr\"odinger
operators $L_j=A_j^+A_j^-+\lambda_j$ where $A_j^{\pm}=\pm
\frac{d}{dx}+f_j(x)$ and $\lambda_j$ is a constant. If $L_jA_j^+=
A_j^+L_{j+1}$ and $A_j^{-}L_j=L_{j+1}A_j^-$, then the $A_j^{\pm}$ can
be used to map eigenstates of $L_j$ into those of $L_{j+1}$ and
vice versa. Therefore, if we know the spectrum of $L_j$ we also
know that of $L_{j+1}$. This technique to construct new exactly solvable
Schr\"odinger operators from old ones is known as the factorization method.
An interesting situation arises when one imposes a kind of periodic
boundary condition on the chain of operators $L_j$, namely if one
requires $L_{j+N}=L_j+\mu$ and $\lambda_{j+N}=\lambda_j+\mu$ for some
parameter $\mu$. If we denote $L_1$ by $j_0$ and define
\be
j_+=A_1^+\ldots A_N^+, \hspace{1cm} j_-=A_N^- \ldots A_1^-,
\ee
then the following relations hold
\ba \label{zz08}
[j_0,j_+] & =  &  \mu j_+,  \nn \\ \nn
[j_0,j_-] & =  &  -\mu j_-,   \\ \nn
j_+j_- & = & \prod_{k=1}^N (j_0-\lambda_k),  \\
j_-j_+ & = & \prod_{k=1}^N (j_0-\lambda_k+\mu).
\ea
In particular, for $N=3$, the commutator of $j_+$ and $j_-$ will be
a quadratic polynomial in $j_0$, exactly as in the $W_3^{(2)}$ algebra.
The Hamiltonian $j_0$ in (\ref{zz08}) is $-\frac{d^2}{dx^2}+f_1^2+f_1'+
\lambda_1$, where $f_1$ satisfies the following set of coupled differential
equations
\ba \label{zz09}
-f_1'+f_1^2 + \lambda_1 & = & f_2^2 + f_2' + \lambda_2,  \nn \\
-f_2'+f_2^2 + \lambda_2 & = & f_3^2 + f_3' + \lambda_3,  \nn \\
-f_3'+f_3^2 + \lambda_3 & = & f_1^2 + f_1' + \lambda_1+\mu.  \nn \\
\ea
To solve these, first notice that the sum of these three equations
is $2(f_1'+f_2'+f_3')+\mu=0$, from which one derives $f_2+f_3=-f_1-\mu x/2+k$,
with $k$ some integration constant. Next, the second equation in (\ref{zz09})
can be rewritten as $(f_3-f_2)(f_3+f_2)+ (f_3+f_2)' +\lambda_3-\lambda_2=0$,
or equivalently $(f_3-f_2)(-f_1-\mu x/2+k) + (-f_1'-\mu/2) +
\lambda_3-\lambda_2=0$. We thus have one equation for $f_2+f_3$ and one
for $f_2-f_3$ in terms of $f_1$. Solving these for $f_2$ and $f_3$
gives upon substituting these back into the differential equations
(\ref{zz09}) a differential equation for $f_1$, which turns out to
be the Painlev\'e-IV equation. The corresponding potential in $j_0$
is then a one-gap potential. It would be interesting to see what role
the representations of $W_3^{(2)}$ play in the spectrum of $j_0$.

The periodicity conditions for the operators $L_j$ allow for a natural
$q$-deformation \cite{Spir}. The corresponding Schr\"odinger
operators have a spectrum generating algebra which is a $q$-deformation
of (\ref{zz08}), and in particular for $N=3$ one finds a $q$-deformation
of $W_3^{(2)}$. The issue whether one can $q$-deform arbitrary
finite $W$-algebras is an entirely different story, which we will not
discuss here, but it is amusing to see that a $q$-deformation of
$W_3^{(2)}$ can still occur in simple quantum mechanical systems.

\setcounter{equation}{0}
%
\section*{Appendix}
In this appendix we discuss the finite $W$-algebras that can be obtained
from $sl_4$. Using the standard constructions developed in this
chapter we calculate their relations and the quantum Miura maps.

\subsection*{$4= 2+ 1 + 1$}
The basis of $sl_4$ we use to study this quantum algebra is
\be \label{basis211}
r_at_a=\left(\begin{array}{cccc}
\frac{1}{2}r_{10}-\frac{1}{8}r_8 -\frac{1}{8}r_9 & r_{11} &
r_{12} & r_{15} \\
r_5 & \frac{3}{8}r_8-\frac{1}{8}r_9 & r_7 & r_{14} \\
r_4 & r_6 & -\frac{1}{8}r_8+\frac{3}{8}r_9 & r_{13} \\
r_1 & r_2 & r_3 &
-\frac{1}{2}r_{10}-\frac{1}{8}r_8 -\frac{1}{8}r_9
\end{array} \right).
\ee
The $sl_2$ embedding is given by $t_+=t_{15}$, $t_0=t_{10}$ and
$t_-=\hf t_1$. The nilpotent algebra ${\G}_+$ is spanned by
$\{t_{13},t_{14},t_{15}\}$, ${\G}_0$ by $\{t_4,\ldots,t_{12}\}$ and
${\G}_-$ by $\{t_1,t_2,t_3\}$. The $d_1$ cohomology of
$\Omega_{red}$ is generated by $\hj^a$ for $a=1, \ldots ,9$.
Representatives that are exactly $d$-closed, are given by
$W(\hj^a)=\hj^a$ for $a=4,\ldots,9$, and by
\ba \label{reps211}
W(\hj^1) & = & \hj^1 + \hj^4\hj^{12}+\hj^5\hj^{11}+\deel{1}{4}
\hj^{10}\hj^{10}+\deel{\hbar}{2}\hj^{10}, \nn \\ {}
W(\hj^2) & = & \hj^2 + \hj^6\hj^{12} + \hf\hj^8\hj^{11} +
\hf\hj^{10}\hj^{11}+\hbar\hj^{11}, \nn \\ {}
W(\hj^3) & = & \hj^3 + \hj^7\hj^{11} + \hf\hj^9\hj^{12} +
\hf\hj^{12}\hj^{10}+\deel{\hbar}{2}\hj^{12}.
\ea
Introduce a new basis of fields as follows:
\ba \label{newgens}
U & = & \frac{1}{4} (\Whj{8}+\Whj{9}), \nn \\ {}
H & = & \frac{1}{4} (\Whj{8}-\Whj{9}), \nn \\ {}
F & = & -\Whj{7}, \nn \\ {}
E & = & -\Whj{6}, \nn \\ {}
G_1^- & = & -\Whj{3},  \nn \\ {}
G_1^+ & = & \Whj{2}, \nn \\ {}
G_2^- & = & \Whj{5}, \nn \\ {}
G_2^+ & = & \Whj{4}, \nn \\ {}
C & = & \Whj{1}+\hf EF+\hf FE + H^2+\hf U^2 + 2\hbar U.
\ea
If we compute the commutators of these expressions, we find that
$C$ commutes with everything, $\{E,F,H\}$ form a $sl_2$
subalgebra and $G_i^{\pm}$ are spin $\hf$ representations for
this $sl_2$ subalgebra. $U$ represents an extra $u(1)$ charge.
The nonvanishing commutators, with $\hbar$ dependence,  are
\ba \label{alg211}
[E,F] & = & 2\hbar H , \nn \\ {}
[H,E] & = & \hbar E , \nn \\ {}
[H,F] & = & -\hbar F , \nn \\ {}
[U,G_1^{\pm}] & = & \hbar G_1^{\pm}, \nn \\ {}
[U,G_2^{\pm}] & = & -\hbar G_2^{\pm}, \nn \\ {}
[H,G_i^{\pm}] & = & \pm\deel{\hbar}{2} G_i^{\pm}, \nn \\ {}
[E,G_i^-] & = & \hbar G_i^+, \nn \\ {}
[F,G_i^+] & = & \hbar G_i^-, \nn \\ {}
[G_1^+,G_2^+] & = & -2\hbar E(U+\hbar), \nn \\ {}
[G_1^-,G_2^-] & = & 2\hbar F(U+\hbar), \nn \\ {}
[G_1^+,G_2^-] & = & \hbar(-C+EF+FE+2H^2+\deel{3}{2}U^2+2HU)+\hbar^2
(2H+3U), \nn \\ {}
[G_1^-,G_2^+] & = & \hbar(C-EF-FE-2H^2-\deel{3}{2}U^2+2HU)+\hbar^2
(2H-3U).
\ea
Let us also present the quantum Miura transformation for this
algebra. In this case, ${\G}_0=sl_3\oplus u(1)$. Standard
generators of ${\G}_0$ can be easily identified. A generator of
the $u(1)$ is $s=\hf\hj^8+\hf\hj^9+2\hj^{10}$, and the $sl_3$
generators are $e_1=\hj^5$, $e_2=\hj^6$, $e_3=\hj^4$,
$f_1=\hj^{11}$, $f_2=\hj^7$, $f_3=\hj^{12}$,
$h_1=-\hf\hj^8+\hf\hj^{10}$ and $h_2=\hf\hj^8-\hf\hj^9$.
The convention is such that the commutation relations between
$\{e_i,f_i,h_i\}$ are the same as those of corresponding
matrices defined by
\be \label{conv211}
a_i e_i + b_i f_i + c_i
h_i=\hbar\mats{c_1}{a_1}{a_3}{b_1}{c_2-c_1}{a_2}{b_3}{b_2}{-c_2}.
\ee
The quantum Miura transformation reads
\ba \label{miura211}
U & = & \deel{1}{6}(s-2h_2-4h_1), \nn \\ {}
H & = & \hf h_2, \nn \\ {}
F & = & -f_2, \nn \\ {}
E & = & -e_2, \nn \\ {}
G_1^- & =  & -f_2f_1-\deel{1}{3}(s-2h_2-h_1+3\hbar)f_3, \nn \\ {}
G_1^+ & = & e_2f_3+\deel{1}{3}(s+h_2-h_1+3\hbar)f_1, \nn \\ {}
G_2^- & = & e_1, \nn \\ {}
G_2^+ & = & e_3, \nn \\ {}
C & = & (\deel{1}{24}s^2+\deel{\hbar}{2}s) +
\hf(e_1f_1+f_1e_1+e_2f_2+f_2e_2+e_3f_3+f_3e_3) \nn \\ {}
& & +
\deel{1}{3} (h_1^2+h_1h_2+h_2^2).
\ea
In $C$ we again recognize the second Casimir of $sl_3$. It is a
general feature of finite $W$-algebras that they contain a
central element $C$, whose Miura transform contains the second
Casimir of ${\G}_0$. $C$ is the finite counterpart of the energy
momentum tensor that every infinite $W$-algebra possesses.

\subsection*{$4= 2+2$}
A convenient basis to study this case is
\be \label{basis22}
r_at_a=\left(\begin{array}{cccc} \frac{r_6}{4}+\frac{r_{10}}{2}
& \frac{r_5}{2}+\frac{r_8}{2} & r_{12} & r_{14} \\
\frac{r_7}{2}+\frac{r_9}{2} & -\frac{r_6}{4}-\frac{r_{11}}{2} &
r_{13} & r_{15} \\ r_1 & r_2 & \frac{r_6}{4}-\frac{r_{10}}{2} &
\frac{r_5}{2}-\frac{r_8}{2} \\ r_3 & r_4 &
\frac{r_7}{2}-\frac{r_9}{2} & -\frac{r_6}{4}+\frac{r_{11}}{2}
\end{array} \right).
\ee
The $sl_2$ embedding is given by $t_+=t_{12}+t_{15}$,
$t_0=t_{10}-t_{11}$ and $t_-=\hf(t_1+t_4)$. The subalgebra ${\G}_+$
is spanned by $\{t_{12},\ldots,t_{15}\}$, ${\G}_0$ by
$\{t_5,\ldots,t_{11}\}$ and ${\G}_-$ by $\{t_1,\ldots,t_4\}$. The
$d_1$ cohomology of $\Omega_{red}$ is generated by
$\hj^1,\ldots,\hj^7$. The $d$-closed representatives are
$W(\hj^a)=\hj^a$ for $a=5,6,7$, and
\ba \label{reps22}
W(\hj^1) & = & \hj^1-\hv\hj^5\hj^9+\hv\hj^7\hj^8+\hv\hj^8\hj^9+
 \hv\hj^{10}\hj^{10}+\deel{3\hbar}{4}\hj^{10}-\deel{\hbar}{4}
 \hj^{11},\nn \\ {}
 W(\hj^2) & = &
 \hj^2+\hv\hj^5\hj^{10}+\hv\hj^5\hj^{11}-\hv\hj^6\hj^8+
 \hv\hj^8\hj^{10}-\hv\hj^{10}\hj^{11}+\deel{\hbar}{2}\hj^8,
 \nn \\ {}
 W(\hj^3) & = &
 \hj^3+\hv\hj^6\hj^9-\hv\hj^7\hj^{10}-\hv\hj^7\hj^{11}+
 \hv\hj^9\hj^{10}-\hv\hj^9\hj^{11}+\deel{\hbar}{2}\hj^9,
 \nn \\ {}
W(\hj^4) & = &
\hj^4+\hv\hj^5\hj^9-\hv\hj^7\hj^8+\hv\hj^8\hj^9+
\hv\hj^{11}\hj^{11}+\deel{\hbar}{4}\hj^{10}-
\deel{3\hbar}{4}\hj^{11}.
\ea
To display the properties of the algebra as clearly as possible,
we introduce a new basis of fields
\ba \label{newgens22}
H & = & \hf \Whj{6}, \nn \\ {}
E & = & -\Whj{7}, \nn \\ {}
F & = & -\Whj{5}, \nn \\ {}
G^+ & = & \Whj{3}, \nn \\ {}
G^0 & = & \Whj{1}-\Whj{4}, \nn \\ {}
G^- & = & -\Whj{2}, \nn \\ {}
C & = & \Whj{1}+\Whj{4}+\deel{1}{8}\Whj{6}\Whj{6}+
 \hv\Whj{5}\Whj{7} \nn \\ {}
 & & +\hv\Whj{7}\Whj{5}+\deel{\hbar}{4}\Whj{6}.
\ea
Here, $C$ is the by now familiar central element, $\{E,H,F\}$
form an $sl_2$ algebra and $\{G^+,G^0,G^-\}$ form a spin $1$
representation with respect to this $sl_2$ algebra. The
nonvanishing commutators are
\ba \label{alg22}
\nn[E,F] & = & 2\hbar H, \nn \\ {}
[H,E] & = & \hbar E, \nn \\ {}
[H,F] & = & -\hbar F, \nn \\ {}
[E,G^0] & = & 2\hbar G^+, \nn \\ {}
[E,G^-] & = & \hbar G^0, \nn \\ {}
[F,G^+] & = & \hbar G^0, \nn \\ {}
[F,G^0] & = & 2\hbar G^-, \nn \\ {}
[H,G^+] & = & \hbar G^+, \nn \\ {}
[H,G^-] & = & -\hbar G^-, \nn \\ {}
[G^0,G^+] & = & \hbar(-CE+EH^2+\hf EEF+\hf EFE)-2\hbar^3 E,
\nn \\ {}
[G^0,G^-] & = & \hbar(-CF+FH^2+\hf FFE+\hf FEF)-2\hbar^3 F,
\nn \\ {}
[G^+,G^-] & = & \hbar(-CH+H^3+\hf HEF+\hf HFE)-2\hbar^3 H.
\ea
Since ${\G}_0=sl_2\oplus sl_2 \oplus u(1)$, the quantum Miura
transformation expresses this algebra in term of generators
$\{e_1,h_1,f_1\}$, $\{e_2,h_2,f_2\}$, $s$ of ${\G}_0$. The relation
between these generators and the $\hj^a$ are:
$s=\hj^{10}-\hj^{11}$, $h_1=\hf(\hj^6+\hj^{10}+\hj^{11})$,
$h_2=\hf(\hj^6-\hj^{10}-\hj^{11})$, $e_1=\hf(\hj^7+\hj^9)$,
$e_2=\hf(\hj^7-\hj^9)$, $f_1=\hf(\hj^5+\hj^8)$ and
$f_2=\hf(\hj^5-\hj^8)$. The commutation relations for these are
$[e_1,f_1]=\hbar h_1$, $[h_1,e_1]=2\hbar e_1$,
$[h_1,f_1]=-2\hbar f_1$, and similar for $\{e_2,h_2,f_2\}$.
For the quantum Miura transformation one
then finds
\ba \label{miura22}
H & = & \hf(h_1+h_2), \nn \\ {}
E & = & -e_1-e_2, \nn \\ {}
F & = & -f_1-f_2, \nn \\ {}
G^+ & = & \hf e_1h_2-\hf e_2h_1 +\hv s(e_1-e_2)+\hbar(e_1-e_2),
\nn \\ {}
G^0 & = & f_1e_2-f_2e_1+\hv s (h_1-h_2)+\hbar(h_1-h_2), \nn \\ {}
G^- & = & \hf f_1h_2 -\hf f_2 h_1 -\hv s (f_1-f_2)-\hbar
(f_1-f_2), \nn \\ {}
C & = & (\deel{1}{8}s^2+\hbar s) +
\hf(e_1f_1+f_1e_1+e_2f_2+f_2e_2) +
\hv(h_1^2 +h_2^2).
\ea
The infinite-dimensional version of this algebra is one of the
`covariantly coupled' algebras that have been studied in
\cite{tjark2}. The finite algebra (\ref{alg22})
is almost a Lie algebra. If we assign particular values to $C$
and to the second Casimir $C_2=(H^2+\hf EF+\hf FE)$ of the $sl_2$
subalgebra spanned by $\{E,H,F\}$, then (\ref{alg22}) reduces to
a Lie algebra. For a generic choice of the values of $C$ and
$C_2$ this Lie algebra is isomorphic to $sl_2 \oplus sl_2$. An
interesting question is, whether similar phenomena occur
for different covariantly coupled algebras.

\subsection*{$4=3+1$}

The last nontrivial nonprincipal $sl_2$ embedding we consider is
$\underline{4}_4\simeq \underline{3}_2\oplus \underline{1}_2$. We choose yet
another basis
\be \label{basis31}
r_at_a = \left( \begin{array}{cccc}
\frac{r_{5}}{12}-\frac{r_{6}}{3} & r_8 & r_{11} & r_{12} \\
r_4 & -\frac{r_5}{4} & r_{13} & r_{14} \\
\frac{r_3}{2}-\frac{r_9}{2} & r_{10} &
\frac{r_5}{12}+\frac{r_6}{6}+\frac{r_7}{2} & r_{15} \\
r_1 & r_2 & \frac{r_3}{2}+\frac{r_9}{2} &
\frac{r_5}{12}+\frac{r_6}{6}-\frac{r_7}{2}
\end{array} \right),
\ee
in terms of which the $sl_2$ embedding is $t_+=t_{11}+t_{15}$,
$t^0=-3t_6+t_7$ and $t_-=2t_3$. The subalgebra ${\G}_+$ is
generated by $\{t_{11},\ldots,t_{15}\}$, ${\G}_-$ is generated by
$\{t_1,t_2,t_3,t_9,t_{10}\}$, and ${\G}_0$ is generated by
$\{t_4,\ldots,t_8\}$. The $d_1$ cohomology is generated by
$\hj^1,\ldots,\hj^5$, and $d$-closed representatives are given
by $\Whj{4}=\hj^4$, $\Whj{5}=\hj^5$, and
\ba \label{reps31}
\Whj{1} & = & \hj^1+\www{1}{6}{3}{6}-\www{1}{12}{6}{3}
+\www{1}{4}{7}{3}+\hj^4\hj^{10}+\www{1}{4}{6}{9}-\www{1}{4}{7}{9}-
\www{1}{3}{4}{5}\hj^8 \nn \\ {}
& &
-\www{1}{3}{4}{6}\hj^8-\www{1}{108}{6}{6}
\hj^6+\www{1}{12}{6}{7}\hj^7+\www{31\hbar}{12}{4}{8}-
\deel{3\hbar}{4}\hj^9+\www{5\hbar}{48}{6}{6}
\nn \\ {}
& &
+\www{\hbar}{6}{7}{6}-\www{3\hbar}{16}{7}{7}-
\deel{3\hbar^2}{8}\hj^7-\deel{7\hbar^2}{24}\hj^6,
\nn \\ {}
\Whj{2} & = & \hj^2-\www{1}{2}{3}{8}-\www{1}{3}{5}{10}-
\www{1}{6}{6}{10}+\www{1}{2}{7}{10}-\www{1}{2}{8}{9}+
\deel{3\hbar}{2}\hj^{10}+
\www{1}{9}{5}{5}\hj^8 \nn \\ {}
& &
+\www{1}{9}{5}{6}\hj^8+
\www{1}{36}{6}{6}\hj^8-\www{1}{4}{7}{7}\hj^8-
\hbar\hj^5\hj^8-\www{\hbar}{2}{6}{8}-\www{\hbar}{2}{7}{8}+
2\hbar^2\hj^8, \nn \\ {}
\Whj{3} & = & \hj^3 +\hj^4\hj^8+\www{1}{12}{6}{6}+
\www{1}{4}{7}{7}+\deel{\hbar}{2}\hj^7-\deel{\hbar}{2}\hj^6.
\ea
We introduce a new basis
\ba \label{newgens31}
U & = & \deel{1}{4}\Whj{5}, \nn \\ {}
G^+ & = & \Whj{4}, \nn \\ {}
G^- & = & \Whj{2}, \nn \\ {}
S & = & \Whj{1}, \nn \\ {}
C & = &
\Whj{3}+\deel{1}{24}\Whj{5}\Whj{5}-\deel{\hbar}{2}\Whj{5}.
\ea
In this case, the fields are not organized according to $sl_2$
representations, because the centralizer of this $sl_2$
embedding in $sl_4$ does not contain an $sl_2$. Again $C$ is a
central element, and the nonvanishing commutators are
\ba \label{alg31}
[U,G^+] & = & \hbar G^+ , \nn \\ {}
[U,G^-] & = & -\hbar G^- , \nn \\ {}
[S,G^+] & = & \hbar G^+ (-\deel{2}{3} C+\deel{20}{9}U^2-
\deel{43\hbar}{9}U+\deel{29\hbar^2}{27}), \nn \\ {}
[S,G^-] & = & \hbar  (\deel{2}{3} C-\deel{20}{9}U^2+
\deel{43\hbar}{9}U-\deel{29\hbar^2}{27})G^-, \nn \\ {}
[G^+,G^-] & = & \hbar S-\deel{4\hbar}{3}CU+\deel{3\hbar^2}{4}C
    +\deel{88\hbar}{27}U^3-\deel{17\hbar^2}{2}U^2+
    \deel{25\hbar^2}{6}U.
\ea
This is the first example where the brackets are no longer
quadratic, but contain third order terms.
For the sake of completeness, let us also give the quantum Miura
transformation for this algebra. We identify generators
$\{e,f,h\}$,$s_1$,$s_2$ of $sl_2\oplus u(1) \oplus u(1)={\G}_0$ via
$f=\hj^8$, $e=\hj^4$, $h=\deel{1}{3}(\hj^5-\hj^6)$,
$s_1=\deel{1}{3}(2\hj^6+\hj^5)$ and $s_2=\hj^7$. The only
nontrivial commutators between these five generators are
$[e,f]=\hbar h$, $[h,e]=2\hbar e$ and $[h,f]=-2\hbar f$.
The quantum
Miura transformation now reads
\ba \label{miura31}
U & = & \deel{1}{4}s_1+\deel{1}{2}h, \nn \\ {}
G^+ & = & e, \nn \\ {}
G^- & = & (\deel{1}{4}s_1^2+\deel{1}{2}s_1 h + \deel{1}{4}h^2
-\deel{1}{4}s_2^2-\deel{\hbar}{2}(3s_1+3h+s_2)+2\hbar^2)f,
\nn \\ {}
S & = & -\deel{1}{12} e(8s_1+4h-31\hbar)f-\deel{1}{108}
(s_1-h)^3+\deel{1}{12}(s_1-h)s_2^2+\deel{5\hbar}{48}(s_1-h)^2
\nn \\ {}
& & + \deel{\hbar}{6}s_2(s_1-h)-\deel{3\hbar}{16}s_2^2-
\deel{3\hbar^2}{8}s_2-\deel{7\hbar^2}{24}(s_1-h),
\nn \\ {}
C & = & (\deel{1}{2}ef+\deel{1}{2}fe+\deel{1}{4}h^2)+
(\deel{1}{4}s_2^2+\deel{\hbar}{2}s_2)+
(\deel{1}{8}s_1^2-\hbar s_1).
\ea
This completes our list of finite quantum $W$-algebras from
$sl_4$.

\noindent
ACKNOWLEDGEMENTS

We would like to thank CERN for hospitality while part of this work was
being done. JdB is sponsored in part by NSF grant no.PHY9309888.


\end{document}